\documentclass[a4paper,11pt]{article}
\pdfoutput=1 
\date{}
\usepackage{jheppub}
\usepackage[utf8]{inputenc}
\usepackage[english]{babel}
\usepackage{amsmath,amssymb,amsfonts,amsxtra, mathrsfs,graphics,graphicx,amsthm,epsfig, youngtab,bm,longtable,float,tikz,empheq}
\usepackage[margin=0.5cm]{caption}
\usepackage{thmtools}

\allowdisplaybreaks[1]
\usetikzlibrary{positioning}
\usetikzlibrary{automata}
\usetikzlibrary{arrows}
\usetikzlibrary{calc}
\usetikzlibrary{decorations.markings}
\usetikzlibrary{decorations.pathreplacing}
\usetikzlibrary{intersections}
\usetikzlibrary{positioning}
\usetikzlibrary{topaths}
\usetikzlibrary{shapes.geometric}
\usetikzlibrary{shapes.misc}
\tikzset{cf-group/.style = {
    shape = rounded rectangle, minimum size=1.0cm,
    rotate=90,
    rounded rectangle right arc = none,
    draw}}
\tikzset{cross/.style={path picture={ 
  \draw[black]
(path picture bounding box.south east) -- (path picture bounding box.north west) (path picture bounding box.south west) -- (path picture bounding box.north east);
}}}

\tikzset{unode/.style=
{black, circle,draw,thick,fill=black!100 ,minimum size=1mm}}

\tikzset{sunode/.style=
{black, circle,draw,thick,fill=yellow!100 ,minimum size=1mm}}

\tikzset{fnode/.style=
{black, rectangle,draw,thick,minimum size=1mm}}

\tikzset{afnode/.style=
{blue,rectangle,draw,thick,minimum size=1mm}}

\tikzset{cross/.style={cross out, draw=black, minimum size=4mm, inner sep=0pt, outer sep=0pt}}

\newcommand{\be}{\begin{equation}}
\newcommand{\ee}{\end{equation}}
\newcommand{\ba}{\begin{array}}
\newcommand{\ea}{\end{array}} 
\newcommand{\bi}{\begin{itemize}}
\newcommand{\ei}{\end{itemize}}
\def\vec#1{\bm{#1}}
\def\bea#1\eea{\allowdisplaybreaMs \begin{align}#1\end{align}}
 \newcommand{\ben}{\begin{enumerate}}
\newcommand{\een}{\end{enumerate}}
\newcommand{\bean}{\begin{eqnarray*}}
\newcommand{\eean}{\end{eqnarray*}}
\newcommand{\eref}[1]{(\ref{#1})}

\newcommand{\nn}{\nonumber}

\newcommand{\tr}{\mathrm{Tr}}

\newcommand{\BZ}{\mathbb{Z}}

\newcommand{\comment}[1]{}

\newcommand{\CT}{{\cal T}}
\newcommand{\CD}{{\cal D}}
\newcommand{\CM}{{\cal M}}
\newcommand{\CO}{{\cal O}}

\newcommand{\CN}{{\cal N}}

\newcommand{\CX}{{\cal X}}

\newcommand{\frsu}{\mathfrak{su}}
\newcommand{\fru}{\mathfrak{u}}

\newcommand{\frg}{\mathfrak{g}}

\newcommand{\wt}{\widetilde}

\newcommand{\ch}{\cosh \pi}
\newcommand{\s}{\sigma}

\newcommand{\Secref}[1]{Section~\ref{#1}}

\newcommand{\Appref}[1]{Appendix~\ref{#1}}

\newcommand{\figref}[1]{Fig.~\ref{#1}}
\renewcommand{\eqref}[1]{(\ref{#1})}

\title{Exploring Seiberg-like N-alities with Eight Supercharges}
\author{Anindya Dey}
\affiliation{Department of Physics and Astronomy, Johns Hopkins University, 3400 North Charles Street,
Baltimore, MD 21218, USA}
\emailAdd{anindya.hepth@gmail.com}
\abstract{We show that a large subclass of 3d $\mathcal{N}=4$ quiver gauge theories consisting of unitary and special unitary gauge nodes 
with only fundamental/bifundamental matter have multiple Seiberg-like IR duals. 
A generic quiver $\mathcal{T}$ in this subclass has a non-zero number of balanced special unitary gauge nodes and it is 
a good theory in the Gaiotto-Witten sense. We refer to this phenomenon as \textit{IR N-ality} and the set of mutually IR dual theories 
as the \textit{N-al set} associated with the quiver $\mathcal{T}$. 
Starting from $\mathcal{T}$, we construct a sequence of dualities by step-wise implementing a set 
of quiver mutations which act locally on the gauge nodes. The associated N-al theories can then be read off 
from this duality sequence.
The quiver $\mathcal{T}$ generically has an emergent Coulomb branch global symmetry in the IR, such that the rank of the IR symmetry is 
always greater than the rank visible in the UV. We show that there exists at least one theory in the 
N-al set for which the rank of the IR symmetry precisely matches the rank of the UV symmetry. 
In certain special cases that we discuss in this work, the correct emergent symmetry algebra itself may be read off from this preferred theory (or theories)
in addition to the correct rank. Finally, we give a recipe for constructing the 3d mirror associated with a given N-al set and show 
how the emergent Coulomb branch symmetry of $\mathcal{T}$ is realized as a UV-manifest Higgs branch symmetry 
of the 3d mirror. This paper is the second in a series of four papers on 3d $\mathcal{N}=4$ Seiberg-like dualities, preceded by the work
\cite{Dey:2022pqr}.}

\begin{document}
\maketitle

\section{A brief summary of the paper}

\subsection{Background}

UV/IR dualities in $d\leq 6$ space-time dimensions provide one of the most powerful tools for probing non-perturbative/strongly-coupled 
physics in Quantum Field Theories. Seiberg duality \cite{Seiberg:1994pq} is one of the classic examples of an IR duality for a pair of $\CN=1$ gauge 
theories in four dimensions. There are dualities in lower space-time dimensions that share certain broad features of the Seiberg duality 
and are often referred to as \textit{Seiberg-like dualities}. In $d=3$, a rich family of such dualities exist for $\CN \geq 2$ gauge theories with 
Chern-Simons levels \cite{Aharony:1997gp, Giveon:2008zn}.

Seiberg-like dualities in 3d $\CN=4$ theories without any Chern-Simons terms have been far less studied in the literature. The known 
examples involve theories which are ugly and bad \cite{Kapustin:2010mh, Yaakov:2013fza} in the Gaiotto-Witten sense \cite{Gaiotto:2008ak}, and in the latter case the dual should be understood as a low energy effective field theory on a certain singular loci of the moduli space of the original theory \cite{Assel:2017jgo}.

Recently, a family of Seiberg-like dualities were proposed for pairs of 3d $\CN=4$ gauge theories \cite{Dey:2022pqr} where both theories 
are good in the Gaiotto-Witten sense. The theory on one side of the duality is a $U(N)$ SQCD with $N_f\geq 2N-1$ fundamental 
hypermultiplets and $P \geq 1$ hypermultiplets in the determinant representation of $U(N)$. The latter hypermultiplets 
are referred to as Abelian hypermultiplets, since they are only charged under the central $U(1)$ subgroup of the 
$U(N)$ gauge group. These theories will be denoted as $\CT^N_{N_f, P}$ and the IR duality associated with 
the theory $\CT^N_{N_f, P}$ will be denoted as $\CD^N_{N_f,P}$. The quiver representation of $\CT^N_{N_f, P}$
is given below (the reader is referred to \Secref{quiv-notation} for details on the quiver notation): 
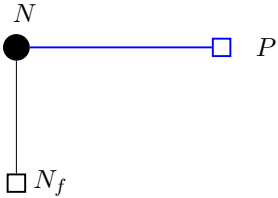
\begin{center}
\scalebox{.9}{\begin{tikzpicture}
\node[unode] (1) at (0,0){};
\node[fnode] (2) at (0,-2){};
\node[afnode] (3) at (3,0){};
\draw[-] (1) -- (2);
\draw[-, thick, blue] (1)-- (3);
\node[text width=0.1cm](10) at (0, 0.5){$N$};
\node[text width=1.5cm](11) at (1, -2){$N_f$};
\node[text width=1cm](12) at (4, 0){$P$};
\end{tikzpicture}}
\end{center}

One of the main results of \cite{Dey:2022pqr} was to show that the dualities $\CD^N_{N_f,P}$ are non-trivial in the following 
parameter ranges of $N_f$ and $P$ for a fixed $N$: $(N_f=2N+1, P=1)$, $(N_f=2N, P\geq 1)$ and $(N_f=2N-1, P\geq 1)$. 
These dualities are reviewed in \Secref{Duality-rev} and summarized in Table \ref{Tab: Review}. Here, we list the 
gauge groups and the matter content of the dual pairs in each case:

\begin{itemize}

\item \textbf{Duality $\CD^N_{2N+1,1}$:} The dual pair involves the theory $\CT^N_{2N+1,1}$ and 
an $SU(N+1)$ theory with $N_f=2N+1$ fundamental flavors. 

\item \textbf{Duality $\CD^N_{2N,P}$:} This is given by the self-duality of the theory $\CT=\CT^N_{2N,P}$.

\item \textbf{Duality $\CD^N_{2N-1,P}$:} The dual pair involves the theory $\CT^N_{2N-1,P}$  and a $U(1) \times U(N-1)$ gauge theory with $1$ and $2N-1$ fundamental hypers respectively, and $P$ Abelian hypermultiplet with charges $(1, -(N-1))$ under the $U(1) \times U(N-1)$ gauge group.

\end{itemize}

Given the dualities $\CD^N_{N_f,P}$ for the parameter ranges mentioned above, one can use them to construct 
IR dualities involving more complicated quiver gauge theories. Starting from an appropriate quiver gauge theory $\CT$
that we will specify momentarily, one can implement these dualities locally at different gauge nodes in a step-wise 
fashion. This generates a duality sequence and an associated set of quiver gauge theories that are all IR dual to the original theory. 
We will call this phenomenon an \textit{IR N-ality} and the set of theories IR dual to each other as an \textit{N-al set} which includes 
$\CT$. The local operations at the gauge nodes will be referred to as \textit{quiver mutations} 
in the spirit of similar operations for Seiberg duality. 

In this paper, we will study IR N-ality associated with a quiver $\CT$ of the following type.
The quiver $\CT$ has both unitary as well as special unitary gauge nodes with only fundamental/bifundamental matter,
such that the number of balanced special unitary gauge nodes in the quiver is non-zero\footnote{We will define the balance parameter 
of an $SU(N_\alpha)$ gauge node as $e^{(SU)}_\alpha = N^{\rm fund}_\alpha + N^{\rm bif}_\alpha - (2N_\alpha-1)$, where $N^{\rm fund}_\alpha, 
N^{\rm bif}_\alpha$ are fundamental and bifundamental hypers associated with the node. Therefore, an 
$SU(N_\alpha)$ gauge node is balanced if $N^{\rm fund}_\alpha + N^{\rm bif}_\alpha = (2N_\alpha-1)$. In contrast, we will 
define the balance parameter of an $U(N_\alpha)$ gauge node as $e^{(U)}_\alpha = N^{\rm fund}_\alpha + N^{\rm bif}_\alpha - 2N_\alpha$.
}. 
We will demand in addition that the quiver $\CT$ be good in the 
Gaiotto-Witten sense. We will refer to this class of quivers as class $\CX$. With this quiver $\CT$ in class $\CX$ as a starting point, 
we will implement the allowed quiver mutations to generate the duality sequence and read off the N-al set of quivers. 
It turns out that a generic theory in the N-al set does not belong to class $\CX$. It is still a quiver gauge theory with unitary and 
special unitary gauge nodes, but the matter sector involves 
a non-zero number of Abelian hypermultiplets in addition to the fundamental/bifundamental hypermultiplets.  
We will present a general recipe for constructing the IR N-ality, and explicitly work out the N-al set for a subclass of 
quivers in $\CX$ i.e. linear quivers with a single special unitary gauge node. We will also give a recipe for constructing the 
3d mirror associated with the N-al set of a generic $\CT \subset \CX$ and determine it explicitly for the examples studied in this paper.

In principle, one may start from any theory in the N-al set and proceed to generate the others by implementing appropriate 
quiver mutations. However, the quiver $\CT \subset \CX$ is a natural starting point in the following sense. A generic quiver 
in an N-al set containing $\CT$ has an emergent Coulomb branch global symmetry in the IR such that the rank of the IR 
symmetry is always greater than the rank manifest in the UV. The quiver $\CT$ has the minimum UV-manifest rank in 
the N-al set that it belongs to. It turns out that the duality sequence constructed with $\CT$ as the starting point 
leads to at least one theory for which the UV-manifest rank matches the rank of the emergent IR symmetry. For a large class of examples, which 
includes the quiver families discussed in this paper, the full emergent symmetry algebra can be read off from this special quiver (or quivers)
belonging to the N-al set. \\

The paper is organized in the following fashion. In the rest of this section, we summarize the main results of the paper. 
In \Secref{recipe-gen}, we present the general recipe for constructing the N-al set starting from the theory $\CT$, after briefly 
reviewing the quiver notation and the dualities $\CD^N_{N_f,P}$. In \Secref{N-al-2node}, we implement this procedure and explicitly work out the N-al set for the case where $\CT$ is a linear quiver with a single balanced special unitary gauge node and a single unitary gauge node. 
We extend the construction to the case where $\CT$ is a linear quiver with three gauge nodes -- a single 
balanced special unitary node and two unitary gauge nodes, in \Secref{N-al-3node}. 
In \Secref{3dmirr} we give a recipe for constructing the 3d mirror of a given 
N-al set and work out several examples. The appendices contain various computational details of the results that 
appear in the main text.

\subsection{Summary of the main results}

\subsection*{Systematics of N-ality: Recipe and Examples}

One of the main results of this paper is to present a systematic procedure for constructing the N-al set of theories given a 
quiver $\CT$ in class $\CX$, and implement the procedure in simple examples where $\CT$ is a linear quiver involving a single 
balanced special unitary node. The general construction is discussed in \Secref{recipe-gen0}. 
As mentioned earlier, the theories in the N-al set can be read off from a duality sequence generated 
by the action of certain quiver mutations on $\CT$. The allowed quiver mutations, each of which arises from one of the dualities 
listed above (and reviewed in \Secref{Duality-rev}) are classified in \Secref{recipe-gen0}. 

The first example, discussed in \Secref{N-al-2node}, involves the following 3-parameter family of linear quivers:

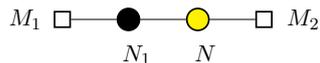
\begin{center}
\scalebox{0.8}{\begin{tikzpicture}
\node[fnode] (1) {};
\node[unode] (2) [right=.75cm  of 1]{};
\node[sunode] (3) [right=.75cm of 2]{};
\node[fnode] (4) [right=0.75 cm of 3]{};
\draw[-] (1) -- (2);
\draw[-] (2)-- (3);
\draw[-] (3) -- (4);
\node[text width=.1cm](10) [left=0.5 cm of 1]{$M_1$};
\node[text width=.2cm](11) [below=0.1cm of 2]{$N_1$};
\node[text width=.1cm](12) [below=0.1cm of 3]{$N$};
\node[text width=.1cm](13) [right=0.1cm of 4]{$M_2$};
\end{tikzpicture}}
\end{center}

The above is a linear quiver with a $U(N_1) \times SU(N)$ gauge group. The $SU(N)$ gauge node is 
balanced which implies that the integers $(M_1, M_2, N_1,N)$ satisfy the constraint $N_1 + M_2 =2N-1$, 
and $U(N_1)$ node is either overbalanced or balanced i.e. $M_1 + N = 2N_1 + e$ with $e \geq 0$. The 
N-ality and the associated duality sequence depends crucially on the specific value of $e$. We consider 
the two following sub-cases:

\begin{enumerate}

\item \textbf{Overbalanced unitary node ($e \neq 0$):} The N-ality and the duality sequence is discussed in \Secref{Unbal-2node}. For 
any $e>2$, the N-ality is simply a duality given in \figref{IRdual-Ex6a}. For $e=1$, one obtains a triality shown in 
\figref{IRdual-Ex6b}, while for $e=2$ one also obtains a different triality shown in \figref{IRdual-Ex6c}.

\item  \textbf{Balanced unitary node ($e = 0$):} The N-ality, discussed in \Secref{Bal-2node}, is yet another triality, and 
the associated duality sequence is given in \figref{IRdual-Ex6d}.

\end{enumerate}

The second example, discussed in \Secref{N-al-3node}, involves the following 4-parameter family of linear quivers:
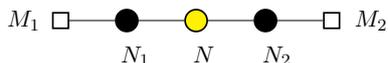
\begin{center}
\scalebox{0.8}{\begin{tikzpicture}
\node[fnode] (1) {};
\node[unode] (2) [right=.75cm  of 1]{};
\node[sunode] (3) [right=.75cm of 2]{};
\node[unode] (4) [right=0.75 cm of 3]{};
\node[fnode] (5) [right=0.75 cm of 4]{};
\draw[-] (1) -- (2);
\draw[-] (2)-- (3);
\draw[-] (3) -- (4);
\draw[-] (4) -- (5);
\node[text width=.1cm](10) [left=0.5 cm of 1]{$M_1$};
\node[text width=.2cm](11) [below=0.1cm of 2]{$N_1$};
\node[text width=.1cm](12) [below=0.1cm of 3]{$N$};
\node[text width=.1cm](13) [below=0.1cm of 4]{$N_2$};
\node[text width=.1cm](14) [right=0.1cm of 5]{$M_2$};
\end{tikzpicture}}
\end{center}

The above is a linear quiver with a $U(N_1) \times SU(N) \times U(N_2)$ gauge group. 
The $SU(N)$ gauge node is balanced i.e. $N_1 + N_2=2N-1$, while the unitary gauge nodes can be 
either balanced or overbalanced with $M_1 + N =2N_1 +e_1$ and $M_2 + N =2N_2 +e_2$ such 
that $e_1, e_2 \geq 0$. The N-ality and the associated duality sequence depends crucially on the specific 
values of the doublet of in integers $(e_1, e_2)$ as follows: 
\begin{enumerate}

\item  \textbf{Overbalanced unitary nodes ($e_1 \neq 0, e_2 \neq 0$):} For $e_1 > 2, e_2 >2$, the N-ality 
is simply a duality as shown in \figref{IRdual-Ex4a} with details discussed in \Secref{Unbal-3node}. The special 
cases of $e_1, e_2=1,2$ can be worked out in a fashion similar to the 2-node quiver.

\item  \textbf{ A single balanced node ($e_1 = 0, e_2 \neq 0$):}  For $e_2 >2$, the N-ality 
is a triality shown in \figref{IRdual-Ex4b} with details discussed in \Secref{Bal1-3node}. The special cases 
of $e_2=1,2$ can again be worked out in a fashion similar to the 2-node quiver.

\item  \textbf{Balanced unitary nodes ($e_1 = 0, e_2 = 0$):} The N-ality in this case, discussed in \Secref{Bal2-3node}, 
is a hexality with the associated duality sequence given in \figref{IRdual-Ex4c}. 

\end{enumerate}

In each case listed above, the quiver mutations may be implemented in terms of the sphere partition function.  
For the 2-node and the 3-node quiver, the relevant computational details can be found in \Appref{2node-pf} 
and \Appref{3node-pf} respectively.\\

\subsection*{Emergent global symmetry and N-ality}

We mentioned earlier that a quiver $\CT$ in class $\CX$ generically has an emergent Coulomb branch symmetry 
in the IR.  The second main result of this paper is to show that there exists at least a single quiver in the N-al set 
for which the rank of this emergent symmetry of $\CT$ becomes manifest.
In the special case where $\CT$ is a linear quiver, one can in fact read off the correct symmetry algebra itself from the 
aforementioned quiver (or quivers). From the examples presented in this paper, we will explicitly confirm this fact. 

Consider first the case of the 2-node quiver discussed above. The Coulomb branch symmetry manifest in the UV corresponds 
to the $\fru(1)$ topological symmetry associated with the $U(N_1)$ gauge node. The IR emergent symmetry 
of the theory is discussed in \Secref{MO-2node} and may be summarized as follows:

\begin{enumerate}

\item \textbf{Overbalanced unitary node ($e \neq 0$):} In this case $\frg_{\rm C} = \fru(1) \oplus \fru(1)$, where the two $\fru(1)$ factors correspond to the topological symmetry for the $U(N_1)$ gauge node and a monopole operator associated with the $SU(N)$ gauge node. 

\item \textbf{Balanced unitary node ($e=0$):} In this case $\frg_{\rm C} =\frsu(2) \oplus \frsu(2) \oplus \fru(1)$, where the origin of the various factors in terms of the monopole operators and the topological symmetry generator is explained in \Secref{MO-2node}. 

\end{enumerate}

\begin{figure}
\begin{center}
\begin{tabular}{ccccc}
\scalebox{0.6}{\begin{tikzpicture}
\node[fnode] (1) at (0,0){};
\node[unode] (2) at (2,0){};
\node[sunode] (3) at (4,0){};
\node[fnode] (4) at (6,0){};
\draw[-] (1) -- (2);
\draw[-] (2)-- (3);
\draw[-] (3) -- (4);
\node[text width=.1cm](20) [left=0.5 cm of 1]{$M_1$};
\node[text width=.2cm](21) [below=0.1cm of 2]{$N_1$};
\node[text width=.1cm](22) [below=0.1cm of 3]{$N$};
\node[text width=.1cm](23) [right=0.1cm of 4]{$M_2$};
\node[text width=.2cm](20) at (4,-2){$(\CT)$};
\end{tikzpicture}}
&\qquad
&\scalebox{0.6}{\begin{tikzpicture}
\node[fnode] (1) at (0,0){};
\node[unode] (2) at (2,0){};
\node[unode] (3) at (4,0){};
\node[fnode] (4) at (6,0){};
\draw[-] (1) -- (2);
\draw[-] (2)-- (3);
\draw[-] (3) -- (4);
\draw[-, thick, blue] (2)--(2,1.5);
\draw[-, thick, blue] (3)--(4,1.5);
\draw[-, thick, blue] (2,1.5)--(4,1.5);
\node[text width=2 cm](10) at (3, 2){\footnotesize{$(N_1, -N+1)$}};
\node[text width=.1cm](20) [left=0.5 cm of 1]{$M_1$};
\node[text width=0.5 cm](21) [below=0.1cm of 2]{$N_1$};
\node[text width=1 cm](22) [below=0.1cm of 3]{$N-1$};
\node[text width=.1cm](23) [right=0.1cm of 4]{$M_2$};
\node[text width=.2cm](20) at (4,-2){$(\CT^\vee_1)$};
\end{tikzpicture}}
&\qquad
&\scalebox{0.6}{\begin{tikzpicture}
\node[fnode] (1) at (0,0){};
\node[unode] (2) at (2,0){};
\node[unode] (3) at (4,0){};
\node[fnode] (4) at (6,0){};
\node[unode] (6) at (2,3){};
\node[fnode] (7) at (0,3){};
\node[](8) at (2, 1.5){};
\draw[-] (1) -- (2);
\draw[-] (2)-- (3);
\draw[-] (3) -- (4);
\draw[-] (6) --(7);
\draw[-, thick, blue] (2)--(6);
\node[text width=.1cm](20) [left=0.5 cm of 1]{$M_1$};
\node[text width=2 cm](21) [below=0.1cm of 2]{$N_1-1$};
\node[text width=2 cm](22) [below=0.1cm of 3]{$N-1$};
\node[text width=.1cm](23) [right=0.1cm of 4]{$M_2$};
\node[text width=.1cm](24) [right=0.1cm of 6]{1};
\node[text width=.1cm](25) [left=0.5 cm of 7]{1};
\node[text width=2 cm](26) [right=0.1 cm of 8]{\footnotesize{$(1, -N_1+1)$}};
\node[text width=.2cm](40) [below=1cm of 2]{$(\CT^\vee_2)$};
\end{tikzpicture}}
\end{tabular}
\end{center}
\caption{\footnotesize{Triality of quivers for the case of a balanced unitary gauge node in the 2-node quiver.}}
\label{triality-1}
\end{figure}
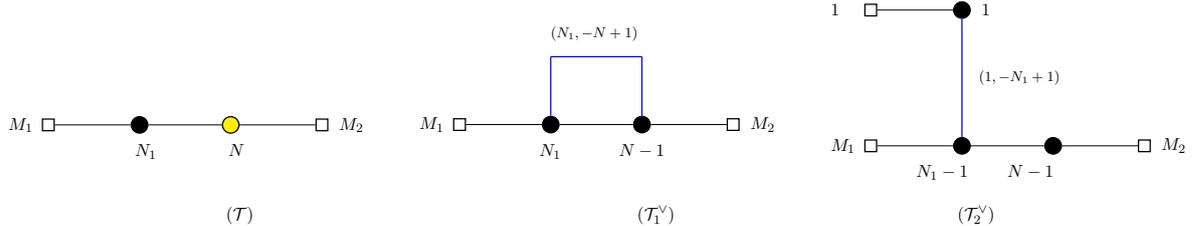

For the case of the balanced unitary node i.e. $e=0$, the duality sequence in \figref{IRdual-Ex6d} leads to a triality of quivers. 
The quivers in the N-al set (with N=3) are given in \figref{triality-1}. The quiver $(\CT^\vee_2)$ is the theory at which the duality sequence terminates. Firstly, note that the rank of the Coulomb branch symmetry for $(\CT^\vee_2)$ manifest in the UV precisely matches that of $\frg_{\rm C}$ for the theory $\CT$ . 
Secondly, one may read off $\frg_{\rm C}$ itself from the quiver $(\CT^\vee_2)$ using the following argument. Recall that in a good linear 
quiver with unitary gauge nodes, an array of $k$ balanced nodes contributes an $\frsu(k+1)$ factor to the Coulomb branch symmetry 
while every overbalanced node contributes a $\fru(1)$ factor \cite{Gaiotto:2008ak}. Given a generic quiver with unitary gauge nodes, which can be constructed using a set of linear quivers glued together by gauging flavor symmetries, the Coulomb branch symmetry is expected to be of the form: $\frg_{\rm C} = \oplus_i \,\frg^i_{\rm C}$, where $i$ labels the constituent linear quivers and $\frg^i_{\rm C}$ is the Coulomb branch symmetry of the $i$-th linear quiver. Applying this argument to $(\CT^\vee_2)$, where the $U(1)$ and the $U(N-1)$ gauge nodes are balanced while the $U(N_1-1)$ node is overbalanced, 
one can read off the Coulomb branch symmetry as $\frg_{\rm C}=\frsu(2) \oplus \frsu(2) \oplus \fru(1)$.

For the case $e \neq 0$, the emergent IR symmetry can be similarly read off from the duality sequence associated with the given value 
of $e$. We refer the reader to \Secref{Unbal-2node} for further details. \\

Consider the case of the 3-node quiver next. The Coulomb branch symmetry manifest in the UV corresponds 
to the $\fru(1)\oplus \fru(1)$ topological symmetry associated with the $U(N_1)$ and the $U(N_2)$ gauge nodes respectively. 
The IR emergent symmetry of the theory is discussed in \Secref{MO-3node} and may be summarized as follows:

\begin{enumerate}

\item \textbf{Overbalanced unitary nodes ($e_1 \neq 0, e_2 \neq 0$):}  In this case $\frg_{\rm C}= \fru(1)\oplus \fru(1)\oplus \fru(1)$, 
where two of the $\fru(1)$ factors correspond to the topological symmetries of the unitary gauge nodes and the third arises from 
a monopole operator charged under the $SU(N)$ gauge node. 

\item \textbf{ A single balanced node ($e_1 = 0, e_2 \neq 0$):}  In this case $\frg_{\rm C}=\frsu(2) \oplus \frsu(2) \oplus \fru(1)\oplus \fru(1)$, 
where the origin of the various factors in terms of monopole operators and topological symmetry generators is discussed in \Secref{MO-3node}.

\item \textbf{Balanced unitary nodes ($e_1 = 0, e_2 = 0$):}  In this case, there is a further enhancement of the global symmetry with 
$\frg_{\rm C}= \frsu(2) \oplus \frsu(2) \oplus \frsu(4) \oplus \fru(1)$, as discussed in \Secref{MO-3node}.

\end{enumerate}

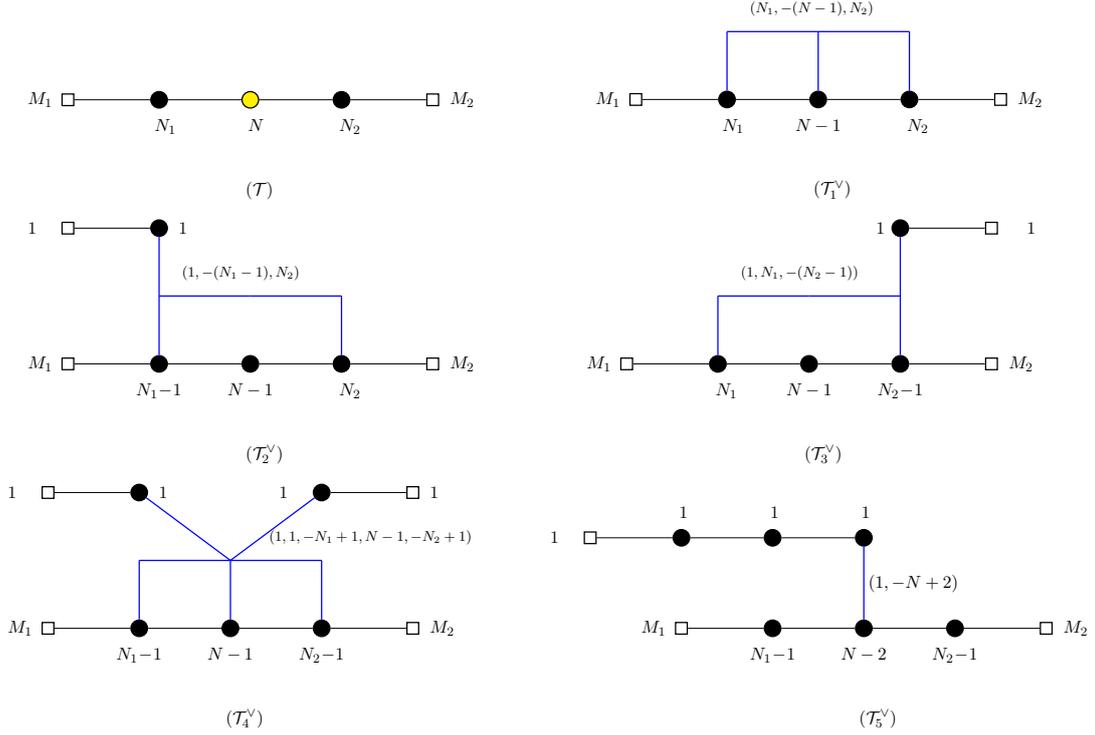
\begin{figure}
\begin{center}
\begin{tabular}{ccc}
\scalebox{0.6}{\begin{tikzpicture}
\node[fnode] (1) at (0,0){};
\node[unode] (2) at (2,0){};
\node[sunode] (3) at (4,0){};
\node[unode] (4) at (6,0){};
\node[fnode] (5) at (8,0){};
\draw[-] (1) -- (2);
\draw[-] (2)-- (3);
\draw[-] (3) -- (4);
\draw[-] (4) --(5);
\node[text width=.1cm](10) [left=0.5 cm of 1]{$M_1$};
\node[text width=.2cm](11) [below=0.1cm of 2]{$N_1$};
\node[text width=.1cm](12) [below=0.1cm of 3]{$N$};
\node[text width=.1cm](13) [below=0.1cm of 4]{$N_2$};
\node[text width=.1cm](14) [right=0.1cm of 5]{$M_2$};
\node[text width=.2cm](20) at (4,-2){$(\CT)$};
\end{tikzpicture}}
& \qquad
&\scalebox{0.6}{\begin{tikzpicture}
\node[fnode] (1) at (0,0){};
\node[unode] (2) at (2,0){};
\node[unode] (3) at (4,0){};
\node[unode] (4) at (6,0){};
\node[fnode] (5) at (8,0){};
\draw[-] (1) -- (2);
\draw[-] (2)-- (3);
\draw[-] (3) -- (4);
\draw[-] (4) --(5);
\draw[-, thick, blue] (2)--(2,1.5);
\draw[-, thick, blue] (3)--(4,1.5);
\draw[-, thick, blue] (4)--(6,1.5);
\draw[-, thick, blue] (2,1.5)--(4,1.5);
\draw[-, thick, blue] (4,1.5)--(6,1.5);
\node[text width=3 cm](10) at (4, 2){\footnotesize{$(N_1, -(N-1), N_2)$}};
\node[text width=.1cm](10) [left=0.5 cm of 1]{$M_1$};
\node[text width=.2cm](11) [below=0.1cm of 2]{$N_1$};
\node[text width= 1cm](12) [below=0.1cm of 3]{$N-1$};
\node[text width=.1cm](13) [below=0.1cm of 4]{$N_2$};
\node[text width=.1cm](14) [right=0.1cm of 5]{$M_2$};
\node[text width=.2cm](20) at (4,-2){$(\CT^\vee_1)$};
\end{tikzpicture}}\\
\scalebox{0.6}{\begin{tikzpicture}
\node[fnode] (1) at (0,0){};
\node[unode] (2) at (2,0){};
\node[unode] (3) at (4,0){};
\node[unode] (4) at (6,0){};
\node[fnode] (5) at (8,0){};
\node[unode] (6) at (2,3){};
\node[fnode] (7) at (0,3){};
\draw[-] (1) -- (2);
\draw[-] (2)-- (3);
\draw[-] (3) -- (4);
\draw[-] (4) --(5);
\draw[-] (6) --(7);
\draw[-, thick, blue] (2)--(6);
\draw[-, thick, blue] (4)--(6,1.5);
\draw[-, thick, blue] (2,1.5)--(4,1.5);
\draw[-, thick, blue] (4,1.5)--(6,1.5);
\node[text width=3 cm](40) at (4, 2){\footnotesize{$(1, -(N_1-1), N_2)$}};
\node[text width=.1cm](20) [left=0.5 cm of 1]{$M_1$};
\node[text width=1 cm](21) [below=0.1cm of 2]{$N_1-1$};
\node[text width=1 cm](22) [below=0.1cm of 3]{$N-1$};
\node[text width=0.1 cm](23) [below=0.1cm of 4]{$N_2$};
\node[text width=.1cm](24) [right=0.1cm of 5]{$M_2$};
\node[text width=.1cm](25) [right=0.1cm of 6]{1};
\node[text width=.1cm](26) [left=0.5 cm of 7]{1};
\node[text width=.2cm](20) at (4,-2){$(\CT^\vee_2)$};
\end{tikzpicture}}
&\qquad
& \scalebox{0.6}{\begin{tikzpicture}
\node[fnode] (1) at (0,0){};
\node[unode] (2) at (2,0){};
\node[unode] (3) at (4,0){};
\node[unode] (4) at (6,0){};
\node[fnode] (5) at (8,0){};
\node[unode] (6) at (6,3){};
\node[fnode] (7) at (8,3){};
\draw[-] (1) -- (2);
\draw[-] (2)-- (3);
\draw[-] (3) -- (4);
\draw[-] (4) --(5);
\draw[-] (6) --(7);
\draw[-, thick, blue] (4)--(6);
\draw[-, thick, blue] (2)--(2,1.5);
\draw[-, thick, blue] (2,1.5)--(4,1.5);
\draw[-, thick, blue] (4,1.5)--(6,1.5);
\node[text width=3 cm](40) at (4, 2){\footnotesize{$(1, N_1, -(N_2-1))$}};
\node[text width=.1cm](20) [left=0.5 cm of 1]{$M_1$};
\node[text width=0.1 cm](21) [below=0.1cm of 2]{$N_1$};
\node[text width=1 cm](22) [below=0.1cm of 3]{$N-1$};
\node[text width=1 cm](23) [below=0.1cm of 4]{$N_2-1$};
\node[text width=.1cm](24) [right=0.1cm of 5]{$M_2$};
\node[text width=.1cm](25) [left=0.1cm of 6]{1};
\node[text width=.1cm](26) [right=0.5 cm of 7]{1};
\node[text width=.2cm](20) at (4,-2){$(\CT^\vee_3)$};
\end{tikzpicture}}\\
\scalebox{0.6}{\begin{tikzpicture}
\node[fnode] (1) at (0,0){};
\node[unode] (2) at (2,0){};
\node[unode] (3) at (4,0){};
\node[unode] (4) at (6,0){};
\node[fnode] (5) at (8,0){};
\node[unode] (6) at (2,3){};
\node[fnode] (7) at (0,3){};
\node[unode] (8) at (6,3){};
\node[fnode] (9) at (8,3){};
\draw[-] (1) -- (2);
\draw[-] (2)-- (3);
\draw[-] (3) -- (4);
\draw[-] (4) --(5);
\draw[-] (6) --(7);
\draw[-] (8) --(9);
\draw[-, thick, blue] (6)--(4,1.5);
\draw[-, thick, blue] (3)--(4,1.5);
\draw[-, thick, blue] (8)--(4,1.5);
\draw[-, thick, blue] (2)--(2,1.5);
\draw[-, thick, blue] (4)--(6,1.5);
\draw[-, thick, blue] (2,1.5)--(4,1.5);
\draw[-, thick, blue] (4,1.5)--(6,1.5);
\node[text width=4.5cm](19) at (7.1, 2){\footnotesize{$(1,1, -N_1+1,N-1,-N_2+1)$}};
\node[text width=.2cm](12) at (4,-2){$(\CT^\vee_4)$};
\node[text width=.1cm](20) [left=0.5 cm of 1]{$M_1$};
\node[text width=1 cm](21) [below=0.1cm of 2]{$N_1-1$};
\node[text width=1 cm](22) [below=0.1cm of 3]{$N-1$};
\node[text width=1 cm](23) [below=0.1cm of 4]{$N_2-1$};
\node[text width=.1cm](24) [right=0.1cm of 5]{$M_2$};
\node[text width=.1cm](25) [right=0.1cm of 6]{1};
\node[text width=.1cm](26) [left=0.5 cm of 7]{1};
\node[text width=.1cm](27) [left=0.5 cm of 8]{1};
\node[text width=.1cm](28) [right=0.1 cm of 9]{1};
\end{tikzpicture}}
& \qquad
& \scalebox{0.6}{\begin{tikzpicture}
\node[fnode] (1) at (0,0){};
\node[unode] (2) at (2,0){};
\node[unode] (3) at (4,0){};
\node[unode] (4) at (6,0){};
\node[fnode] (5) at (8,0){};
\node[unode] (6) at (4,2){};
\node[unode] (7) at (2,2){};
\node[unode] (8) at (0,2){};
\node[fnode] (9) at (-2,2){};
\draw[-] (1) -- (2);
\draw[-] (2)-- (3);
\draw[-] (3) -- (4);
\draw[-] (4) --(5);
\draw[-] (6) --(7);
\draw[-] (8) --(7);
\draw[-] (8) --(9);
\draw[thick, blue] (6)-- (3);
\node[text width=.1cm](40) [left=0.5 cm of 1]{$M_1$};
\node[text width=1 cm](41) [below=0.1cm of 2]{$N_1-1$};
\node[text width=1 cm](42) [below=0.1cm of 3]{$N-2$};
\node[text width=1 cm](43) [below=0.1cm of 4]{$N_2-1$};
\node[text width=.1cm](44) [right=0.1cm of 5]{$M_2$};
\node[text width=.1cm](45) [above=0.1cm of 6]{1};
\node[text width=.1cm](46) [above=0.1 cm of 7]{1};
\node[text width=.1cm](47) [above=0.1 cm of 8]{1};
\node[text width=.1cm](48) [left=0.5 cm of 9]{1};
\node[text width=2cm](21) at (5.1, 1){$(1,-N+2)$};
\node[text width=.2cm](20) at (4,-2){$(\CT^\vee_5)$};
\end{tikzpicture}}
\end{tabular}
\end{center}
\caption{\footnotesize{Hexality of quivers for the case where both unitary gauge nodes in the 3-node quiver are balanced.}}
\label{hexality-1}
\end{figure}

For the case where both unitary nodes are balanced i.e. $e_1 = 0, e_2 = 0$, the duality sequence in \figref{IRdual-Ex4c} leads to a hexality of 
quivers which are shown in \figref{hexality-1}. Note that the rank of the Coulomb branch symmetry manifest in the UV increases monotonically along the duality 
sequence in this case. The quiver $(\CT^\vee_5)$ is the theory at which the duality sequence terminates. To begin with, the rank of the Coulomb branch symmetry manifest in the UV for $(\CT^\vee_5)$ precisely matches that of $\frg_{\rm C}$ for the theory $\CT$. The quiver $(\CT^\vee_5)$ involves a pair of balanced unitary nodes - $U(N_1-1)$ and $U(N_2-1)$, separated by an overbalanced node $U(N-2)$ which in turn is attached to a tail of three balanced $U(1)$ nodes. Using our intuition for linear quivers as discussed above, the two isolated balanced nodes contribute an $\frsu(2)$ factor each to the Coulomb branch global symmetry while the overbalanced node gives a $\fru(1)$ factor. Finally, the tail of three balanced $U(1)$ nodes gives an $\frsu(4)$ factor. One can therefore read off the Coulomb branch symmetry $\frg_{\rm C}=\frsu(2) \oplus \frsu(2) \oplus \frsu(4) \oplus \fru(1)$ from the quiver $(\CT^\vee_5)$, where $\frg_{\rm C}$ is indeed the emergent symmetry algebra of the theory $\CT$ as noted above.

For the cases $(e_1 \neq 0, e_2 \neq 0)$ and $(e_1=0, e_2 \neq 0)$, the emergent global symmetry can be similarly read off from the duality sequence associated with the quiver $\CT$ for the given values of $e_1, e_2$. We refer the reader to \Secref{Unbal-3node}-\Secref{Bal1-3node} for further details. \\

\subsection*{3d mirror of an N-al sequence: Recipe and Examples}

$\CN=4$ theories in three space-time dimensions have yet another IR duality which exchanges the Higgs and the Coulomb branches of the 
moduli space in the deep IR -- 3d mirror symmetry \cite{Intriligator:1996ex}. The quiver gauge theory $\CT$ and the associated N-al set 
of quivers will therefore have a common 3d mirror. In \Secref{gen-mirr}, we discuss a general recipe for constructing this 3d mirror, 
using the technology of $S$-type operations introduced in \cite{Intriligator:1996ex}.
For a completely generic quiver gauge theory $\CT$ in class $\CX$, the 3d mirror is not guaranteed to be a Lagrangian 
theory. However, for the case of $\CT$ being a linear quiver, one can show that the 3d mirror is indeed Lagrangian and 
one can explicitly work out the quiver, given the duality sequence associated with $\CT$. In \Secref{2node-mirr} and \Secref{3node-mirr}, 
we apply the recipe of \Secref{gen-mirr} to the two-node linear quiver and the three-node linear quiver respectively and construct 
the corresponding 3d mirrors. The computational details of the construction can be found in \Appref{Mirr-pf-S}.\\

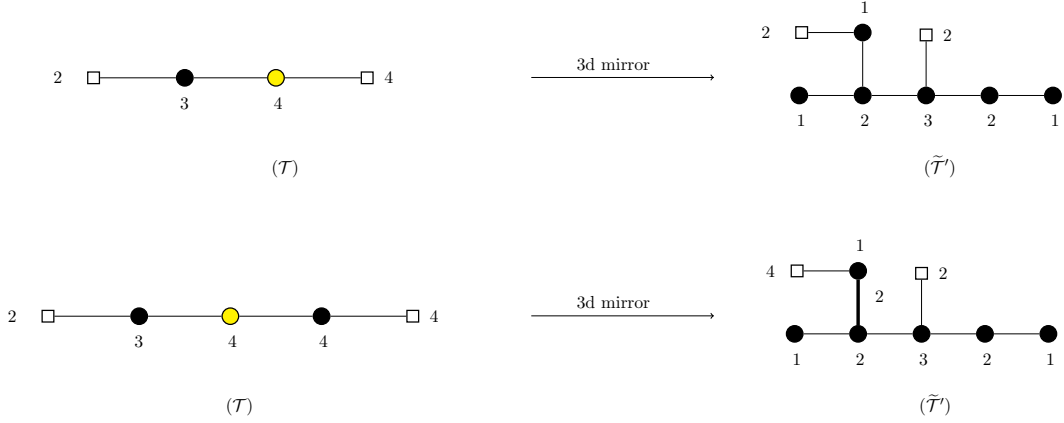
\begin{figure}
\begin{center}
\begin{tabular}{ccc}
\scalebox{0.6}{\begin{tikzpicture}
\node[fnode] (1) at (0,0){};
\node[unode] (2) at (2,0){};
\node[sunode] (3) at (4,0){};
\node[fnode] (4) at (6,0){};
\draw[-] (1) -- (2);
\draw[-] (2)-- (3);
\draw[-] (3) -- (4);
\node[text width=.1cm](20) [left=0.5 cm of 1]{$2$};
\node[text width=.2cm](21) [below=0.1cm of 2]{$3$};
\node[text width=.1cm](22) [below=0.1cm of 3]{$4$};
\node[text width=.1cm](23) [right=0.1cm of 4]{$4$};
\node[text width=.2cm](20) at (4,-2){$(\CT)$};
\end{tikzpicture}}
&\qquad \scalebox{.6}{\begin{tikzpicture}
\draw[->] (3,0) -- (7, 0);
\node[text width=2 cm](29) at (5, 0.3) {3d mirror};
\node[](30) at (5, -2.2) {};
\end{tikzpicture}} \qquad 
&\scalebox{0.6}{\begin{tikzpicture}
\node[unode] (1) {};
\node[unode] (2) [right=1cm  of 1]{};
\node[unode] (3) [right=1 cm of 2]{};
\node[unode] (4) [right=1 cm of 3]{};
\node[unode] (5) [right=1 cm of 4]{};
\node[unode] (6) [above=1 cm of 2]{};
\node[fnode] (7) [above=1 cm of 3]{};
\node[fnode] (8) [left= 1 cm of 6]{};
\draw[-] (1) -- (2);
\draw[-] (2)-- (3);
\draw[-] (3) -- (4);
\draw[-] (4) -- (5);
\draw[-] (2) -- (6);
\draw[-] (3) -- (7);
\draw[-] (6) -- (8);
\node[text width=.1cm](10) [below=0.1 cm of 1]{$1$};
\node[text width= .1cm](11) [below=0.1cm of 2]{$2$};
\node[text width= .1cm](12) [below=0.1cm of 3]{$3$};
\node[text width=.1cm](13) [below=0.1cm of 4]{$2$};
\node[text width=.1cm](14) [below=0.1cm of 5]{$1$};
\node[text width=.1cm](15) [above=0.1cm of 6]{$1$};
\node[text width=.1cm](16) [right=0.1cm of 7]{$2$};
\node[text width=.1cm](17) [left=0.5cm of 8]{$2$};
\node[text width=0.1cm](20)[below=1 cm of 3]{$(\wt{\CT}')$};
\end{tikzpicture}}\\
\qquad
&\qquad
&\qquad\\
\scalebox{0.6}{\begin{tikzpicture}
\node[fnode] (1) at (0,0){};
\node[unode] (2) at (2,0){};
\node[sunode] (3) at (4,0){};
\node[unode] (4) at (6,0){};
\node[fnode] (5) at (8,0){};
\draw[-] (1) -- (2);
\draw[-] (2)-- (3);
\draw[-] (3) -- (4);
\draw[-] (4) --(5);
\node[text width=.1cm](10) [left=0.5 cm of 1]{$2$};
\node[text width=.2cm](11) [below=0.1cm of 2]{$3$};
\node[text width=.1cm](12) [below=0.1cm of 3]{$4$};
\node[text width=.1cm](13) [below=0.1cm of 4]{$4$};
\node[text width=.1cm](14) [right=0.1cm of 5]{$4$};
\node[text width=.2cm](20) at (4,-2){$(\CT)$};
\end{tikzpicture}}
&\qquad \scalebox{.6}{\begin{tikzpicture}
\draw[->] (3,0) -- (7, 0);
\node[text width=2 cm](29) at (5, 0.3) {3d mirror};
\node[](30) at (5, -2.2) {};
\end{tikzpicture}} \qquad 
& \scalebox{0.6}{\begin{tikzpicture}
\node[unode] (1) {};
\node[unode] (2) [right=1cm  of 1]{};
\node[unode] (3) [right=1 cm of 2]{};
\node[unode] (4) [right=1 cm of 3]{};
\node[unode] (5) [right=1 cm of 4]{};
\node[unode] (6) [above=1 cm of 2]{};
\node[fnode] (7) [above=1 cm of 3]{};
\node[fnode] (8) [left= 1 cm of 6]{};
\node[](9) [above=0.5 cm of 2]{};
\draw[-] (1) -- (2);
\draw[-] (2)-- (3);
\draw[-] (3) -- (4);
\draw[-] (4) -- (5);
\draw[line width=0.75mm] (2) -- (6);
\draw[-] (3) -- (7);
\draw[-] (6) -- (8);
\node[text width=.1cm](10) [below=0.1 cm of 1]{$1$};
\node[text width= .1cm](11) [below=0.1cm of 2]{$2$};
\node[text width= .1cm](12) [below=0.1cm of 3]{$3$};
\node[text width=.1cm](13) [below=0.1cm of 4]{$2$};
\node[text width=.1cm](14) [below=0.1cm of 5]{$1$};
\node[text width=.1cm](15) [above=0.1cm of 6]{$1$};
\node[text width=.1cm](16) [right=0.1cm of 7]{$2$};
\node[text width=.1cm](17) [left=0.3cm of 8]{$4$};
\node[text width=.1cm](18) [right=0.1cm of 9]{$2$};
\node[text width=0.1cm](20)[below=1 cm of 3]{$(\wt{\CT}')$};
\end{tikzpicture}}
\end{tabular}
\end{center}
\caption{\footnotesize{Top row: 3d mirror for a 2-node linear quiver $\CT$. Bottom row: 3d mirror for a 3-node linear quiver $\CT$.}}
\label{3dmirr-1}
\end{figure}

The emergent Coulomb branch symmetry for the linear quivers $\CT$ is realized in the corresponding 3d mirrors $\wt{\CT}'$ as a Higgs 
branch global symmetry which is manifest in the UV. One can readily check this for the two-node and the 
three-node quivers. Let us consider a specific example of the two-node linear quiver with $e=0$, where $M_1=2$, $N_1=3$, $N=4$, and $M_2=4$. 
The corresponding 3d mirror $\wt{\CT}'$ is shown in the top right corner of \figref{3dmirr-1}. Recall that, $\frg^{(\CT)}_{\rm C}=\frsu(2) \oplus \frsu(2) \oplus \fru(1)$, while from the quiver $\wt{\CT}'$ one can read off $\frg^{(\wt{\CT}')}_{\rm H}=\frsu(2) \oplus \frsu(2) \oplus \fru(1)$. 
One can similarly consider the case of a three-node quiver with $e_1=e_2=0$, where $M_1=2$, $N_1=3$, $N=4$, $N_2=4$, and $M_2=4$. 
The 3d mirror is given by the quiver $\wt{\CT}'$ in the bottom right corner of \figref{3dmirr-1}. 
The Higgs branch symmetry of $\wt{\CT}'$ can be read off as $\frg^{(\wt{\CT}')}_{\rm H}=\frsu(2) \oplus \frsu(2) \oplus \frsu(4) \oplus \fru(1)$, 
which evidently matches the emergent Coulomb branch symmetry of the theory $\CT$ as we discussed above.

\subsection*{IR Dualities involving quivers without Abelian hypermultiplets}

As an interesting byproduct of the construction of IR N-ality presented in this paper, one can obtain IR 
dualities for a pair of theories which both belong to class $\CX$. For the subclass of linear quivers that 
we study in this paper, this implies IR dualities between such linear quivers with unitary and special 
unitary gauge groups. We obtain an explicit example of such an IR duality from the duality sequence 
in \figref{IRdual-Ex6c} associated with the triality of the two-node quiver with $e=2$. The dual pair is 
given as follows:\\

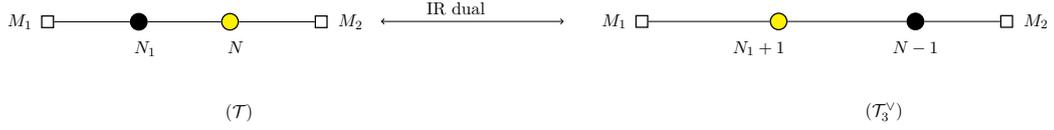
\begin{center}
\begin{tabular}{ccc}
\scalebox{0.6}{\begin{tikzpicture}
\node[fnode] (1) at (0,0){};
\node[unode] (2) at (2,0){};
\node[sunode] (3) at (4,0){};
\node[fnode] (4) at (6,0){};
\draw[-] (1) -- (2);
\draw[-] (2)-- (3);
\draw[-] (3) -- (4);
\node[text width=.1cm](20) [left=0.5 cm of 1]{$M_1$};
\node[text width=.2cm](21) [below=0.1cm of 2]{$N_1$};
\node[text width=.1cm](22) [below=0.1cm of 3]{$N$};
\node[text width=.1cm](23) [right=0.1cm of 4]{$M_2$};
\node[text width=.2cm](20) at (4,-2){$(\CT)$};
\end{tikzpicture}}
& \scalebox{.6}{\begin{tikzpicture}
\draw[<->] (3,0) -- (7, 0);
\node[text width= 2cm](29) at (5, 0.3) {IR dual};
\node[](30) at (5, -2.2) {};
\end{tikzpicture}}
&\scalebox{0.6}{\begin{tikzpicture}
\node[fnode] (1) at (-1,0){};
\node[sunode] (2) at (2,0){};
\node[unode] (3) at (5,0){};
\node[fnode] (4) at (7,0){};
\draw[-] (1) -- (2);
\draw[-] (2)-- (3);
\draw[-] (3) -- (4);
\node[text width=.1cm](20) [left=0.5 cm of 1]{$M_1$};
\node[text width=2 cm](21) [below=0.1cm of 2]{$N_1+1$};
\node[text width=1 cm](22) [below=0.1cm of 3]{$N-1$};
\node[text width=.1cm](23) [right=0.1cm of 4]{$M_2$};
\node[text width=.2cm](20) at (4,-2){$(\CT^\vee_3)$};
\end{tikzpicture}}
\end{tabular}
\end{center}

The theory dual to $\CT$ is also a two-node linear quiver with gauge group $SU(N_1 +1) \times U(N-1)$. 
In addition, one can check that the unitary gauge node in the dual theory is also overbalanced with $e=2$.  
The duality holds for generic labels in the quiver $\CT$ with the constraint $e=2$, which implies that the above is a 
2-parameter family of dualities.

\section{From duality to N-ality: Quivers and mutations}\label{recipe-gen}

In this section, we will present the general construction of the N-al set of theories starting from a given
3d $\CN=4$ quiver gauge theory of class $\CX$, after introducing the quiver notation used in this 
work and briefly reviewing the Seiberg-like dualities discussed in \cite{Dey:2022pqr}.

\subsection{Quiver notation} \label{quiv-notation}

Let us begin by discussing the quiver notation that was introduced in \cite{Dey:2022pqr} and will be 
used for the rest of this paper. We will consider quiver gauge theories with $SU(N)$ and 
$U(N)$ gauge nodes, hypermultiplets in the fundamental/bifundamental respresentation 
of both types of gauge nodes, and a number of hypermultiplets in the determinant/anti-determinant 
representations of the unitary gauge nodes. A generic gauge theory of this type will be 
denoted by a quiver diagram of the following form:

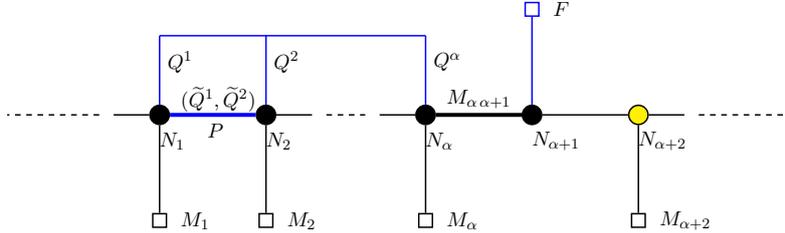
\begin{center}
\scalebox{0.7}{\begin{tikzpicture}
\node[] (100) at (-3,0) {};
\node[] (1) at (-1,0) {};
\node[unode] (2) at (0,0) {};
\node[text width=.2cm](31) at (0.1,-0.5){$N_1$};
\node[unode] (3) at (2,0) {};
\node[text width=.2cm](32) at (2.1,-0.5){$N_2$};
\node[] (4) at (3,0) {};
\node[] (5) at (4,0) {};
\node[unode] (6) at (5,0) {};
\node[text width=.2cm](33) at (5.1,-0.5){$N_{\alpha}$};
\node[unode] (7) at (7,0) {};
\node[text width=.2cm](34) at (7.1,-0.5){$N_{\alpha+1}$};
\node[text width=.2cm](35) at (5.5,0.3){$M_{\alpha\,\alpha+1}$};
\node[sunode] (8) at (9,0) {};
\node[text width=.2cm] (40) at (9.1,-0.5) {$N_{\alpha+2}$};
\node[fnode] (9) at (9,-2) {};
\node[text width=.2cm] (40) at (9.5,-2) {$M_{\alpha+2}$};
\node[] (10) at (10,0) {};
\node[] (11) at (12,0) {};
\node[afnode] (12) at (7,2) {};
\node[text width=.2cm](36) at (7.5,2){$F$};
\node[fnode] (20) at (0,-2) {};
\node[fnode] (21) at (2,-2) {};
\node[fnode] (22) at (5,-2) {};
\node[text width=.2cm](23) at (0.5,-2){$M_1$};
\node[text width=.2cm](24) at (2.5,-2){$M_2$};
\node[text width=.2cm](25) at (5.5,-2){$M_\alpha$};
\draw[thick] (1) -- (2);
\draw[line width=0.75mm, blue] (2) -- (3);
\node[text width=.2cm](50) at (0.5,0.3){$(\wt{Q}^1,\wt{Q}^2)$};
\node[text width=.2cm](51) at (1,-0.3){$P$};
\draw[thick] (3) -- (4);
\draw[thick,dashed] (4) -- (5);
\draw[thick] (5) -- (6);
\draw[line width=0.75mm] (6) -- (7);
\draw[thick] (7) -- (8);
\draw[thick,blue] (7) -- (12);
\draw[thick] (8) -- (9);
\draw[thick] (8) -- (10);
\draw[thick,dashed] (10) -- (11);
\draw[thick,dashed] (1) -- (100);
\draw[thick, blue] (2) -- (0,1.5);
\draw[thick, blue] (0,1.5) -- (5,1.5);
\draw[thick, blue] (2,1.5) -- (3);
\draw[thick, blue] (5,1.5) -- (6);
\draw[thick] (2) -- (20);
\draw[thick] (3) -- (21);
\draw[thick] (6) -- (22);
\node[text width=.2cm](15) at (0.25,1){$Q^1$};
\node[text width=.2cm](16) at (2.25,1){$Q^2$};
\node[text width=.2cm](17) at (5.25,1){$Q^\alpha$};
\end{tikzpicture}}
\end{center}

The gauge group and the matter content of the theory can be read off from the quiver diagram
using the following dictionary:

\begin{enumerate}

\item A node \scalebox{0.7}{\begin{tikzpicture} \node[unode] (1) at (0,0){};\end{tikzpicture}} labelled $N$ denotes a $U(N)$ vector multiplet.

\item A node \scalebox{0.7}{ \begin{tikzpicture} \node[sunode] (1) at (0,0){};\end{tikzpicture}} labelled $N$ denotes an $SU(N)$ 
vector multiplet.

\item A box \scalebox{0.7}{\begin{tikzpicture} \node[fnode] (1) at (0,0){};\end{tikzpicture}} labelled $F$ attached to a gauge node
denotes $F$ hypermultiplets in the fundamental representation of the gauge group associated to that node.

\item A thin black line between two gauge nodes represents a bifundamental hypermultiplet. A collection of 
$M$ bifundamental hypermultiplets between the same pair of gauge nodes is denoted by 
a thick black line with the label $M$.

\item A box \scalebox{0.7}{\begin{tikzpicture} \node[afnode] (1) at (0,0){};\end{tikzpicture}} labelled $F$ 
attached to a unitary gauge node denotes $F$ hypermultiplets in the determinant representation.

\item A thin blue line connecting multiple unitary gauge nodes denotes an Abelian hypermultiplet in the 
determinant/anti-determinant representation of those gauge nodes. We will show the 
charges $\{Q^i\}$ of the Abelian hypermultiplet explicitly in the quiver diagram where 
$Q^i= \pm N_i$, with $N_i$ being the rank (and therefore the label) of the $i$-th unitary 
gauge node. A collection of $P$ such hypermultiplets connecting the same set of nodes 
will be denoted by a thick blue line with the label $P$.

\end{enumerate}

\subsection{Dualities in the $\CT^N_{N_f,P}$ theories: A review}\label{Duality-rev}

In this section, we will briefly review the Seiberg-like dualities for the theories $\CT^N_{N_f,P}$ 
-- a $U(N)$ SQCD with $N_f$ fundamental flavors and $P$ Abelian hypermultiplets. In \cite{Dey:2022pqr},
it was demonstrated that such dualities, denoted as $\CD^N_{N_f,P}$, exist for the parameter 
ranges $(N_f=2N+1, P=1)$, $(N_f=2N, P\geq 1)$ and $(N_f=2N-1, P\geq 1)$ for a given $N$. 
The dual pairs in each case, as well as the 3d mirror theory associated with the pair, are 
summarized in Table \ref{Tab: Review}.

\begin{table}[htbp]
\begin{center}
{%
\begin{tabular}{|c|c|c|c|}
\hline
Duality &Theory $\CT$ & IR dual  $\CT^\vee$ & 3d mirror \\
\hline \hline 
$\CD^N_{2N+1,1}$
&\scalebox{.6}{\begin{tikzpicture}
\node[unode] (1) at (0,0){};
\node[fnode] (2) at (0,-2){};
\node[afnode] (3) at (3,0){};
\draw[-] (1) -- (2);
\draw[-, thick, blue] (1)-- (3);
\node[text width=0.1cm](10) at (0, 0.5){$N$};
\node[text width=1.5cm](11) at (1, -2){$2N+1$};
\node[text width=1cm](12) at (4, 0){$1$};
\end{tikzpicture}}
&\scalebox{.6}{\begin{tikzpicture}
\node[sunode] (1) at (0,0){};
\node[fnode] (2) at (0,-2){};
\draw[-] (1) -- (2);
\node[text width=1 cm](10) at (0, 0.5){$N+1$};
\node[text width=1.5cm](11) at (1, -2){$2N+1$};
\end{tikzpicture}}
&  \scalebox{0.6}{\begin{tikzpicture}
\node[unode] (1) {};
\node[unode] (2) [right=.5cm  of 1]{};
\node[unode] (3) [right=.5cm of 2]{};
\node[unode] (4) [right=1cm of 3]{};
\node[] (5) [right=0.5cm of 4]{};
\node[unode] (6) [right=1 cm of 4]{};
\node[unode] (7) [right=1cm of 6]{};
\node[unode] (8) [right=0.5cm of 7]{};
\node[unode] (9) [right=0.5cm of 8]{};
\node[unode] (13) [above=0.5cm of 4]{};
\node[fnode] (14) [left=0.5cm of 13]{};
\node[fnode] (15) [above=0.5cm of 6]{};
\node[text width=0.1cm](41)[below=0.2 cm of 1]{1};
\node[text width=0.1cm](42)[below=0.2 cm of 2]{2};
\node[text width=0.1cm](43)[below=0.2 cm of 3]{3};
\node[text width=0.1cm](44)[below=0.2 cm of 4]{$N$};
\node[text width=0.1cm](45)[below=0.2 cm of 6]{$N$};
\node[text width=0.1cm](46)[below=0.2 cm of 7]{3};
\node[text width=0.1cm](47)[below=0.2 cm of 8]{2};
\node[text width=0.1cm](48)[below=0.2 cm of 9]{1};
\node[text width=0.1cm](49)[above=0.2 cm of 13]{1};
\node[text width=0.1cm](50)[left=0.2 cm of 14]{1};
\node[text width=0.1cm](51)[right=0.2 cm of 15]{1};
\draw[-] (1) -- (2);
\draw[-] (2)-- (3);
\draw[dashed] (3) -- (4);
\draw[-] (4) --(6);
\draw[dashed] (6) -- (7);
\draw[-] (7) -- (8);
\draw[-] (8) --(9);
\draw[-] (4) -- (13);
\draw[-] (13) -- (14);
\draw[-] (6) -- (15);
\end{tikzpicture}}\\
\hline
$\CD^N_{2N,P}$
&\scalebox{.6}{\begin{tikzpicture}
\node[unode] (1) at (0,0){};
\node[fnode] (2) at (0,-2){};
\node[afnode] (3) at (3,0){};
\draw[-] (1) -- (2);
\draw[-, thick, blue] (1)-- (3);
\node[text width=0.1cm](10) at (0, 0.5){$N$};
\node[text width=1.5cm](11) at (1, -2){$2N$};
\node[text width=1cm](12) at (4, 0){$P$};
\end{tikzpicture}} 
& \scalebox{.6}{\begin{tikzpicture}
\node[unode] (1) at (0,0){};
\node[fnode] (2) at (0,-2){};
\node[afnode] (3) at (3,0){};
\draw[-] (1) -- (2);
\draw[-, thick, blue] (1)-- (3);
\node[text width=0.1cm](10) at (0, 0.5){$N$};
\node[text width=1.5cm](11) at (1, -2){$2N$};
\node[text width=1cm](12) at (4, 0){$P$};
\end{tikzpicture}}
 & \scalebox{0.6}{\begin{tikzpicture}
\node[unode] (1) {};
\node[unode] (2) [right=.5cm  of 1]{};
\node[unode] (3) [right=.5cm of 2]{};
\node[unode] (4) [right=1cm of 3]{};
\node[unode] (5) [right=1cm of 4]{};
\node[unode] (6) [right=1 cm of 5]{};
\node[unode] (7) [right=1cm of 6]{};
\node[unode] (8) [right=0.5cm of 7]{};
\node[unode] (9) [right=0.5cm of 8]{};
\node[unode] (13) [above= 1cm of 5]{};
\node[fnode] (14) [below=1 cm of 5]{};
\node[unode] (15) [left=1 cm of 13]{};
\node[unode] (16) [left=1 cm of 15]{};
\node[fnode] (17) [left=1 cm of 16]{};
\node[text width=0.1cm](41)[below=0.2 cm of 1]{1};
\node[text width=0.1cm](42)[below=0.2 cm of 2]{2};
\node[text width=0.1cm](43)[below=0.2 cm of 3]{3};
\node[text width=1cm](44)[below=0.2 cm of 4]{$N-1$};
\node[text width=0.1cm](45)[below=0.2 cm of 5]{$N$};
\node[text width=1cm](46)[below=0.2 cm of 6]{$N-1$};
\node[text width=0.1cm](47)[below=0.2 cm of 7]{3};
\node[text width=0.1cm](48)[below=0.2 cm of 8]{2};
\node[text width=0.1cm](49)[below=0.1 cm of 9]{1};
\node[text width=0.1cm](50)[right=0.1 cm of 14]{1};
\node[text width=0.1cm](51)[left=0.2 cm of 17]{1};
\draw[-] (1) -- (2);
\draw[-] (2)-- (3);
\draw[dashed] (3) -- (4);
\draw[-] (4) --(5);
\draw[-] (5) --(6);
\draw[dashed] (6) -- (7);
\draw[-] (7) -- (8);
\draw[-] (8) --(9);
\draw[-] (5) -- (13);
\draw[-] (13) -- (15);
\draw[dashed] (16) -- (15);
\draw[-] (16) -- (17);
\draw[-] (5) -- (14);
\node[text width=0.1cm](25)[above=0.2 cm of 13]{1};
\node[text width=0.1cm](26)[above=0.2 cm of 15]{1};
\node[text width=0.1cm](27)[above=0.2 cm of 16]{$1$};
\end{tikzpicture}}\\
\hline
$\CD^N_{2N-1,P}$
& \scalebox{.6}{\begin{tikzpicture}
\node[unode] (1) at (0,0){};
\node[fnode] (2) at (0,-2){};
\node[afnode] (3) at (3,0){};
\draw[-] (1) -- (2);
\draw[-, thick, blue] (1)-- (3);
\node[text width=0.1cm](10) at (0, 0.5){$N$};
\node[text width=1.5cm](11) at (1, -2){$2N-1$};
\node[text width=1cm](12) at (4, 0){$P$};
\end{tikzpicture}}
 & \scalebox{.6}{\begin{tikzpicture}
\node[unode] (1) at (0,0){};
\node[fnode] (2) at (0,-2){};
\node[unode] (3) at (4,0){};
\node[fnode] (4) at (4,-2){};
\draw[-] (1) -- (2);
\draw[line width=0.75mm, blue] (1)-- (3);
\draw[-] (3)-- (4);
\node[text width=2.3cm](10) at (2.0, 0.2){$(1, -(N-1))$};
\node[text width=2cm](11) at (3.0, -0.3){$P$};
\node[text width=0.1 cm](20) at (0,0.5){$1$};
\node[text width=0.1 cm](21) at (0.5,-2){$1$};
\node[text width=1 cm](22) at (4, 0.5){$N-1$};
\node[text width=1.5 cm](23) at (5, -2){$2N-1$};
\end{tikzpicture}}
 &  \scalebox{0.6}{\begin{tikzpicture}
 \node[unode] (1) {};
\node[unode] (2) [right=.5cm  of 1]{};
\node[unode] (3) [right=.5cm of 2]{};
\node[unode] (4) [right=1cm of 3]{};
\node[] (5) [right=0.5cm of 4]{};
\node[unode] (6) [right=1 cm of 4]{};
\node[unode] (7) [right=1cm of 6]{};
\node[unode] (8) [right=0.5cm of 7]{};
\node[unode] (9) [right=0.5cm of 8]{};
\node[unode] (13) [above=0.5cm of 4]{};
\node[unode] (14) [left=0.5cm of 13]{};
\node[fnode] (15) [above=0.5cm of 6]{};
\node[unode] (16) [left=0.5cm of 14]{};
\node[fnode] (17) [left=0.5cm of 16]{};
\node[fnode] (18) [above=1 cm of 13]{};
\node[text width=0.1cm](41)[below=0.2 cm of 1]{1};
\node[text width=0.1cm](42)[below=0.2 cm of 2]{2};
\node[text width=0.1cm](43)[below=0.2 cm of 3]{3};
\node[text width=1cm](44)[below=0.2 cm of 4]{$N-1$};
\node[text width=1cm](45)[below=0.2 cm of 6]{$N-1$};
\node[text width=0.1cm](46)[below=0.2 cm of 7]{3};
\node[text width=0.1cm](47)[below=0.2 cm of 8]{2};
\node[text width=0.1cm](48)[below=0.2 cm of 9]{1};
\node[text width=0.1cm](49)[right=0.1 cm of 15]{1};
\node[text width=0.1cm](50)[left=0.2 cm of 17]{1};
\node[text width=0.1cm](51)[right=0.2 cm of 18]{1};
\draw[-] (1) -- (2);
\draw[-] (2)-- (3);
\draw[dashed] (3) -- (4);
\draw[-] (4) --(6);
\draw[dashed] (6) -- (7);
\draw[-] (7) -- (8);
\draw[-] (8) --(9);
\draw[-] (4) -- (13);
\draw[-] (13) -- (14);
\draw[-] (6) -- (15);
\draw[dashed] (14) -- (16);
\draw[-] (16) -- (17);
\draw[-] (18) -- (13);
\node[text width=0.1cm](25)[above=0.2 cm of 13]{1};
\node[text width=0.1cm](26)[above=0.2 cm of 14]{1};
\node[text width=0.1cm](27)[above=0.2 cm of 16]{1};
\end{tikzpicture}}\\
\hline
\end{tabular}}
\end{center}
\caption{\footnotesize{Summary of the IR dualities for the $\CT^N_{N_f,P}$ theories, and the 3d mirror associated with each dual pair. 
For $\CD^N_{2N,P}$ and $\CD^N_{2N-1,P}$, the tail of $U(1)$ gauge nodes in the 3d mirror has precisely $P$ gauge nodes. }}
\label{Tab: Review}
\end{table}

The IR dualities listed above can be checked using a variety of supersymmetric observables. In \cite{Dey:2022pqr}, they were checked explicitly using 
the three-sphere partition function and the Coulomb/Higgs branch Hilbert Series. In this paper, we will make use of the three-sphere 
partition function extensively to construct the duality sequences\footnote{We would like to emphasize that one could have equally well used any other supersymmetric observable, like the 3d superconformal index, for this purpose.}.
We will therefore also review the partition function identities associated with the dualities below in a form that will be useful later in the paper. 
For further details on these dualities, we refer the reader to Section 3 and Section 4 of \cite{Dey:2022pqr}.

\begin{enumerate}

\item \textbf{Duality $\CD^N_{2N+1,1}$:} The dual pair $(\CT, \CT^\vee)$ involves the theory $\CT=\CT^N_{2N+1,1}$ 
and the theory $\CT^\vee$ -- an $SU(N+1)$ theory with $N_f=2N+1$ fundamental flavors.
Let $\vec m$ and $m_{\rm ab}$ be the real masses associated with the fundamental hypers 
and the Abelian hypermultiplet respectively in the $\CT^N_{2N+1,1}$ theory, and $\eta$ be the FI parameter. Then the duality 
translates into the following identity:
\begin{align}\label{Id-1}
& Z^{\CT^{N}_{2N+1,1}}(\vec m, m_{\rm ab}=\tr \vec m, \eta=0) = Z^{SU(N+1),\,2N +1}(\vec m), \nn \\
& Z^{\CT^{N}_{2N+1,1}}(\vec m, m_{\rm ab}=\tr \vec m, \eta) =  \int \, [d \vec \s] \, \frac{e^{2\pi i \eta\, \tr \vec \s}}{\ch{(\tr \vec \s - \tr \vec m)}}\,\frac{\prod_{1\leq j<k \leq N}\, \sinh^2{\pi(\s_j- \s_k)}}{\prod^{N}_{j=1} \prod^{2N+1}_{i=1}\ch{(\s_j- m_i)}}, \nn \\
& Z^{SU(N+1),\,2N +1}(\vec m)= \int \, [d \vec s] \, \frac{\delta(\tr \vec s)\,\prod_{1\leq j<k \leq N+1}\, \sinh^2{\pi(s_j- s_k)}}{\prod^{N+1}_{j=1} \prod^{2N+1}_{i=1}\ch{(\s_j- m_i)}},
\end{align}
where $\tr \vec m= \sum^{2N+1}_{i=1}\,m_i$. The topological $\fru(1)$ symmetry of the $\CT^N_{2N+1,1}$ theory is mapped to an emergent 
$\fru(1)$ IR symmetry in the $SU(N+1)$ theory, which explains why the FI parameter has to be turned off for the partition functions to agree.
The map of the $\fru(1)_B$ symmetry of the $SU(N+1)$ theory across the duality can be read off from the partition function identity.

\item \textbf{Duality $\CD^N_{2N,P}$:} This is the self-duality of the theory $\CT=\CT^N_{2N,P}$.
Let $\vec m$ and $\vec m_{\rm ab}$ denote the real masses for the $2N$ fundamental hypers and the $P$ Abelian hypers respectively. 
The associated partition function identity is given as:
\be \label{Id-2a}
Z^{\CT^N_{2N,P}}(\vec m, \vec{m}_{\rm ab}, \eta) = Z^{\CT^N_{2N,P}}(\vec m, -\vec{m}_{\rm ab}, -\eta),
\ee
where $\vec m$ obey the constraint $ \sum^{2N}_{i=1}\, m_i=0$. 
In particular, for $P=1$, the identity can be put into the following form after a simple change of variables:
\begin{align}\label{Id-2b}
Z^{\CT^N_{2N,1}}(\vec m, m_{\rm ab}= \tr \vec m, \eta) = &   \int \,  [d \vec \s] \, \frac{e^{-2\pi i \eta \tr \vec \s}}{\ch{(\tr \vec \s - \tr \vec m')}}\,Z^{\CT^N_{2N,0}}_{\rm 1-loop} (\vec \s, \vec m') \nn \\
= & Z^{\CT^N_{2N,1}}(\vec m', m'_{\rm ab}= \tr \vec m', -\eta),
\end{align}
where the masses $\vec m$ are unconstrained and the masses $\vec m'$ are related to $\vec m$ as follows:
\be
m'_a = m_a - \frac{1}{N}\,\tr \vec m, \qquad a=1,\ldots, 2N,
\ee
and $\tr \vec m = \sum^{2N}_{i=1}\, m_i$, $\tr \vec m' = \sum^{2N}_{i=1}\, m'_i = - \tr \vec m$. An immediate consequence of the duality 
$\CD^N_{2N,P}$ is the self-duality of an $SU(N)$ gauge theory with $2N$ flavors, which can be seen at the level of the partition 
function by integrating over the FI parameter $\eta$ on both sides of the identity \eref{Id-2a} or the identity \eref{Id-2b}. 

\item  \textbf{Duality $\CD^N_{2N-1,P}$:} The dual pair $(\CT, \CT^\vee)$ involves the theory $\CT=\CT^N_{2N-1,P}$ and the theory $\CT^\vee$ -- a $U(1) \times U(N-1)$ gauge theory with $1$ and $2N-1$ fundamental hypers respectively plus $P$ Abelian hypermultiplet with charges $(1, -(N-1))$ under the $U(1) \times U(N-1)$ gauge group.
Let $\vec m$ and $\vec m_{\rm ab}$ denote the real masses for the $2N-1$ fundamental hypers 
and the $P$ Abelian hypers respectively. The associated partition function identity is given as:
\begin{align}\label{Id-3}
Z^{\CT^N_{2N-1,P}}(\vec m, \vec{m}_{\rm{Ab}};\eta) =  &\int \,d{\s'}\,[d \vec \s]\,\frac{e^{2\pi i \eta\,(\s' - \tr \vec \s)}\,Z^{\CT^{N-1}_{2N-1,0}}_{\rm 1-loop} (\vec \s, \vec m)}{\ch{(\s'- \tr m)}\,\prod^P_{l=1}\,\ch{(\s'-\tr \vec \s -m^l_{\rm{Ab}})} } \nn \\
=& Z^{\CT^\vee}(\vec m, m'=\tr \vec m,\vec m_{\rm{Ab}};\eta, -\eta),
\end{align}
where $\vec m, m', \vec m_{\rm{Ab}}$ denote the real masses for the hypers in the fundamental 
representation of the $U(N-1)$ gauge node, the single hyper in the fundamental of $U(1)$ 
gauge node, and the Abelian hypers charged under both the $U(1)$ and the $U(N-1)$ nodes respectively.
The parameters $\vec m$ and $\vec m_{\rm{Ab}}$ are unconstrained in the above identity.

For $P>1$, the $\fru(1)$ topological symmetry of the theory $\CT^N_{2N-1, P}$ is enhanced to $\fru(1) \oplus \fru(1)$
in the IR. On the dual side, this enhanced symmetry is manifest in the UV as topological symmetries of two unitary gauge 
nodes -- $U(1)$ and $U(N-1)$. For $P=1$, the Coulomb branch symmetry of $\CT=\CT^N_{2N-1, 1}$ is enhanced to 
$\frsu(2) \oplus \fru(1)$. The rank of the symmetry algebra can be read off from the UV Lagrangian on the dual side 
as before. In addition, one observes that the quiver has a single balanced gauge node and a single overbalanced 
gauge node. Using the standard intuition of enhancement of Coulomb branch symmetries for balanced nodes in linear 
quivers \cite{Gaiotto:2008ak}, one may therefore infer from the quiver $\CT^\vee$ that the IR symmetry algebra is enhanced to a 
$\frsu(2) \oplus \fru(1)$.

\end{enumerate}

\subsection{Constructing the duality sequence: The general recipe}\label{recipe-gen0}

Consider a quiver gauge theory in class $\CX$. A theory in this class is specified by a graph 
of the form discussed in \Secref{quiv-notation} where the gauge nodes (unitary and special unitary) are connected 
by fundamental/bifundamental hypermultiplets only. Additionally, at least one of the special unitary nodes must be balanced, 
while the unitary nodes may be balanced or overbalanced. Finally, we will also impose the condition that the quiver is 
good in the Gaiotto-Witten sense \footnote{For a non-linear graph, the balance of individual gauge nodes does not guarantee that the quiver is good, it will depend on the specifics of the graph in question.}.

Given the quiver gauge theory $\CT$, one can construct a duality sequence by implementing step-wise a set of 
quiver mutations, which are built out of the dualities summarized in Table \ref{Tab: Review}. We will first discuss the 
quiver mutations individually and then discuss the construction of the duality sequence using these mutations.

\subsection*{Mutation $I$}

For a given balanced special unitary gauge node in $\CT$, one can use the duality $\CD^N_{2N+1,1}$ 
in Table \ref{Tab: Review} (in reverse) locally to replace the special unitary node with a unitary factor with the same number of 
fundamental hypers and a single Abelian hypermultiplet. We will refer to this operation as \textit{mutation $I$} 
and denote it as $\CO_{I}$:
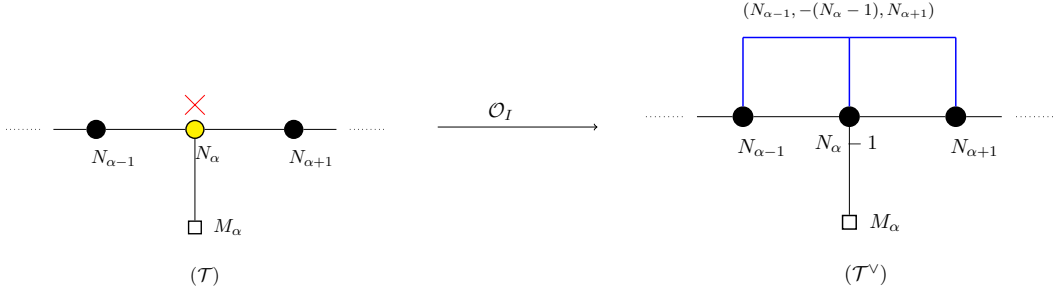
\begin{center}
\begin{tabular}{ccc}
\scalebox{0.65}{\begin{tikzpicture}
\node[] (1) at (1,0){};
\node[] (100) at (0,0){};
\node[unode] (2) at (2,0){};
\node[sunode] (3) at (4,0){};
\node[unode] (4) at (6,0){};
\node[] (5) at (7,0){};
\node[] (200) at (8,0){};
\node[cross, red] (6) at (4,0.5){};
\node[fnode] (8) at (4,-2){};
\draw[-] (1) -- (2);
\draw[-] (2)-- (3);
\draw[-] (3) -- (4);
\draw[-] (4) --(5);
\draw[-] (3) --(8);
\draw[-, dotted] (1) -- (100);
\draw[-,dotted] (5) -- (200);
\node[text width=.2cm](11) [below=0.1cm of 2]{$N_{\alpha-1}$};
\node[text width=.2cm](12) at (4.1, -0.5){$N_\alpha$};
\node[text width=.2cm](13) [below=0.1cm of 4]{$N_{\alpha+1}$};
\node[text width=.2cm](14) [right=0.1cm of 8]{$M_\alpha$};
\node[text width=.2cm](20) at (4,-3){$(\CT)$};
\end{tikzpicture}}
& \scalebox{.7}{\begin{tikzpicture}
\draw[->] (0,0) -- (3,0);
\node[text width=0.1cm](29) at (1, 0.3) {$\CO_{I}$};
\node[](30) at (0, -3) {};
\end{tikzpicture}}
&\scalebox{0.7}{\begin{tikzpicture}
\node[] (1) at (1,0){};
\node[] (100) at (0,0){};
\node[unode] (2) at (2,0){};
\node[unode] (3) at (4,0){};
\node[unode] (4) at (6,0){};
\node[] (5) at (7,0){};
\node[] (200) at (8,0){};
\node[fnode] (8) at (4,-2){};
\draw[-] (1) -- (2);
\draw[-] (2)-- (3);
\draw[-] (3) -- (4);
\draw[-] (4) --(5);
\draw[-] (3) --(8);
\draw[-, thick, blue] (2)--(2,1.5);
\draw[-, thick, blue] (3)--(4,1.5);
\draw[-, thick, blue] (4)--(6,1.5);
\draw[-, thick, blue] (2,1.5)--(4,1.5);
\draw[-, thick, blue] (4,1.5)--(6,1.5);
\node[text width=4 cm](10) at (4, 2){\footnotesize{$(N_{\alpha-1}, -(N_\alpha -1), N_{\alpha +1})$}};
\draw[-, dotted] (1) -- (100);
\draw[-,dotted] (5) -- (200);
\node[text width=.2cm](11) [below=0.1cm of 2]{$N_{\alpha-1}$};
\node[text width=1.5cm](12) at (4.1, -0.5){$N_\alpha-1$};
\node[text width=.2cm](13) [below=0.1cm of 4]{$N_{\alpha+1}$};
\node[text width=.2cm](14) [right=0.1cm of 8]{$M_\alpha$};
\node[text width=.2cm](20) at (4,-3){$(\CT^\vee)$};
\end{tikzpicture}}
\end{tabular}
\end{center}

Since $SU(N_\alpha)$ is a balanced node, the labels of the gauge and the flavor nodes obey the 
constraint: $N_{\alpha-1} + N_{\alpha +1} +M_\alpha =2N_\alpha-1$. Since a subgroup of the flavor symmetry 
of the $SU(N_\alpha)$ node is gauged, the Abelian hyper is charged under all the neighboring unitary gauge nodes to which 
the $SU(N_\alpha)$ node is connected. \\

In terms of the sphere partition function, the quiver mutation can be realized as follows. The 
duality $\CD^N_{2N+1,1}$ is implemented by using the identity \eref{Id-1} locally 
for the $SU(N_\alpha)$ gauge node in the quiver $\CT$:
\begin{align}\label{Id-1-main}
Z^{(\CT)}&=  \int \, [d \vec s_\alpha] \, \frac{\delta(\tr \vec s_\alpha)\, Z_{\rm vec}(\vec s_\alpha)\, \Big[ \ldots \Big]}{\prod_{j,i,k}\,\ch{(s^j_\alpha - \s^i_{\alpha-1})} \,\ch{(s^j_\alpha - \s^k_{\alpha+1})}\, \prod_{j,a}\, \ch{(s^j_\alpha - m^a_\alpha)}} \nn \\
&=  \int \, [d \vec \s_\alpha] \, \frac{Z_{\rm vec}(\vec \s_\alpha)\, \Big[ \ldots \Big]}{\prod_{j,i,k}\,\ch{(\s^j_\alpha - \s^i_{\alpha-1})} \,\ch{(\s^j_\alpha - \s^k_{\alpha+1})}\,\prod_{j,a}\, \ch{(\s^j_\alpha - m^a_\alpha)}} \nn \\
& \qquad \times \frac{1}{\ch{(-\tr \vec \s_\alpha + \tr \vec \s_{\alpha-1} + \tr \vec \s_{\alpha+1} + \tr \vec m_\alpha)}} \nn \\
&= Z^{(\CT^\vee)}(\eta^\vee_\alpha=0, \ldots), 
\end{align}
where $\Big[ \ldots \Big]$ denotes the terms in the partition function independent of the $SU(N_\alpha)$ node 
and the $U(N_\alpha -1)$ node respectively, $\tr \vec m_\alpha = \sum^{M_\alpha}_{a=1}\,m^a_\alpha$, 
and $\eta_\alpha$ is the FI parameter of the $U(N_\alpha -1)$ gauge node in $\CT^\vee$. 
The charges for the Abelian hypermultiplet in $\CT^\vee$ can be simply read off
from the partition function -- the charge vector $\vec Q$ will have only three non-vanishing entries, i.e.
$\vec Q = (0,\ldots, N_{\alpha-1}, - (N_\alpha-1), N_{\alpha+1}, \ldots,0)$. \\

For a generic (non-linear) quiver, the mutation $I$ can be readily extended as follows: 

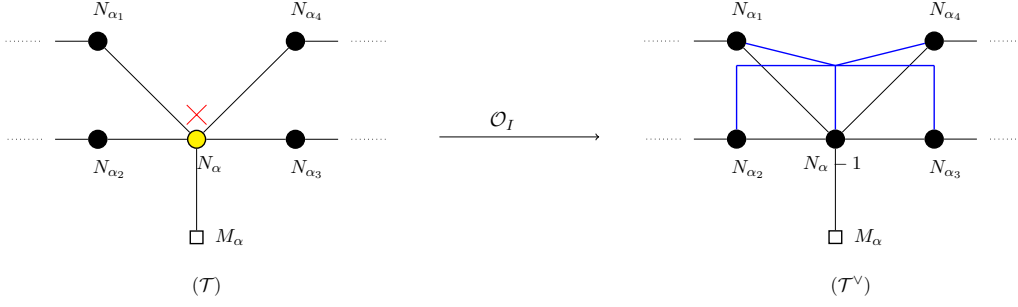
\begin{center}
\begin{tabular}{ccc}
\scalebox{0.65}{\begin{tikzpicture}
\node[] (1) at (1,0){};
\node[] (100) at (0,0){};
\node[unode] (2) at (2,0){};
\node[sunode] (3) at (4,0){};
\node[unode] (4) at (6,0){};
\node[unode] (51) at (2,2){};
\node[] (52) at (0,2){};
\node[] (53) at (1,2){};
\node[unode] (61) at (6,2){};
\node[] (62) at (7,2){};
\node[] (63) at (8,2){};
\node[] (5) at (7,0){};
\node[] (200) at (8,0){};
\node[cross, red] (6) at (4,0.5){};
\node[fnode] (8) at (4,-2){};
\draw[-] (1) -- (2);
\draw[-] (2)-- (3);
\draw[-] (3) -- (4);
\draw[-] (4) --(5);
\draw[-] (3) --(8);
\draw[-] (3) --(51);
\draw[-] (3) --(61);
\draw[-, dotted] (1) -- (100);
\draw[-,dotted] (5) -- (200);
\draw[-, dotted] (52) -- (53);
\draw[-] (51) -- (53);
\draw[-, dotted] (62) -- (63);
\draw[-] (61) -- (62);
\node[text width=.2cm](11) [below=0.1cm of 2]{$N_{\alpha_2}$};
\node[text width=.2cm](12) at (4.1, -0.5){$N_\alpha$};
\node[text width=.2cm](13) [below=0.1cm of 4]{$N_{\alpha_3}$};
\node[text width=.2cm](14) [right=0.1cm of 8]{$M_\alpha$};
\node[text width=.2cm](15) [above=0.1cm of 51]{$N_{\alpha_1}$};
\node[text width=.2cm](16) [above=0.1cm of 61]{$N_{\alpha_4}$};
\node[text width=.2cm](20) at (4,-3){$(\CT)$};
\end{tikzpicture}}
& \scalebox{.7}{\begin{tikzpicture}
\draw[->] (0,0) -- (3,0);
\node[text width=0.1cm](29) at (1, 0.3) {$\CO_{I}$};
\node[](30) at (0, -3) {};
\end{tikzpicture}}
& \scalebox{0.65}{\begin{tikzpicture}
\node[] (1) at (1,0){};
\node[] (100) at (0,0){};
\node[unode] (2) at (2,0){};
\node[unode] (3) at (4,0){};
\node[unode] (4) at (6,0){};
\node[unode] (51) at (2,2){};
\node[] (52) at (0,2){};
\node[] (53) at (1,2){};
\node[unode] (61) at (6,2){};
\node[] (62) at (7,2){};
\node[] (63) at (8,2){};
\node[] (5) at (7,0){};
\node[] (200) at (8,0){};
\node[fnode] (8) at (4,-2){};
\draw[-] (1) -- (2);
\draw[-] (2)-- (3);
\draw[-] (3) -- (4);
\draw[-] (4) --(5);
\draw[-] (3) --(8);
\draw[-] (3) --(51);
\draw[-] (3) --(61);
\draw[-, dotted] (1) -- (100);
\draw[-,dotted] (5) -- (200);
\draw[-, dotted] (52) -- (53);
\draw[-] (51) -- (53);
\draw[-, dotted] (62) -- (63);
\draw[-] (61) -- (62);
\draw[-, thick, blue] (2)--(2,1.5);
\draw[-, thick, blue] (3)--(4,1.5);
\draw[-, thick, blue] (4)--(6,1.5);
\draw[-, thick, blue] (2,1.5)--(4,1.5);
\draw[-, thick, blue] (4,1.5)--(6,1.5);
\draw[-, thick, blue] (4,1.5)--(51);
\draw[-, thick, blue] (4,1.5)--(61);
\node[text width=.2cm](11) [below=0.1cm of 2]{$N_{\alpha_2}$};
\node[text width=1.5cm](12) at (4.1, -0.5){$N_\alpha -1$};
\node[text width=.2cm](13) [below=0.1cm of 4]{$N_{\alpha_3}$};
\node[text width=.2cm](14) [right=0.1cm of 8]{$M_\alpha$};
\node[text width=.2cm](15) [above=0.1cm of 51]{$N_{\alpha_1}$};
\node[text width=.2cm](16) [above=0.1cm of 61]{$N_{\alpha_4}$};
\node[text width=.2cm](20) at (4,-3){$(\CT^\vee)$};
\end{tikzpicture}}
\end{tabular}
\end{center}

In the quiver $\CT$, the $SU(N_\alpha)$ node is balanced, i.e. $\sum_i \, N_{\alpha_i} + M_\alpha = 2N_\alpha -1$, where the sum 
extends over the neighboring nodes of $SU(N_\alpha)$ connected to $SU(N_\alpha)$ by bifundamental hypers. The charge vector 
$\vec Q$ in the theory $\CT^\vee$ will have non-vanishing entries for the $U(N_\alpha -1)$ node and its neighboring nodes, i.e. 
$\vec Q = (0,\ldots, N_{\alpha_1},  N_{\alpha_2}, - (N_\alpha-1), N_{\alpha_3}, N_{\alpha_4},\ldots,0)$.\\

 
 \subsection*{Mutation $I'$}
 
 Consider a gauge node $U(N)$ with $2N+1$ fundamental/bifundamental hypers plus a single Abelian hyper, 
where the Abelian hyper can be charged under any of the other unitary gauge node in the linear quiver. 
We can use the duality $\CD^N_{2N+1,1}$ in Table \ref{Tab: Review} locally at the gauge node to replace the unitary factor with 
a special unitary node. We will refer to this operation as \textit{mutation $I'$} and denote it as $\CO_{I'}$:\\

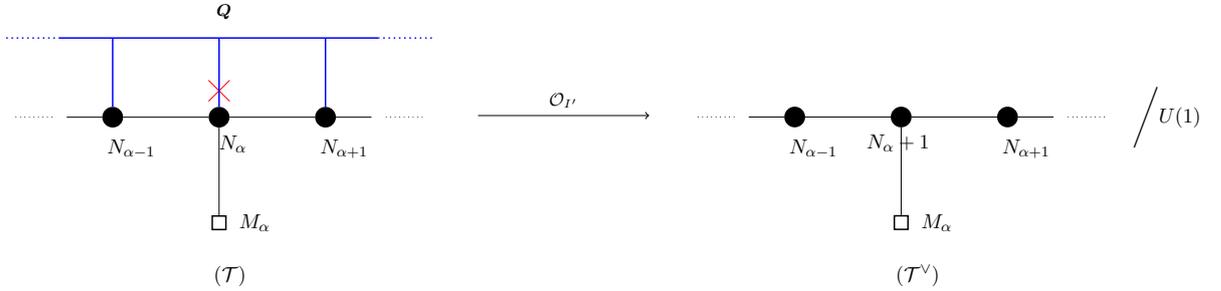
\begin{center}
\begin{tabular}{ccc}
\scalebox{0.7}{\begin{tikzpicture}
\node[] (1) at (1,0){};
\node[] (100) at (0,0){};
\node[unode] (2) at (2,0){};
\node[unode] (3) at (4,0){};
\node[unode] (4) at (6,0){};
\node[] (5) at (7,0){};
\node[] (200) at (8,0){};
\node[fnode] (8) at (4,-2){};
\node[cross, red] (6) at (4,0.5){};
\draw[-] (1) -- (2);
\draw[-] (2)-- (3);
\draw[-] (3) -- (4);
\draw[-] (4) --(5);
\draw[-] (3) --(8);
\draw[-, thick, blue] (2)--(2,1.5);
\draw[-, thick, blue] (3)--(4,1.5);
\draw[-, thick, blue] (4)--(6,1.5);
\draw[-, thick, blue] (2,1.5)--(4,1.5);
\draw[-, thick, blue] (4,1.5)--(6,1.5);
\draw[dotted, thick, blue] (0,1.5)--(1,1.5);
\draw[-, thick, blue] (1,1.5)--(2,1.5);
\draw[dotted, thick, blue] (8,1.5)--(7,1.5);
\draw[-, thick, blue] (6,1.5)--(7,1.5);
\node[text width=0.1 cm](10) at (4, 2){\footnotesize{$\vec Q$}};
\draw[-, dotted] (1) -- (100);
\draw[-,dotted] (5) -- (200);
\node[text width=.2cm](11) [below=0.1cm of 2]{$N_{\alpha-1}$};
\node[text width=0.2cm](12) at (4.1, -0.5){$N_\alpha$};
\node[text width=.2cm](13) [below=0.1cm of 4]{$N_{\alpha+1}$};
\node[text width=.2cm](14) [right=0.1cm of 8]{$M_\alpha$};
\node[text width=.2cm](20) at (4,-3){$(\CT)$};
\end{tikzpicture}}
& \scalebox{.75}{\begin{tikzpicture}
\draw[->] (0,0) -- (3,0) node   [midway,above  ] {\footnotesize {$\CO_{I'}$}};
\node[](1) at (0,-3.0){};
\end{tikzpicture}}
& \scalebox{0.7}{\begin{tikzpicture}
\node[] (1) at (1,0){};
\node[] (100) at (0,0){};
\node[unode] (2) at (2,0){};
\node[unode] (3) at (4,0){};
\node[unode] (4) at (6,0){};
\node[] (5) at (7,0){};
\node[] (200) at (8,0){};
\node[fnode] (8) at (4,-2){};
\draw[-] (1) -- (2);
\draw[-] (2)-- (3);
\draw[-] (3) -- (4);
\draw[-] (4) --(5);
\draw[-] (3) --(8);
\node[text width=0.3 cm](10) at (8.5,0){$\Bigg/ U(1)$};
\draw[-, dotted] (1) -- (100);
\draw[-,dotted] (5) -- (200);
\node[text width=.2cm](11) [below=0.1cm of 2]{$N_{\alpha-1}$};
\node[text width=1.5cm](12) at (4.1, -0.5){$N_\alpha +1$};
\node[text width=.2cm](13) [below=0.1cm of 4]{$N_{\alpha+1}$};
\node[text width=.2cm](14) [right=0.1cm of 8]{$M_\alpha$};
\node[text width=.2cm](20) at (4,-3){$(\CT^\vee)$};
\end{tikzpicture}}
\end{tabular}
\end{center}

The labels of the gauge and the flavor nodes obey the constraint: $N_{\alpha-1} + N_{\alpha +1} +M_\alpha =2N_\alpha +1$. 
In contrast to the quiver obtained via mutation $I$ above, the Abelian hypermultiplet in $(\CT)$ may be charged 
under any of the unitary gauge nodes and not just the neighboring gauges of $U(N_\alpha)$. On the RHS, the quiver 
$(\CT^\vee)$ is obtained by replacing $N_\alpha \to N_\alpha +1$ (keeping all the other gauge/flavor nodes unchanged), 
and gauging a specific $\fru(1)$ subalgebra of the topological symmetry algebra, corresponding to the generator:
\be
J_G = \sum_a\, \Big(\frac{Q_a}{N_a}\Big)\, J_a  + J_{\alpha -1} + J_{\alpha+1} - J_\alpha, 
\ee
where $J_i$ denotes the topological symmetry generator associated with the gauge group $U(N_i)$, and 
 the index $a$ runs over all the unitary gauge nodes in $(\CT)$ under which the Abelian hypermultiplet is 
charged, in addition to $U(N_\alpha)$. The gauging operation amounts to ungauging a $U(1)$ subgroup of the 
gauge group and is denoted by $``\Bigg/ U(1)"$ in the quiver.\\

In terms of the sphere partition function, the quiver mutation can be realized as follows. Using a simple change of variables, 
one can rewrite the identity \eref{Id-1} in the following fashion:
\begin{align}
& Z^{\CT^{N}_{2N+1,1}}(\vec m, m_{\rm ab}, \eta=0) = Z^{SU(N+1),\,2N +1}(\vec m^\vee), \nn \\
& m^\vee_i = m_i - \frac{\tr \vec m - m_{\rm ab}}{N+1}, \qquad i=1, \ldots, 2N+1,
\end{align}
where $\vec m$ denote the real masses for the fundamental hypers with $\tr \vec m= \sum^{2N+1}_{i=1}\,m_i$, and $m_{\rm ab}$ is the real 
mass for the Abelian hyper. Using this identity locally for the $U(N_\alpha)$ gauge node in the quiver $\CT$, we obtain:
\begin{align}\label{Id-1a-main}
Z^{(\CT)}&=  \int \, [d \vec s_\alpha] \, \frac{Z_{\rm vec}(\vec s_\alpha)\, \Big[ \ldots \Big]}{\prod_{\beta= \alpha \pm 1}\,\prod_{j,i}\,\ch{(s^j_\alpha - \s^i_{\beta})} \, \prod_{j,a}\,\ch{(s^j_\alpha - m^a_\alpha)}\, \ch{(\tr \vec s_\alpha + \sum_a \frac{Q_a}{N_a}\, \tr \vec\s_a)}} \nn \\
&=  \int \, [d \vec \s_\alpha] \, \frac{ \delta \Big(\tr \vec \s_\alpha \Big)\, Z_{\rm vec}(\vec \s_\alpha)\, \Big[ \ldots \Big]}{\prod_{\beta= \alpha \pm 1}\,\prod_{j,i}\,\ch{(\s^j_\alpha - \s^i_{\beta} +\delta)} \,\prod_{j,a}\, \ch{(\s^j_\alpha - m^a_\alpha + \delta)}},
\end{align}
where $\delta=\frac{1}{N_\alpha +1}\Big(\tr \vec m_\alpha + \tr \s_{\alpha -1} + \tr \s_{\alpha +1} + \sum_a \frac{Q_a}{N_a}\, \tr \vec\s_a \Big)$, 
$\vec s_\alpha$ and $\vec \s_\alpha$ live in the Cartan subalgebra of $U(N_\alpha)$ and $U(N_\alpha +1)$ respectively. 

Now, consider the change of variables: 
\be
\s^j_\alpha \to \s^j_\alpha - \delta, \qquad \tr \vec \s_\alpha \to \tr \vec \s_\alpha -\tr \vec m_\alpha - \tr \s_{\alpha -1} - \tr \s_{\alpha +1} - \sum_a \frac{Q_a}{N_a}\, \tr \vec\s_a.
\ee
The partition function identity can then be written as
\begin{align}\label{Id-1b-main}
Z^{(\CT)}= & \int \, [d \vec \s_\alpha] \, \frac{ \delta \Big(\tr \vec \s_\alpha -\tr \vec m_\alpha - \tr \s_{\alpha -1} - \tr \s_{\alpha +1} - \sum_a \frac{Q_a}{N_a}\, \tr \vec\s_a \Big)\, Z_{\rm vec}(\vec \s_\alpha)\, \Big[ \ldots \Big]}{\prod_{\beta= \alpha \pm 1}\,\prod_{j,i}\,\ch{(\s^j_\alpha - \s^i_{\beta})} \,\prod_{j,a}\, \ch{(\s^j_\alpha - m^a_\alpha)}} \nn \\
=& Z^{(\CT^\vee)},
\end{align}
where $\CT^\vee$ can be readily identified as the quiver above, and the precise $\fru(1)$ subalgebra to be ungauged can be read off 
from the argument of the delta function in the matrix model. Note that in the special case where 
$\tr \s_{\alpha -1} + \tr \s_{\alpha +1} + \sum_a \frac{Q_a}{N_a}\, \tr \vec\s_a =0$, the ungauging operation simply gives an 
$SU(N_\alpha +1)$ node. In this case, the operation $\CO_{I'}$ simply reduces to the inverse of the $\CO_I$ operation studied above. \\

For a generic (non-linear) quiver, the mutation I$'$ can be readily extended as follows: 

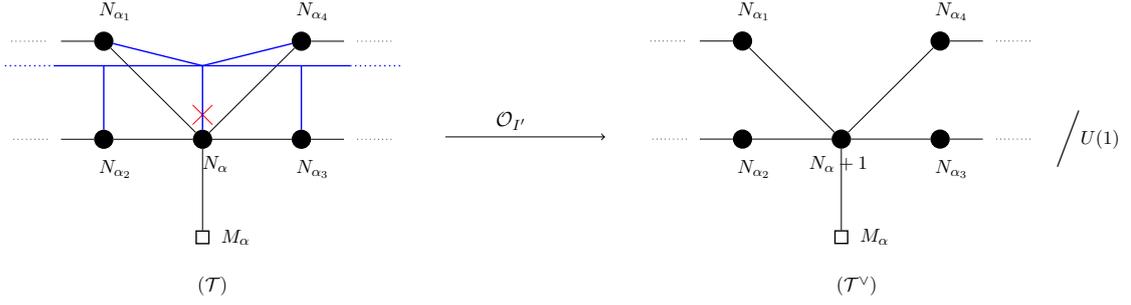
\begin{center}
\begin{tabular}{ccc}
\scalebox{0.65}{\begin{tikzpicture}
\node[] (1) at (1,0){};
\node[] (100) at (0,0){};
\node[unode] (2) at (2,0){};
\node[unode] (3) at (4,0){};
\node[unode] (4) at (6,0){};
\node[unode] (51) at (2,2){};
\node[] (52) at (0,2){};
\node[] (53) at (1,2){};
\node[unode] (61) at (6,2){};
\node[] (62) at (7,2){};
\node[] (63) at (8,2){};
\node[] (5) at (7,0){};
\node[] (200) at (8,0){};
\node[cross, red] (6) at (4,0.5){};
\node[fnode] (8) at (4,-2){};
\draw[-] (1) -- (2);
\draw[-] (2)-- (3);
\draw[-] (3) -- (4);
\draw[-] (4) --(5);
\draw[-] (3) --(8);
\draw[-] (3) --(51);
\draw[-] (3) --(61);
\draw[-, dotted] (1) -- (100);
\draw[-,dotted] (5) -- (200);
\draw[-, dotted] (52) -- (53);
\draw[-] (51) -- (53);
\draw[-, dotted] (62) -- (63);
\draw[-] (61) -- (62);
\draw[dotted, thick, blue] (0,1.5)--(1,1.5);
\draw[-, thick, blue] (1,1.5)--(2,1.5);
\draw[-, thick, blue] (2)--(2,1.5);
\draw[-, thick, blue] (3)--(4,1.5);
\draw[-, thick, blue] (4)--(6,1.5);
\draw[-, thick, blue] (2,1.5)--(4,1.5);
\draw[-, thick, blue] (4,1.5)--(6,1.5);
\draw[-, thick, blue] (4,1.5)--(51);
\draw[-, thick, blue] (4,1.5)--(61);
\draw[-, thick, blue] (7,1.5)--(6,1.5);
\draw[dotted, thick, blue] (8,1.5)--(7,1.5);
\node[text width=.2cm](11) [below=0.1cm of 2]{$N_{\alpha_2}$};
\node[text width=.2cm](12) at (4.1, -0.5){$N_\alpha$};
\node[text width=.2cm](13) [below=0.1cm of 4]{$N_{\alpha_3}$};
\node[text width=.2cm](14) [right=0.1cm of 8]{$M_\alpha$};
\node[text width=.2cm](15) [above=0.1cm of 51]{$N_{\alpha_1}$};
\node[text width=.2cm](16) [above=0.1cm of 61]{$N_{\alpha_4}$};
\node[text width=.2cm](20) at (4,-3){$(\CT)$};
\end{tikzpicture}}
& \scalebox{.7}{\begin{tikzpicture}
\draw[->] (0,0) -- (3,0);
\node[text width=0.1cm](29) at (1, 0.3) {$\CO_{I'}$};
\node[](30) at (0, -3) {};
\end{tikzpicture}}
& \scalebox{0.65}{\begin{tikzpicture}
\node[] (1) at (1,0){};
\node[] (100) at (0,0){};
\node[unode] (2) at (2,0){};
\node[unode] (3) at (4,0){};
\node[unode] (4) at (6,0){};
\node[unode] (51) at (2,2){};
\node[] (52) at (0,2){};
\node[] (53) at (1,2){};
\node[unode] (61) at (6,2){};
\node[] (62) at (7,2){};
\node[] (63) at (8,2){};
\node[] (5) at (7,0){};
\node[] (200) at (8,0){};
\node[fnode] (8) at (4,-2){};
\draw[-] (1) -- (2);
\draw[-] (2)-- (3);
\draw[-] (3) -- (4);
\draw[-] (4) --(5);
\draw[-] (3) --(8);
\draw[-] (3) --(51);
\draw[-] (3) --(61);
\draw[-, dotted] (1) -- (100);
\draw[-,dotted] (5) -- (200);
\draw[-, dotted] (52) -- (53);
\draw[-] (51) -- (53);
\draw[-, dotted] (62) -- (63);
\draw[-] (61) -- (62);
\node[text width=.2cm](11) [below=0.1cm of 2]{$N_{\alpha_2}$};
\node[text width=1.5cm](12) at (4.1, -0.5){$N_\alpha +1$};
\node[text width=.2cm](13) [below=0.1cm of 4]{$N_{\alpha_3}$};
\node[text width=.2cm](14) [right=0.1cm of 8]{$M_\alpha$};
\node[text width=.2cm](15) [above=0.1cm of 51]{$N_{\alpha_1}$};
\node[text width=.2cm](16) [above=0.1cm of 61]{$N_{\alpha_4}$};
\node[text width=0.3 cm](10) at (8.5,0){$\Bigg/ U(1)$};
\node[text width=.2cm](20) at (4,-3){$(\CT^\vee)$};
\end{tikzpicture}}
\end{tabular}
\end{center}

In the theory $\CT$, the labels of the gauge and the flavor nodes obey the constraint: $\sum_i\, N_{\alpha_i} +M_\alpha =2N_\alpha +1$. 
The $\fru(1)$ subalgebra to be gauged is given as:
\be
J_G = \sum_a\, \Big(\frac{Q_a}{N_a}\Big)\, J_a  + \sum_i\, J_{\alpha_i}  - J_\alpha, 
\ee
where the first sum extends over all the gauge nodes (aside from the $U(N_\alpha)$ node) under which the Abelian hypermultiplet in 
$\CT$ is charged, and the second sum extends over all the gauge nodes which are connected to $U(N_\alpha)$ by bifundamental 
hypers. 

\subsection*{Mutation $II$}

Consider a gauge node $U(N)$ with $2N$ fundamental/bifundamental hypers plus a single Abelian hyper, 
where the Abelian hyper can be charged under any of the other unitary gauge node in the linear quiver. 
We can use the duality $\CD^N_{2N,1}$ in Table \ref{Tab: Review} locally at the gauge node. 
We will refer to this operation as \textit{mutation $I$I} and denote it as $\CO_{II}$:

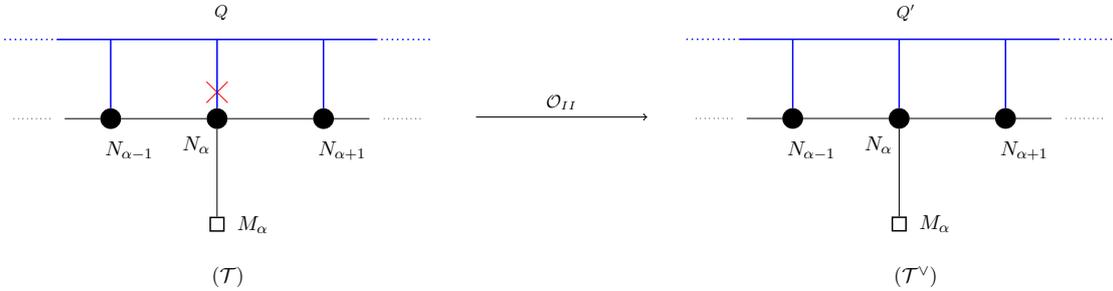
\begin{center}
\begin{tabular}{ccc}
\scalebox{0.7}{\begin{tikzpicture}
\node[] (1) at (1,0){};
\node[] (100) at (0,0){};
\node[unode] (2) at (2,0){};
\node[unode] (3) at (4,0){};
\node[unode] (4) at (6,0){};
\node[] (5) at (7,0){};
\node[cross, red] (6) at (4,0.5){};
\node[] (200) at (8,0){};
\node[fnode] (8) at (4,-2){};
\draw[-] (1) -- (2);
\draw[-] (2)-- (3);
\draw[-] (3) -- (4);
\draw[-] (4) --(5);
\draw[-] (3) --(8);
\draw[-, thick, blue] (2)--(2,1.5);
\draw[-, thick, blue] (3)--(4,1.5);
\draw[-, thick, blue] (4)--(6,1.5);
\draw[-, thick, blue] (2,1.5)--(4,1.5);
\draw[-, thick, blue] (4,1.5)--(6,1.5);
\draw[dotted, thick, blue] (0,1.5)--(1,1.5);
\draw[-, thick, blue] (1,1.5)--(2,1.5);
\draw[dotted, thick, blue] (8,1.5)--(7,1.5);
\draw[-, thick, blue] (6,1.5)--(7,1.5);
\node[text width=0.1 cm](10) at (4, 2){\footnotesize{$Q$}};
\draw[-, dotted] (1) -- (100);
\draw[-,dotted] (5) -- (200);
\node[text width=.2cm](11) [below=0.1cm of 2]{$N_{\alpha-1}$};
\node[text width=1.5cm](12) at (4.1, -0.5){$N_\alpha$};
\node[text width=.2cm](13) [below=0.1cm of 4]{$N_{\alpha+1}$};
\node[text width=.2cm](14) [right=0.1cm of 8]{$M_\alpha$};
\node[text width=.2cm](20) at (4,-3){$(\CT)$};
\end{tikzpicture}}
& \scalebox{.75}{\begin{tikzpicture}
\draw[->] (0,0) -- (3,0) node   [midway,above  ] {\footnotesize {$\CO_{II}$}};
\node[](1) at (0,-3.0){};
\end{tikzpicture}}
& \scalebox{0.7}{\begin{tikzpicture}
\node[] (1) at (1,0){};
\node[] (100) at (0,0){};
\node[unode] (2) at (2,0){};
\node[unode] (3) at (4,0){};
\node[unode] (4) at (6,0){};
\node[] (5) at (7,0){};
\node[] (200) at (8,0){};
\node[fnode] (8) at (4,-2){};
\draw[-] (1) -- (2);
\draw[-] (2)-- (3);
\draw[-] (3) -- (4);
\draw[-] (4) --(5);
\draw[-] (3) --(8);
\draw[-, thick, blue] (2)--(2,1.5);
\draw[-, thick, blue] (3)--(4,1.5);
\draw[-, thick, blue] (4)--(6,1.5);
\draw[-, thick, blue] (2,1.5)--(4,1.5);
\draw[-, thick, blue] (4,1.5)--(6,1.5);
\draw[dotted, thick, blue] (0,1.5)--(1,1.5);
\draw[-, thick, blue] (1,1.5)--(2,1.5);
\draw[dotted, thick, blue] (8,1.5)--(7,1.5);
\draw[-, thick, blue] (6,1.5)--(7,1.5);
\node[text width=0.1 cm](10) at (4, 2){\footnotesize{$Q'$}};
\draw[-, dotted] (1) -- (100);
\draw[-,dotted] (5) -- (200);
\node[text width=.2cm](11) [below=0.1cm of 2]{$N_{\alpha-1}$};
\node[text width=1.5cm](12) at (4.1, -0.5){$N_\alpha$};
\node[text width=.2cm](13) [below=0.1cm of 4]{$N_{\alpha+1}$};
\node[text width=.2cm](14) [right=0.1cm of 8]{$M_\alpha$};
\node[text width=.2cm](20) at (4,-3){$(\CT^\vee)$};
\end{tikzpicture}}
\end{tabular}
\end{center}

The labels of the gauge and the flavor nodes obey the constraint: $N_{\alpha-1} + N_{\alpha +1} +M_\alpha =2N_\alpha$.
On the left, the blue line (solid and dashed) denotes a single Abelian hypermultiplet associated with the charge vector 
$\vec Q = (Q_1, \ldots, Q_{\alpha-1}, N_\alpha, Q_{\alpha+1}, \ldots, Q_L)$, with $L$ being the number of gauge nodes in the quiver. 
Note that any number of entries in $\vec Q$ (except for the $\alpha$-th entry) can be zero. For the quiver on the right, the gauge and 
flavor nodes remain the same, while the single Abelian hypermultiplet is now associated with a charge vector $\vec Q'=(-Q_1, \ldots, -Q_{\alpha-2}, -Q_{\alpha-1} - N_{\alpha-1}, N_\alpha, -Q_{\alpha+1} - N_{\alpha+1} , -Q_{\alpha+2}, \ldots, -Q_L)$. One can check that this operation 
squares to an identity operation. \\

In terms of the sphere partition function, the quiver mutation can be realized as follows. The 
duality $\CD^N_{2N,1}$ is implemented by using the identity \eref{Id-2b} locally for the $U(N_\alpha)$ gauge node in the quiver $\CT$ :
\begin{align}\label{Id-2-main}
Z^{(\CT)}&=  \int \, [d \vec s_\alpha] \, \frac{ e^{2\pi i \eta_\alpha\, \tr \vec s_\alpha}\, Z_{\rm vec}(\vec s_\alpha)\, \Big[ \ldots \Big]}{\prod_{\beta= \alpha \pm 1}\,\prod_{j,i}\,\ch{(s^j_\alpha - \s^i_{\beta})} \, \prod_{j,a}\,\ch{(s^j_\alpha - m^a_\alpha)}\, \ch{(\tr \vec s_\alpha + \sum_a \frac{Q_a}{N_a}\, \tr \vec\s_a)}} \nn \\
&=  \int \, [d \vec \s_\alpha] \, \frac{e^{2\pi i \eta_\alpha\, (\tr \vec \s_{\alpha-1} + \tr \vec \s_{\alpha+1})}\,e^{-2\pi i \eta_\alpha\, \tr \vec \s_\alpha}\, Z_{\rm vec}(\vec \s_\alpha)\, \Big[ \ldots \Big]}{\prod_{\beta= \alpha \pm 1}\,\prod_{j,i}\,\ch{(\s^j_\alpha - \s^i_{\beta})} \,\prod_{j,a}\, \ch{(\s^j_\alpha - m^a_\alpha)}} \nn \\
& \qquad \times \frac{1}{\ch{(\tr \vec \s_\alpha - \sum_{\beta= \alpha \pm 1}\, \tr \vec \s_{\beta}  - \sum_a \frac{Q_a}{N_a}\, \tr \vec\s_a - \tr \vec m_\alpha)}} \nn \\
&= Z^{(\CT^\vee)}(\eta^\vee_{\alpha-1}=\eta_{\alpha-1}+ \eta_\alpha, \eta^\vee_\alpha=- \eta_\alpha, \eta^\vee_{\alpha+1}= \eta_{\alpha+1}+ \eta_\alpha, \ldots).
\end{align}
The charge vector $\vec Q'$ associated with the quiver $\CT$ can be read off from the second equality.\\

The above computation can be readily extended for $P>1$. In this case, the mutation preserves the gauge and the flavors nodes 
as before, while mapping $P$ Abelian hypers of charges $\{ \vec Q^l\}_{l=1,\ldots,P}$ to another $P$ Abelian hypers of charges 
$\{ \vec Q'^l\}_{l=1,\ldots,P}$. The map of the charges is given as follows:
\begin{align}
& \vec Q^l \to \vec Q'^l, \qquad \text{for}\,\, l=1,\ldots,P, \nn \\
& \vec Q^l= (Q^l_1, \ldots, Q^l_{\alpha-1}, N_\alpha, Q^l_{\alpha+1}, \ldots, Q^l_L), \nn \\
& \vec Q'^l=(-Q^l_1, \ldots, -Q^l_{\alpha-2}, -Q^l_{\alpha-1} - N_{\alpha-1}, N_\alpha, -Q^l_{\alpha+1} - N_{\alpha+1} , -Q^l_{\alpha+2}, \ldots, -Q^l_L).
\end{align}
Similar to the $P=1$ case, one can check that this operation squares to an identity operation.\\

The extension of the mutation $II$ to a generic quiver assumes the following form:

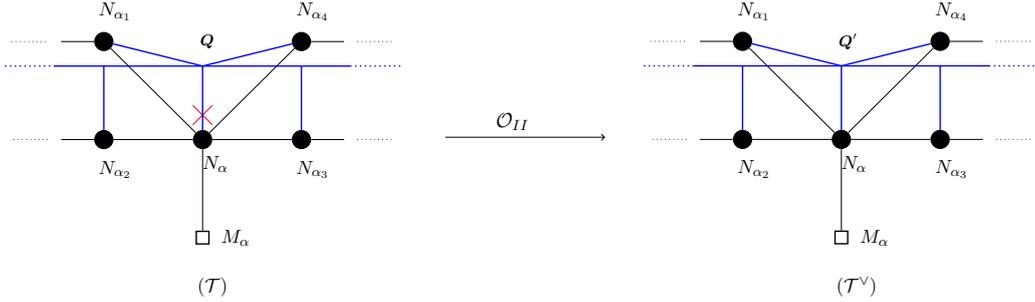
\begin{center}
\begin{tabular}{ccc}
\scalebox{0.65}{\begin{tikzpicture}
\node[] (1) at (1,0){};
\node[] (100) at (0,0){};
\node[unode] (2) at (2,0){};
\node[unode] (3) at (4,0){};
\node[unode] (4) at (6,0){};
\node[unode] (51) at (2,2){};
\node[] (52) at (0,2){};
\node[] (53) at (1,2){};
\node[unode] (61) at (6,2){};
\node[] (62) at (7,2){};
\node[] (63) at (8,2){};
\node[] (5) at (7,0){};
\node[] (200) at (8,0){};
\node[cross, red] (6) at (4,0.5){};
\node[fnode] (8) at (4,-2){};
\draw[-] (1) -- (2);
\draw[-] (2)-- (3);
\draw[-] (3) -- (4);
\draw[-] (4) --(5);
\draw[-] (3) --(8);
\draw[-] (3) --(51);
\draw[-] (3) --(61);
\draw[-, dotted] (1) -- (100);
\draw[-,dotted] (5) -- (200);
\draw[-, dotted] (52) -- (53);
\draw[-] (51) -- (53);
\draw[-, dotted] (62) -- (63);
\draw[-] (61) -- (62);
\draw[dotted, thick, blue] (0,1.5)--(1,1.5);
\draw[-, thick, blue] (1,1.5)--(2,1.5);
\draw[-, thick, blue] (2)--(2,1.5);
\draw[-, thick, blue] (3)--(4,1.5);
\draw[-, thick, blue] (4)--(6,1.5);
\draw[-, thick, blue] (2,1.5)--(4,1.5);
\draw[-, thick, blue] (4,1.5)--(6,1.5);
\draw[-, thick, blue] (4,1.5)--(51);
\draw[-, thick, blue] (4,1.5)--(61);
\draw[-, thick, blue] (7,1.5)--(6,1.5);
\draw[dotted, thick, blue] (8,1.5)--(7,1.5);
\node[text width=.2cm](11) [below=0.1cm of 2]{$N_{\alpha_2}$};
\node[text width=.2cm](12) at (4.1, -0.5){$N_\alpha$};
\node[text width=.2cm](13) [below=0.1cm of 4]{$N_{\alpha_3}$};
\node[text width=.2cm](14) [right=0.1cm of 8]{$M_\alpha$};
\node[text width=.2cm](15) [above=0.1cm of 51]{$N_{\alpha_1}$};
\node[text width=.2cm](16) [above=0.1cm of 61]{$N_{\alpha_4}$};
\node[text width=0.1 cm](10) at (4, 2){\footnotesize{$\vec Q$}};
\node[text width=.2cm](20) at (4,-3){$(\CT)$};
\end{tikzpicture}}
& \scalebox{.7}{\begin{tikzpicture}
\draw[->] (0,0) -- (3,0);
\node[text width=0.1cm](29) at (1, 0.3) {$\CO_{II}$};
\node[](30) at (0, -3) {};
\end{tikzpicture}}
& \scalebox{0.65}{\begin{tikzpicture}
\node[] (1) at (1,0){};
\node[] (100) at (0,0){};
\node[unode] (2) at (2,0){};
\node[unode] (3) at (4,0){};
\node[unode] (4) at (6,0){};
\node[unode] (51) at (2,2){};
\node[] (52) at (0,2){};
\node[] (53) at (1,2){};
\node[unode] (61) at (6,2){};
\node[] (62) at (7,2){};
\node[] (63) at (8,2){};
\node[] (5) at (7,0){};
\node[] (200) at (8,0){};
\node[fnode] (8) at (4,-2){};
\draw[-] (1) -- (2);
\draw[-] (2)-- (3);
\draw[-] (3) -- (4);
\draw[-] (4) --(5);
\draw[-] (3) --(8);
\draw[-] (3) --(51);
\draw[-] (3) --(61);
\draw[-, dotted] (1) -- (100);
\draw[-,dotted] (5) -- (200);
\draw[-, dotted] (52) -- (53);
\draw[-] (51) -- (53);
\draw[-, dotted] (62) -- (63);
\draw[-] (61) -- (62);
\draw[dotted, thick, blue] (0,1.5)--(1,1.5);
\draw[-, thick, blue] (1,1.5)--(2,1.5);
\draw[-, thick, blue] (2)--(2,1.5);
\draw[-, thick, blue] (3)--(4,1.5);
\draw[-, thick, blue] (4)--(6,1.5);
\draw[-, thick, blue] (2,1.5)--(4,1.5);
\draw[-, thick, blue] (4,1.5)--(6,1.5);
\draw[-, thick, blue] (4,1.5)--(51);
\draw[-, thick, blue] (4,1.5)--(61);
\draw[-, thick, blue] (7,1.5)--(6,1.5);
\draw[dotted, thick, blue] (8,1.5)--(7,1.5);
\node[text width=.2cm](11) [below=0.1cm of 2]{$N_{\alpha_2}$};
\node[text width=.2cm](12) at (4.1, -0.5){$N_\alpha$};
\node[text width=.2cm](13) [below=0.1cm of 4]{$N_{\alpha_3}$};
\node[text width=.2cm](14) [right=0.1cm of 8]{$M_\alpha$};
\node[text width=.2cm](15) [above=0.1cm of 51]{$N_{\alpha_1}$};
\node[text width=.2cm](16) [above=0.1cm of 61]{$N_{\alpha_4}$};
\node[text width=0.1 cm](10) at (4, 2){\footnotesize{$\vec Q'$}};
\node[text width=.2cm](20) at (4,-3){$(\CT^\vee)$};
\end{tikzpicture}}
\end{tabular}
\end{center}

The labels of the gauge and the flavor nodes obey the constraint: $\sum_i\, N_{\alpha_i} +M_\alpha =2N_\alpha$. 
The map of the charges is given as follows:
\begin{align}
& \vec Q^l \to \vec Q'^l, \qquad \text{for}\,\, l=1,\ldots,P, \nn \\
& \vec Q^l= (Q^l_1, \ldots, Q^l_{\alpha_1}, Q^l_{\alpha_2}, N_\alpha, Q^l_{\alpha_3}, Q^l_{\alpha_4}, \ldots, Q^l_L), \nn \\
& \vec Q'^l=(-Q^l_1, \ldots, -Q^l_{\alpha_1} - N_{\alpha_1}, -Q^l_{\alpha_2} - N_{\alpha_2}, N_\alpha, -Q^l_{\alpha_3} - N_{\alpha_3} , -Q^l_{\alpha_4} - N_{\alpha_4}, \ldots, -Q^l_L).
\end{align}

\subsection*{Mutation $III$}

Consider a gauge node $U(N)$ with $2N-1$ fundamental/bifundamental hypers plus a single Abelian hyper, 
where the Abelian hyper can be charged under any of the other unitary gauge node in the linear quiver. 
We can use the duality $\CD^N_{2N-1,1}$ in Table \ref{Tab: Review} locally at the gauge node. 
We will refer to this operation as \textit{mutation $I$II} and denote it 
as $\CO_{III}$:

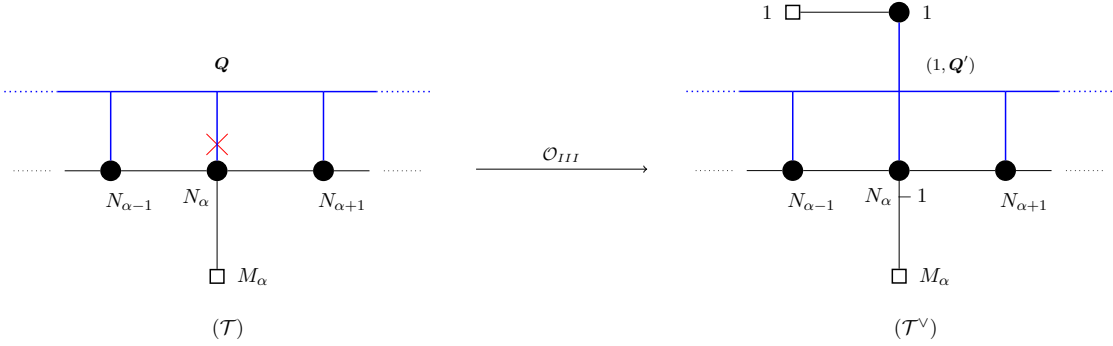
\begin{center}
\begin{tabular}{ccc}
\scalebox{0.7}{\begin{tikzpicture}
\node[] (1) at (1,0){};
\node[] (100) at (0,0){};
\node[unode] (2) at (2,0){};
\node[unode] (3) at (4,0){};
\node[unode] (4) at (6,0){};
\node[] (5) at (7,0){};
\node[cross, red] (6) at (4,0.5){};
\node[] (200) at (8,0){};
\node[fnode] (8) at (4,-2){};
\draw[-] (1) -- (2);
\draw[-] (2)-- (3);
\draw[-] (3) -- (4);
\draw[-] (4) --(5);
\draw[-] (3) --(8);
\draw[-, thick, blue] (2)--(2,1.5);
\draw[-, thick, blue] (3)--(4,1.5);
\draw[-, thick, blue] (4)--(6,1.5);
\draw[-, thick, blue] (2,1.5)--(4,1.5);
\draw[-, thick, blue] (4,1.5)--(6,1.5);
\draw[dotted, thick, blue] (0,1.5)--(1,1.5);
\draw[-, thick, blue] (1,1.5)--(2,1.5);
\draw[dotted, thick, blue] (8,1.5)--(7,1.5);
\draw[-, thick, blue] (6,1.5)--(7,1.5);
\node[text width=0.1 cm](10) at (4, 2){\footnotesize{$\vec Q$}};
\draw[-, dotted] (1) -- (100);
\draw[-,dotted] (5) -- (200);
\node[text width=.2cm](11) [below=0.1cm of 2]{$N_{\alpha-1}$};
\node[text width=1.5cm](12) at (4.1, -0.5){$N_\alpha$};
\node[text width=.2cm](13) [below=0.1cm of 4]{$N_{\alpha+1}$};
\node[text width=.2cm](14) [right=0.1cm of 8]{$M_\alpha$};
\node[text width=.2cm](20) at (4,-3){$(\CT)$};
\end{tikzpicture}}
& \scalebox{.75}{\begin{tikzpicture}
\draw[->] (0,0) -- (3,0) node   [midway,above  ] {\footnotesize {$\CO_{III}$}};
\node[](1) at (0,-3.0){};
\end{tikzpicture}}
& \scalebox{0.7}{\begin{tikzpicture}
\node[] (1) at (1,0){};
\node[] (100) at (0,0){};
\node[unode] (2) at (2,0){};
\node[unode] (3) at (4,0){};
\node[unode] (4) at (6,0){};
\node[] (5) at (7,0){};
\node[] (200) at (8,0){};
\node[fnode] (8) at (4,-2){};
\node[unode] (9) at (4,3){};
\node[fnode] (10) at (2,3){};
\draw[-] (1) -- (2);
\draw[-] (2)-- (3);
\draw[-] (3) -- (4);
\draw[-] (4) --(5);
\draw[-] (3) --(8);
\draw[-] (9) --(10);
\draw[-, thick, blue] (2)--(2,1.5);
\draw[-, thick, blue] (3)--(4,1.5);
\draw[-, thick, blue] (4)--(6,1.5);
\draw[-, thick, blue] (2,1.5)--(4,1.5);
\draw[-, thick, blue] (4,1.5)--(6,1.5);
\draw[dotted, thick, blue] (0,1.5)--(1,1.5);
\draw[-, thick, blue] (1,1.5)--(2,1.5);
\draw[dotted, thick, blue] (8,1.5)--(7,1.5);
\draw[-, thick, blue] (6,1.5)--(7,1.5);
\draw[-, thick, blue] (9)--(4,1.5);
\node[text width=2 cm](30) at (5.5, 2){\footnotesize{$(1,\vec Q')$}};
\draw[-, dotted] (1) -- (100);
\draw[-,dotted] (5) -- (200);
\node[text width=.2cm](11) [below=0.1cm of 2]{$N_{\alpha-1}$};
\node[text width=1.5cm](12) at (4.1, -0.5){$N_\alpha -1$};
\node[text width=.2cm](13) [below=0.1cm of 4]{$N_{\alpha+1}$};
\node[text width=.2cm](14) [right=0.1cm of 8]{$M_\alpha$};
\node[text width=.2cm](15) [right=0.1cm of 9]{1};
\node[text width=.2cm](16) [left=0.1cm of 10]{1};
\node[text width=.2cm](20) at (4,-3){$(\CT^\vee)$};
\end{tikzpicture}}
\end{tabular}
\end{center}

For the quiver $\CT$, the labels of the gauge and the flavor nodes at the node $\alpha$ obey the constraint: 
$N_{\alpha-1} + N_{\alpha +1} +M_\alpha =2N_\alpha -1$, and the blue line (solid and dashed) denotes a 
single Abelian hypermultiplet associated with the charge vector $\vec Q = (Q_1, \ldots,  Q_{\alpha-1}, N_\alpha, Q_{\alpha+1}, \ldots, Q_L)$, 
with $L$ being the number of gauge nodes in the quiver. Note that any number of entries in $\vec Q$ (except for the $\alpha$-th entry) 
can be zero. For the quiver $\CT^\vee$ on the right, the mutation splits the $U(N_\alpha)$ gauge node into a $U(N_\alpha-1)$ and 
a $U(1)$ gauge node as shown, where the latter node has a single fundamental hyper. The single Abelian hypermultiplet of $\CT^\vee$
has the charge $(1,\vec Q')$, where the first entry indicates the charge under the new $U(1)$ gauge node, and $\vec Q'$ is an $L$-dimensional 
charge vector: $\vec Q'= (Q_1, \ldots, Q_{\alpha-2}, Q_{\alpha-1}+ N_{\alpha-1}, - (N_\alpha -1), Q_{\alpha +1}+ N_{\alpha+1},\ldots, Q_L)$. 
Therefore, the charges for the Abelian hypermultiplet transform under the mutation as follows:
\begin{align}
& \vec Q \to (1,\vec Q'), \nn \\
& \vec Q = (Q_1, \ldots,  Q_{\alpha-1}, N_\alpha, Q_{\alpha+1}, \ldots, Q_L), \nn \\
& \vec Q'= (Q_1, \ldots, Q_{\alpha-2}, Q_{\alpha-1}+ N_{\alpha-1}, - (N_\alpha-1), Q_{\alpha +1}+ N_{\alpha+1},\ldots, Q_L).
\end{align}

The $U(N_\alpha -1)$ gauge node has $N_{\alpha-1} + N_{\alpha +1} + M_\alpha=2(N_\alpha -1) +1$ bifundamental/fundamental hypers 
plus a single Abelian multiplet. Therefore, one can implement the $\CO_{I'}$ operation at the gauge node $U(N_\alpha -1)$ of the quiver 
$\CT^\vee$, and check that one gets back the quiver $\CT$. The composition of $\CO_{I'}$ with $\CO_{III}$ therefore gives the identity operation. \\

In terms of the sphere partition function, the quiver mutation can be realized by implementing the identity \eref{Id-3} locally 
for the $U(N_\alpha)$ gauge node in the quiver $\CT$, we obtain:
\begin{align}\label{Id-3-main}
Z^{(\CT)}&=  \int \, [d \vec s_\alpha] \, \frac{ e^{2\pi i \eta_\alpha\, \tr \vec s_\alpha}\, Z_{\rm vec}(\vec s_\alpha)\, \Big[ \ldots \Big]}{\prod_{\beta= \alpha \pm 1}\,\prod_{j,i}\,\ch{(s^j_\alpha - \s^i_{\beta})} \, \prod_{j,a}\,\ch{(s^j_\alpha - m^a_\alpha)}\, \ch{(\tr \vec s_\alpha + \sum_a \frac{Q_a}{N_a}\, \tr \vec\s_a)}} \nn \\
&=  \int \, d\s' \, [d \vec \s_\alpha] \, \frac{e^{2\pi i \eta_\alpha\, (\s' -\tr \vec \s_\alpha)}\, Z_{\rm vec}(\vec \s_\alpha)\, \Big[ \ldots \Big]}{\prod_{\beta= \alpha \pm 1}\,\prod_{j,i}\,\ch{(\s^j_\alpha - \s^i_{\beta})} \,\prod_{j,a}\, \ch{(\s^j_\alpha - m^a_\alpha)}} \nn \\
& \qquad \times \frac{1}{\ch{(\s' - \sum_{\beta= \alpha \pm 1}\, \tr \vec \s_{\beta} - \tr \vec m_\alpha)}\,\ch{(\s' - \tr \vec \s_\alpha + \sum_a \frac{Q_a}{N_a}\, \tr \vec\s_a)}} \nn \\
&= \int \, d\s' \, [d \vec \s_\alpha] \, \frac{e^{2\pi i \eta_\alpha\, (\s' +\tr \vec \s_{\alpha-1} + \tr \vec \s_{\alpha + 1} -\tr \vec \s_\alpha)}\, Z_{\rm vec}(\vec \s_\alpha)\, \Big[ \ldots \Big]}{\prod_{\beta= \alpha \pm 1}\,\prod_{j,i}\,\ch{(\s^j_\alpha - \s^i_{\beta})} \,\prod_{j,a}\, \ch{(\s^j_\alpha - m^a_\alpha)}} \nn \\
& \qquad \times \frac{1}{\ch{(\s'  - \tr \vec m_\alpha)}\,\ch{(\s' - \tr \vec \s_\alpha + \sum_{\beta= \alpha \pm 1}\, \tr \vec \s_{\beta} + \sum_a \frac{Q_a}{N_a}\, \tr \vec\s_a)}} \nn \\
&= Z^{(\CT^\vee)}(\eta^\vee_{\alpha-1}=\eta_{\alpha-1}+ \eta_\alpha, \eta^\vee_\alpha=- \eta_\alpha, \eta^\vee_{\alpha+1}= \eta_{\alpha+1}+ \eta_\alpha, \eta^\vee = \eta_\alpha, \ldots),
\end{align}
where for the third equality, we have redefined the integration variable $\s'$ by a shift. 
The charge vector $\vec Q'$ associated with the quiver $\CT^\vee$ can be read off from the third equality.\\

Now, let us implement $\CO_{I'}$ at the $U(N_\alpha -1)$ gauge node of $\CT^\vee$. Let us first make a change of variable:
$\s' \to \s' - \sum_{\beta= \alpha \pm 1}\, \tr \vec \s_{\beta} + \tr \vec \s_\alpha$. This leads to the following partition function:
\begin{align}
Z^{(\CT^\vee)}&=\int \, d\s' \, [d \vec \s_\alpha] \, \frac{e^{2\pi i \eta_\alpha\, \s' }\, Z_{\rm vec}(\vec \s_\alpha)\, \Big[ \ldots \Big]}{\prod_{\beta= \alpha \pm 1}\,\prod_{j,i}\,\ch{(\s^j_\alpha - \s^i_{\beta})} \,\prod_{j,a}\, \ch{(\s^j_\alpha - m^a_\alpha)}} \nn \\
& \qquad \times \frac{1}{\ch{(\s'  - \sum_{\beta= \alpha \pm 1}\, \tr \vec \s_{\beta} + \tr \vec \s_\alpha - \tr \vec m_\alpha)}\,\ch{(\s' + \sum_a \frac{Q_a}{N_a}\, \tr \vec\s_a)}}.
\end{align}
Using the identity \eref{Id-1b-main} for $\CO_{I'}$, we can replace the $\vec \s_\alpha$-dependent part of the matrix integral:
\begin{align}
Z^{(\CT^\vee)}&=\int \, d\s' \, [d \vec s_\alpha] \, \frac{\delta(\tr \vec s_\alpha -\s')\, Z_{\rm vec}(\vec s_\alpha)\, e^{2\pi i \eta_\alpha\, \s' }\,\Big[ \ldots \Big]}{\prod_{\beta= \alpha \pm 1}\,\prod_{j,i}\,\ch{(\s^j_\alpha - \s^i_{\beta})} \,\prod_{j,a}\, \ch{(\s^j_\alpha - m^a_\alpha)}\, \ch{(\s' + \sum_a \frac{Q_a}{N_a}\, \tr \vec\s_a)}} \nn \\
&= \int \, [d \vec s_\alpha] \, \frac{Z_{\rm vec}(\vec s_\alpha)\, e^{2\pi i \eta_\alpha\, \tr \vec s_\alpha }\,\Big[ \ldots \Big]}{\prod_{\beta= \alpha \pm 1}\,\prod_{j,i}\,\ch{(\s^j_\alpha - \s^i_{\beta})} \,\prod_{j,a}\, \ch{(\s^j_\alpha - m^a_\alpha)}\, \ch{(\tr \vec s_\alpha + \sum_a \frac{Q_a}{N_a}\, \tr \vec\s_a)}} \nn \\
&= Z^{(\CT)}
\end{align}
where we have integrated over $\s'$ in the second step. The second line can be readily identified as the partition function of the quiver $\CT$ that 
we started with above. This shows that the composition of $\CO_{III}$ and $\CO_{I'}$ gives the identity operation on the quiver $\CT$.\\

The above computation can be readily extended for $P>1$. In this case, the mutation gives a quiver where the $U(1)$ node 
and the $U(N_\alpha -1)$ node are connected by $P$ Abelian hypermultiplets, with all the gauge nodes and flavor nodes 
remaining the same as the quiver $\CT^\vee$ above. The $P$ Abelian hypers of charges $\{ \vec Q^l\}_{l=1,\ldots,P}$ on the LHS 
are mapped to another $P$ Abelian hypers of charges $(1,\{\vec Q'^l\}_{l=1,\ldots,P})$ on the RHS. The map of the charges is given as follows:
\begin{align}
& \vec Q^l \to (1,\vec Q'^l), \qquad \text{for}\,\, l=1,\ldots,P, \nn \\
& \vec Q^l= (Q^l_1, \ldots, Q^l_{\alpha-1}, N_\alpha, Q^l_{\alpha+1}, \ldots, Q^l_L), \nn \\
& \vec Q'^l=(Q^l_1, \ldots, Q^l_{\alpha-2}, Q^l_{\alpha-1} + N_{\alpha-1}, -(N_\alpha-1), Q^l_{\alpha+1} + N_{\alpha+1} , Q^l_{\alpha+2}, \ldots, Q^l_L).
\end{align}

One can check that if one implements $\CO_{I'}$ at the gauge node $U(N_\alpha -1)$ of the quiver $\CT^\vee$ (after an appropriate field 
redefinition in the theory), one gets back the quiver $\CT$. The composition of $\CO_{I'}$ with $\CO_{III}$ therefore gives the identity operation, 
as we noted in the $P=1$ case. In terms of the partition function, these operations can be implemented in an analogous fashion as the $P=1$ case. \\

The extension of the mutation $III$ to a non-linear quiver assumes the following form:

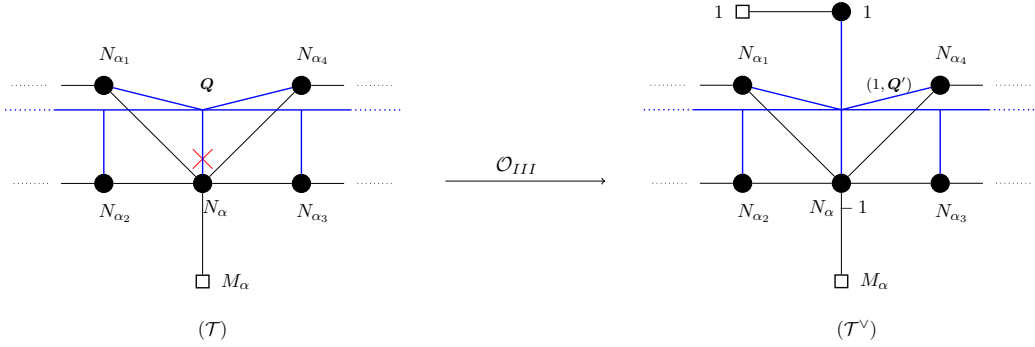
\begin{center}
\begin{tabular}{ccc}
\scalebox{0.65}{\begin{tikzpicture}
\node[] (1) at (1,0){};
\node[] (100) at (0,0){};
\node[unode] (2) at (2,0){};
\node[unode] (3) at (4,0){};
\node[unode] (4) at (6,0){};
\node[unode] (51) at (2,2){};
\node[] (52) at (0,2){};
\node[] (53) at (1,2){};
\node[unode] (61) at (6,2){};
\node[] (62) at (7,2){};
\node[] (63) at (8,2){};
\node[] (5) at (7,0){};
\node[] (200) at (8,0){};
\node[cross, red] (6) at (4,0.5){};
\node[fnode] (8) at (4,-2){};
\draw[-] (1) -- (2);
\draw[-] (2)-- (3);
\draw[-] (3) -- (4);
\draw[-] (4) --(5);
\draw[-] (3) --(8);
\draw[-] (3) --(51);
\draw[-] (3) --(61);
\draw[-, dotted] (1) -- (100);
\draw[-,dotted] (5) -- (200);
\draw[-, dotted] (52) -- (53);
\draw[-] (51) -- (53);
\draw[-, dotted] (62) -- (63);
\draw[-] (61) -- (62);
\draw[dotted, thick, blue] (0,1.5)--(1,1.5);
\draw[-, thick, blue] (1,1.5)--(2,1.5);
\draw[-, thick, blue] (2)--(2,1.5);
\draw[-, thick, blue] (3)--(4,1.5);
\draw[-, thick, blue] (4)--(6,1.5);
\draw[-, thick, blue] (2,1.5)--(4,1.5);
\draw[-, thick, blue] (4,1.5)--(6,1.5);
\draw[-, thick, blue] (4,1.5)--(51);
\draw[-, thick, blue] (4,1.5)--(61);
\draw[-, thick, blue] (7,1.5)--(6,1.5);
\draw[dotted, thick, blue] (8,1.5)--(7,1.5);
\node[text width=.2cm](11) [below=0.1cm of 2]{$N_{\alpha_2}$};
\node[text width=.2cm](12) at (4.1, -0.5){$N_\alpha$};
\node[text width=.2cm](13) [below=0.1cm of 4]{$N_{\alpha_3}$};
\node[text width=.2cm](14) [right=0.1cm of 8]{$M_\alpha$};
\node[text width=.2cm](15) [above=0.1cm of 51]{$N_{\alpha_1}$};
\node[text width=.2cm](16) [above=0.1cm of 61]{$N_{\alpha_4}$};
\node[text width=0.1 cm](10) at (4, 2){\footnotesize{$\vec Q$}};
\node[text width=.2cm](20) at (4,-3){$(\CT)$};
\end{tikzpicture}}
& \scalebox{.7}{\begin{tikzpicture}
\draw[->] (0,0) -- (3,0);
\node[text width=0.1cm](29) at (1, 0.3) {$\CO_{III}$};
\node[](30) at (0, -3) {};
\end{tikzpicture}}
& \scalebox{0.65}{\begin{tikzpicture}
\node[] (1) at (1,0){};
\node[] (100) at (0,0){};
\node[unode] (2) at (2,0){};
\node[unode] (3) at (4,0){};
\node[unode] (4) at (6,0){};
\node[unode] (9) at (4,3.5){};
\node[fnode] (10) at (2,3.5){};
\node[unode] (51) at (2,2){};
\node[] (52) at (0,2){};
\node[] (53) at (1,2){};
\node[unode] (61) at (6,2){};
\node[] (62) at (7,2){};
\node[] (63) at (8,2){};
\node[] (5) at (7,0){};
\node[] (200) at (8,0){};
\node[fnode] (8) at (4,-2){};
\draw[-] (1) -- (2);
\draw[-] (2)-- (3);
\draw[-] (3) -- (4);
\draw[-] (4) --(5);
\draw[-] (3) --(8);
\draw[-] (3) --(51);
\draw[-] (3) --(61);
\draw[-] (9) --(10);
\draw[-, dotted] (1) -- (100);
\draw[-,dotted] (5) -- (200);
\draw[-, dotted] (52) -- (53);
\draw[-] (51) -- (53);
\draw[-, dotted] (62) -- (63);
\draw[-] (61) -- (62);
\draw[dotted, thick, blue] (0,1.5)--(1,1.5);
\draw[-, thick, blue] (1,1.5)--(2,1.5);
\draw[-, thick, blue] (2)--(2,1.5);
\draw[-, thick, blue] (3)--(4,1.5);
\draw[-, thick, blue] (4)--(6,1.5);
\draw[-, thick, blue] (2,1.5)--(4,1.5);
\draw[-, thick, blue] (4,1.5)--(6,1.5);
\draw[-, thick, blue] (4,1.5)--(51);
\draw[-, thick, blue] (4,1.5)--(61);
\draw[-, thick, blue] (9)--(4,1.5);
\draw[-, thick, blue] (7,1.5)--(6,1.5);
\draw[dotted, thick, blue] (8,1.5)--(7,1.5);
\node[text width=.2cm](11) [below=0.1cm of 2]{$N_{\alpha_2}$};
\node[text width=1.5cm](12) at (4.1, -0.5){$N_\alpha -1$};
\node[text width=.2cm](13) [below=0.1cm of 4]{$N_{\alpha_3}$};
\node[text width=.2cm](14) [right=0.1cm of 8]{$M_\alpha$};
\node[text width=.2cm](24) [right=0.1cm of 9]{$1$};
\node[text width=.2cm](25) [left=0.1cm of 10]{$1$};
\node[text width=.2cm](15) [above=0.1cm of 51]{$N_{\alpha_1}$};
\node[text width=.2cm](16) [above=0.1cm of 61]{$N_{\alpha_4}$};
\node[text width=2 cm](30) at (5.5, 2){\footnotesize{$(1,\vec Q')$}};
\node[text width=.2cm](20) at (4,-3){$(\CT^\vee)$};
\end{tikzpicture}}
\end{tabular}
\end{center}

The labels of the gauge and the flavor nodes obey the constraint: $\sum_i\, N_{\alpha_i} +M_\alpha =2N_\alpha -1$. 
The map of the charges is given as follows:
\begin{align}
& \vec Q^l \to (1,\vec Q'^l), \qquad \text{for}\,\, l=1,\ldots,P, \nn \\
& \vec Q^l= (Q^l_1, \ldots, Q^l_{\alpha_1}, Q^l_{\alpha_2}, N_\alpha, Q^l_{\alpha_3}, Q^l_{\alpha_4}, \ldots, Q^l_L), \nn \\
& \vec Q'^l=(Q^l_1, \ldots, Q^l_{\alpha_1} + N_{\alpha_1}, Q^l_{\alpha_2} + N_{\alpha_2},  -(N_\alpha-1), Q^l_{\alpha_3} + N_{\alpha_3} , Q^l_{\alpha_4} + N_{\alpha_4}, \ldots, Q^l_L).
\end{align}

\subsection*{The Duality Sequence}

Given the set of four mutations, we can now discuss the construction of a duality sequence starting from a 
quiver gauge theory $\CT$  in class $\CX$, where $\CT$ is a good theory with at least one balanced 
special unitary node. The precise steps for constructing the duality sequence may be summarized as follows:

\begin{enumerate}

\item Assuming that the quiver $\CT$ has $L$ balanced special unitary gauge nodes, the first step involves implementing mutation $I$ 
at each of the balanced nodes, thereby giving $L$ N-al theories. Note that this quiver mutation will alter the balance condition of the 
gauge nodes which are connected to the balanced node in question.

\item Each of the $L$ theories obtained in step one will have $L-1$ balanced special unitary gauge nodes inherited from $\CT$ as well as 
possibly new balanced special unitary gauge nodes which arise because of mutation $I$. In addition, there will be unitary gauge nodes 
with various balance parameters and a single Abelian hypermultiplet charged under a subset of the gauge nodes. In the second step, 
we implement on each such theory all admissible mutations, with each mutation giving an IR dual theory.

\item One continues with this procedure until no new theories can be generated by the set of mutations. The entire set of theories generated in this fashion 
will be called the $N$-al set.

\end{enumerate}

In the rest of this paper, we will construct such duality sequences for explicit examples of $\CT$. For the sake of 
simplicity, we will restrict ourselves to examples where the quiver $\CT$ is given by a linear graph with a single 
balanced special unitary node and one or more unitary nodes. 
However, the computation can be readily extended to more complicated quiver gauge theories -- for example, 
ones with more than one special unitary nodes and/or with more involved graphs, following 
the general procedure outlined above.

\section{N-ality in a two-node quiver}\label{N-al-2node}

In this section, we will construct an explicit example of the duality sequence starting from a quiver gauge theory $\CT$ 
with two gauge nodes of the following form:\\
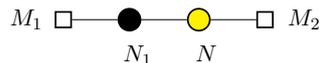
\begin{center}
\scalebox{0.8}{\begin{tikzpicture}
\node[fnode] (1) {};
\node[unode] (2) [right=.75cm  of 1]{};
\node[sunode] (3) [right=.75cm of 2]{};
\node[fnode] (4) [right=0.75 cm of 3]{};
\draw[-] (1) -- (2);
\draw[-] (2)-- (3);
\draw[-] (3) -- (4);
\node[text width=.1cm](10) [left=0.5 cm of 1]{$M_1$};
\node[text width=.2cm](11) [below=0.1cm of 2]{$N_1$};
\node[text width=.1cm](12) [below=0.1cm of 3]{$N$};
\node[text width=.1cm](13) [right=0.1cm of 4]{$M_2$};
\end{tikzpicture}}
\end{center}
The $SU(N)$ node is balanced, i.e. $N_1 + M_2 =2N-1$, while the $U(N_1)$ node can either be balanced or overbalanced, i.e.
$M_1 + N = 2N_1 + e$, with $e \geq 0$. The above quiver therefore represents a 3-parameter family of quiver gauge 
theories. We will first discuss the Coulomb branch global symmetries of the quivers for different ranges of $e$ in 
\Secref{MO-2node}, pointing out the emergent IR symmetry that arises for the respective ranges of $e$. 
Next, we will work out the details of the duality sequence which depend
on the specific value of $e$ -- we will discuss the $e>0$ and $e=0$ cases in \Secref{Unbal-2node} and \Secref{Bal-2node} separately. 
In particular, we will discuss how the emergent IR symmetry of $\CT$ (in addition to the rank of the symmetry) becomes manifest 
in one or more theories of the N-al set.

\subsection{Monopole operators and global symmetry enhancement}\label{MO-2node}

The IR conformal dimensions of the monopole operators of the quiver gauge theory $\CT$ are given as
\begin{align}
\Delta( \vec a, \vec p)=& \frac{M_1}{2}\,\sum^{N_1}_{i=1} |a_i|  + \frac{1}{2}\, \sum_{i,k}\,|a_i -p_k| + \frac{M_2}{2}\,\sum^{N}_{k=1} |p_k| \nn \\
& - \sum_{1 \leq i<j \leq N_1}|a_i-a_j| - \sum_{1 \leq k< l \leq N}|p_k-p_l|, \qquad \sum^N_{k=1}\,p_k=0,
\end{align}
where $\vec a$ and $\vec p$ denote the GNO charges associated with the $U(N_1)$ and the $SU(N)$ node respectively 
for a given monopole operator. To begin with, one can check that that the theory is good in the Gaiotto-Witten sense i.e.  
$\Delta( \vec a, \vec p) > \frac{1}{2}$ for any $\vec a $ and $\vec p$. For a good theory, every monopole operator with $\Delta=1$
corresponds to a conserved current/global symmetry generator in the IR SCFT. The Lie algebra of the enhanced global symmetry group 
of the Coulomb branch can therefore be determined by finding the complete set of $\Delta=1$ monopole operators.

Consider the monopole operators $\vec a = (\pm 1, 0, \ldots, 0), \vec p=0$ -- their conformal
dimensions are 
\be
\Delta(\vec a, \vec p=0) = \frac{M_1 +N-2N_1+2}{2} = \frac{e+2}{2}.
\ee
For the case of a balanced $U(N_1)$ node i.e. $e=0$, the conserved currents associated with these monopole operators
 will combine with the generator of the topological $\fru(1)$ symmetry algebra 
to generate an $\frsu(2)$ algebra. This is the standard enhancement that arises from a single balanced unitary gauge node. 

Next, consider the operator $\vec a =0, \vec p =  (1, 0, \ldots, 0,-1)$, for which
\be
\Delta(\vec a=0, \vec p) = M_2 +N_1 -2N +2 =1,
\ee
which generates a $\fru(1)$ global symmetry for any value of $e \geq 0$. This is the generator that enhances the 
Coulomb branch global symmetry of a balanced $SU(N)$ gauge theory to a $\fru(1)$ in the IR.

In addition to these, there are global symmetry generators 
that arise for the case of $e=0$. To see these, consider the conformal dimensions of the 
following monopole operators :
\begin{align}
& \vec a = (\pm 1, 0, \ldots, 0), \vec p =  (1, 0, \ldots, 0,-1) \,: \,\Delta(\vec a, \vec p) = \frac{e+2}{2}, \\
& \vec a = (1, 0, \ldots, 0,-1), \vec p =  (1, 0, \ldots, 0,-1) \,: \,\,\Delta(\vec a, \vec p) = e+1. 
\end{align}
For $e=0$, the three associated conserved currents will combine to give another $\frsu(2)$ algebra. 
The appearance of this additional $\frsu(2)$ factor is similar to the enhancement of the Coulomb branch 
global symmetry studied in \cite{Gaiotto:2008ak} for a linear quiver of balanced unitary gauge nodes terminating in an orthsosymplectric node.\\

To summarize, the Coulomb branch global symmetry algebra of the quiver $\CT$ is given as 
\begin{enumerate}

\item Overbalanced $U(N_1)$ node : $\frg_{\rm C} = \fru(1) \oplus \fru(1)$, corresponding to the topological
symmetry for the $U(N_1)$ gauge node and the monopole operator associated with the $SU(N)$ gauge node with charge 
$\vec a =0, \vec p = (1,0,\ldots,0,-1)$. 

\item Balanced $U(N_1)$ node : $\frg_{\rm C} =\frsu(2) \oplus \frsu(2) \oplus \fru(1)$, corresponding to the monopole operators 
(and the topological symmetry) listed above.

\end{enumerate}

In comparison, the Higgs branch global symmetry algebra can be directly read off from the quiver, i.e. 
$\frg_{\rm H} = \frsu(M_1) \oplus \frsu(M_2) \oplus \fru(1) \oplus \fru(1)$, which arise from the fundamental hypermultiplets of the 
respective gauge nodes. \\

\subsection{Overbalanced unitary node: Duality and Trialities}\label{Unbal-2node}

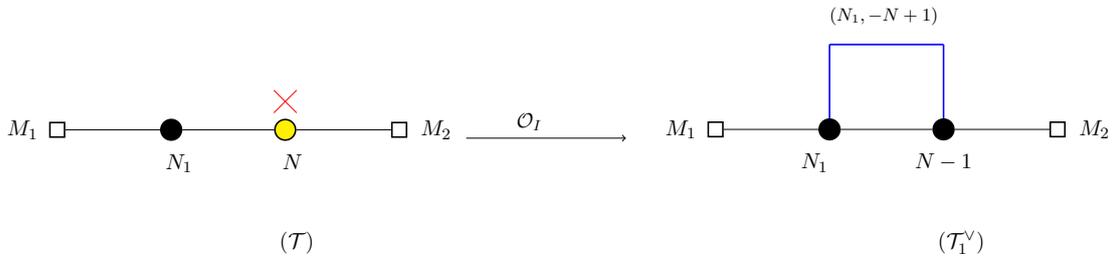
\begin{figure}[htbp]
\begin{center}
\begin{tabular}{ccc}
\scalebox{0.75}{\begin{tikzpicture}
\node[fnode] (1) at (0,0){};
\node[unode] (2) at (2,0){};
\node[sunode] (3) at (4,0){};
\node[fnode] (4) at (6,0){};
\node[cross, red] (5) at (4,0.5){};
\draw[-] (1) -- (2);
\draw[-] (2)-- (3);
\draw[-] (3) -- (4);
\node[text width=.1cm](10) [left=0.5 cm of 1]{$M_1$};
\node[text width=.2cm](11) [below=0.1cm of 2]{$N_1$};
\node[text width=.1cm](12) [below=0.1cm of 3]{$N$};
\node[text width=.1cm](13) [right=0.1cm of 4]{$M_2$};
\node[text width=.2cm](20) at (4,-2){$(\CT)$};
\end{tikzpicture}}
& \scalebox{.7}{\begin{tikzpicture}
\draw[->] (4,0) -- (7, 0);
\node[text width=0.1cm](29) at (5, 0.3) {$\CO_{I}$};
\node[](30) at (5, -2.2) {};
\end{tikzpicture}}
&\scalebox{0.75}{\begin{tikzpicture}
\node[fnode] (1) at (0,0){};
\node[unode] (2) at (2,0){};
\node[unode] (3) at (4,0){};
\node[fnode] (4) at (6,0){};
\draw[-] (1) -- (2);
\draw[-] (2)-- (3);
\draw[-] (3) -- (4);
\draw[-, thick, blue] (2)--(2,1.5);
\draw[-, thick, blue] (3)--(4,1.5);
\draw[-, thick, blue] (2,1.5)--(4,1.5);
\node[text width=2 cm](10) at (3, 2){\footnotesize{$(N_1, -N+1)$}};
\node[text width=.1cm](20) [left=0.5 cm of 1]{$M_1$};
\node[text width=1cm](21) [below=0.1cm of 2]{$N_1$};
\node[text width=1 cm](22) [below=0.1cm of 3]{$N-1$};
\node[text width=.1cm](23) [right=0.1cm of 4]{$M_2$};
\node[text width=.2cm](20) at (4,-2){$(\CT^\vee_1)$};
\end{tikzpicture}}
\end{tabular}
\end{center}
\caption{\footnotesize{Duality for the case $e > 2$. The gauge node at which the mutation I acts is marked by a red cross.}}
\label{IRdual-Ex6a}
\end{figure}

Let us first consider the case where the $U(N)$ node in $\CT$ is overbalanced i.e. $N + M_1 = 2N_1 +e$ with $e >0$. 
As discussed in \Secref{recipe-gen0}, the first step is to implement mutation I at the balanced $SU(N)$ node, which leads 
to the theory $\CT^\vee_1$ in \figref{IRdual-Ex6a}.

One can realize this mutation in terms of the sphere partition function in the  
following fashion. The partition function of the quiver $\CT$ is given as:
\begin{align}\label{PF-CT-01}
Z^{(\CT)}(\vec m^1, \vec m^2; \eta) = & \int\, \prod^{2}_{\gamma=1}\,\Big[d\vec s^{\gamma}\Big] \, Z_{\rm FI}(\vec s^1, \eta) \,\delta(\tr \vec s^2)\, \prod^2_{\gamma=1}\, Z^{\rm vec}_{\rm 1-loop}(\vec s^\gamma) \nn \\
& \times \, Z^{\rm fund}_{\rm 1-loop}(\vec s^1, \vec m^1)\, Z^{\rm bif}_{\rm 1-loop}(\vec s^1, \vec s^{2}, 0)\,
Z^{\rm fund}_{\rm 1-loop}(\vec s^2, \vec m^2), 
\end{align}
where $\vec s^1$ denotes the integration variables associated with the unitary gauge group $U(N_1)$, while $\vec s^2$ denotes 
the integration variable associated with the $SU(N)$ gauge group. The mass parameters for the fundamental hypermultiplets 
at the two gauge nodes are collectively denoted as $\vec m = (\vec m^1, \vec m^2)$, and the FI parameter associated with the $U(N_1)$ gauge node 
is $\eta$. Mutation I can be implemented by isolating the $\vec s^2$-dependent part of the matrix integral, and using the 
identity \eref{Id-1} (after substituting $N \to N-1$), which gives:
\begin{align}
& \int \, \Big[d \vec s^2 \Big] \, \delta(\tr \vec s^2)\, Z^{\rm vec}_{\rm 1-loop}(\vec s^2)\, Z^{\rm bif}_{\rm 1-loop}(\vec s^2, \vec s^1, 0)\, Z^{\rm fund}_{\rm 1-loop}(\vec s^2, \vec m^2) \nn \\
&=   \int \, \Big[d  \vec \s^2 \Big] \, \frac{1}{\ch{(\tr \vec \s^2 - \tr \vec s^1 - \tr \vec m^2)}}\,Z^{\rm vec}_{\rm 1-loop}(\vec \s^2)\, Z^{\rm bif}_{\rm 1-loop}(\vec \s^2, \vec s^1, 0)\,Z^{\rm bif}_{\rm 1-loop}(\vec \s^2, \vec m^2, 0),
\end{align}
where $\vec \s^2$ are the integration variables in the Cartan subalgebra of a $U(N-1)$ group. 
Substituting the above identity in \eref{PF-CT-01}, the partition function for $\CT$ can be written as
\begin{align}\label{PF-CT-01a}
Z^{(\CT)}(\vec m^1, \vec m^2; \eta) = & \int\, \Big[d\vec s^{1}\Big]\, \Big[d  \vec \s^2 \Big] \,Z_{\rm FI}(\vec s^1, \eta) \, Z^{\rm vec}_{\rm 1-loop}(\vec s^1)\,Z^{\rm vec}_{\rm 1-loop}(\vec \s^2) \nn \\
 \times & \,Z^{\rm fund}_{\rm 1-loop}(\vec s^1, \vec m^1)\,Z^{\rm bif}_{\rm 1-loop}(\vec s^1, \vec \s^2, 0)\, Z^{\rm fund}_{\rm 1-loop}(\vec \s^2, \vec m^2)\, Z^{\rm hyper}_{\rm 1-loop}(\vec s^1, \vec \s^2, \vec m^2).
\end{align}
The matrix integral on the RHS of the above equation can be evidently identified with the sphere partition function of the quiver gauge 
theory $\CT^\vee_1$ in \figref{IRdual-Ex6a}, where $Z^{\rm hyper}_{\rm 1-loop}$ is the contribution of a single Abelian hypermultiplet 
with charge $(N_1, -(N-1))$ under the gauge group $U(N_1) \times U(N-1) $, i.e.
\be
Z^{\rm hyper}_{\rm 1-loop}(\vec s^1, \vec \s^2, \vec m^2) = \frac{1}{\ch{( \tr \vec s^1 - \tr \vec \s^2  + \tr \vec m^2)}}. 
\ee
We therefore have the following relation of the two partition functions:
\be \label{pf-duality-1}
\boxed{Z^{(\CT)}(\vec m^1, \vec m^2; \eta)=  Z^{(\CT^\vee_1)}(\vec m^1, \vec m^2, m_{\rm{Ab}}=\tr \vec m^2; \eta, 0).}
\ee

Note that the IR enhanced Coulomb branch symmetry $\fru(1) \oplus \fru(1)$ of $\CT$ is manifest in the dual theory $\CT^\vee_1$ 
as topological symmetries of the two unitary gauge nodes. 
The emergent nature of the symmetry on one side of the duality is indicated by the fact that 
a linear combination of FI parameters in $\CT^\vee_1$ has to be set to zero for the sphere partition functions to agree.\\

\begin{figure}[htbp]
\begin{center}
\begin{tabular}{ccc}
\scalebox{0.75}{\begin{tikzpicture}
\node[fnode] (1) at (0,0){};
\node[unode] (2) at (2,0){};
\node[sunode] (3) at (4,0){};
\node[fnode] (4) at (6,0){};
\node[cross, red] (5) at (4,0.5){};
\draw[-] (1) -- (2);
\draw[-] (2)-- (3);
\draw[-] (3) -- (4);
\node[text width=.1cm](20) [left=0.5 cm of 1]{$M_1$};
\node[text width=.2cm](21) [below=0.1cm of 2]{$N_1$};
\node[text width=.1cm](22) [below=0.1cm of 3]{$N$};
\node[text width=.1cm](23) [right=0.1cm of 4]{$M_2$};
\node[text width=.2cm](20) at (4,-2){$(\CT)$};
\end{tikzpicture}}
& \scalebox{.7}{\begin{tikzpicture}
\draw[->] (4,0) -- (7, 0);
\node[text width=0.1cm](29) at (5, 0.3) {$\CO_{I}$};
\node[](30) at (5, -2.2) {};
\end{tikzpicture}}
&\scalebox{0.75}{\begin{tikzpicture}
\node[fnode] (1) at (0,0){};
\node[unode] (2) at (2,0){};
\node[unode] (3) at (4,0){};
\node[fnode] (4) at (6,0){};
\node[cross, red] (5) at (2, 0.5){};
\draw[-] (1) -- (2);
\draw[-] (2)-- (3);
\draw[-] (3) -- (4);
\draw[-, thick, blue] (2)--(2,1.5);
\draw[-, thick, blue] (3)--(4,1.5);
\draw[-, thick, blue] (2,1.5)--(4,1.5);
\node[text width=2 cm](10) at (3, 2){\footnotesize{$(N_1, -N+1)$}};
\node[text width=.1cm](20) [left=0.5 cm of 1]{$M_1$};
\node[text width=0.5 cm](21) [below=0.1cm of 2]{$N_1$};
\node[text width=1 cm](22) [below=0.1cm of 3]{$N-1$};
\node[text width=.1cm](23) [right=0.1cm of 4]{$M_2$};
\node[text width=.2cm](20) at (4,-2){$(\CT^\vee_1)$};
\end{tikzpicture}}\\
\qquad 
& \qquad
& \scalebox{.7}{\begin{tikzpicture}
\draw[->] (4,-1) -- (4, -3);
\node[text width=0.1cm](29) at (4.2, -2) {$\CO_{II}$};
\end{tikzpicture}}\\
\qquad
& \qquad
&\scalebox{0.75}{\begin{tikzpicture}
\node[fnode] (1) at (0,0){};
\node[unode] (2) at (2,0){};
\node[unode] (3) at (4,0){};
\node[fnode] (4) at (6,0){};
\node[fnode] (6) at (2,2){};
\draw[-] (1) -- (2);
\draw[-] (2)-- (3);
\draw[-] (3) -- (4);
\draw[-, thick, blue] (2)--(6);
\node[text width=.1cm](20) [left=0.5 cm of 1]{$M_1$};
\node[text width=0.5 cm](21) [below=0.1cm of 2]{$N_1$};
\node[text width=1 cm](22) [below=0.1cm of 3]{$N-1$};
\node[text width=.1cm](23) [right=0.1cm of 4]{$M_2$};
\node[text width=.1cm](24) [right=0.1cm of 6]{1};
\node[text width=.2cm](40) at (4,-2){$(\CT^\vee_2)$};
\end{tikzpicture}}
\end{tabular}
\end{center}
\caption{\footnotesize{Triality for the case $e=1$. The gauge node at which a given mutation acts is marked by a red cross at each step.}}
\label{IRdual-Ex6b}
\end{figure}
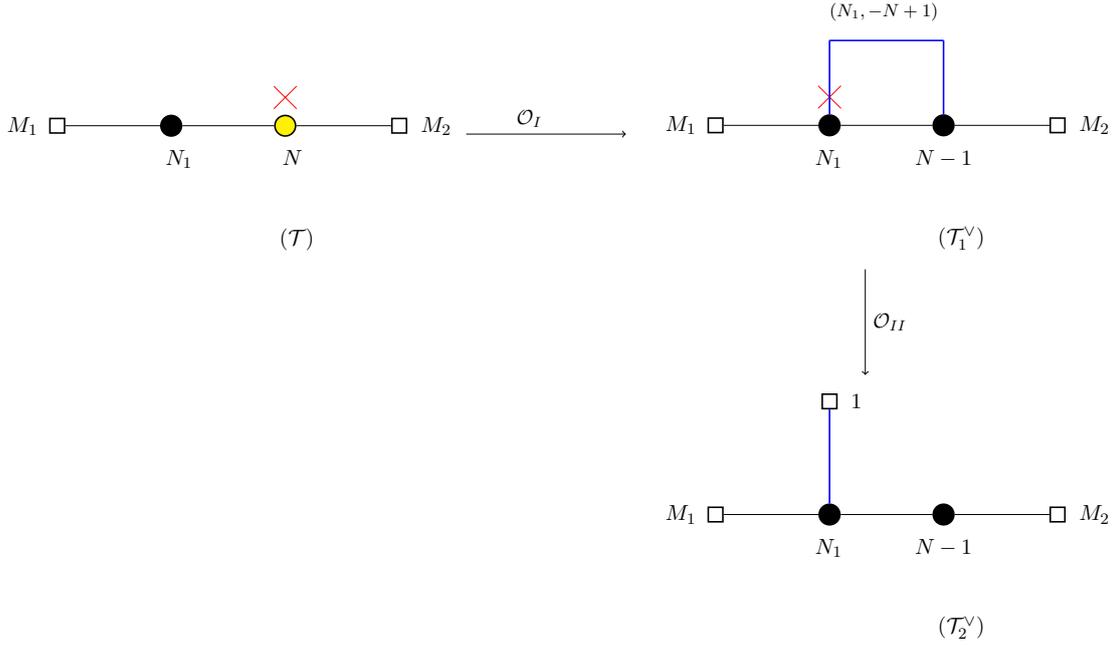

For a generic integer $e > 2$, the N-ality is simply a duality, since we cannot implement any of the other mutations on the $U(N)$ 
node. In this case, for a fixed $e$, we have a 2-parameter family of dualities.
However, for the special cases of $e=1,2$, we can continue further, as we will now demonstrate. 

Consider the case of 
$e=1$ to begin with, as shown in \figref{IRdual-Ex6b}. The first step of obtaining the quiver $\CT^\vee_1$ from $\CT$ 
using mutation I is the same as before. Since $e=1$, the $U(N_1)$ gauge node of $\CT^\vee_1$ has $2N_1$ 
fundamental/bifundamental hypers plus a single Abelian hyper. One can therefore 
implement mutation II on the $U(N_1)$ gauge node which leads to the quiver $\CT^\vee_2$. 
The mutation can again be realized in terms of the partition function and the details of the computation are given in \Appref{2node-pf-ob}. 
The final result is the partition function relation:
\be
\boxed{Z^{(\CT)}(\vec m^1, \vec m^2; \eta)= Z^{(\CT^\vee_1)}(\vec m^1, \vec m^2, m_{\rm{Ab}}=\tr \vec m^2; \eta, 0) = Z^{(\CT^\vee_2)}(\vec m', m'_{\rm{Ab}}; -\eta, \eta),}
\ee
where $\CT^\vee_2$ is the quiver gauge theory in \figref{IRdual-Ex6b}. The duality map for the real masses can be read off from the 
above identity:
\begin{align}
& \vec m'^1 = \vec m^1 - \frac{\tr \vec m^1}{N_1}, \qquad \vec m'^2 = \vec m^2, \\
& m'_{\rm bif}= \frac{\tr \vec m^1}{N_1}, \qquad m'_{\rm{Ab}} = \tr \vec m^2.
\end{align}
Note that the number of independent real mass parameters are $M_1 + M_2$ which live in the Cartan subalgebra of  
$\frg_{\rm H}= \frsu(M_1) \oplus \frsu(M_2) \oplus \fru(1) \oplus \fru(1)$. 
In addition, the IR enhanced Coulomb branch symmetry $\fru(1) \oplus \fru(1)$ of $\CT$ is manifest in the dual theory $\CT^\vee_2$, 
in addition to the theory $\CT^\vee_1$. 
Mutation II squares to an identity as a quiver operation, and therefore applying the mutation II on the $U(N_1)$ gauge node in 
the $\CT^\vee_2$ theory will reproduce the theory $\CT^\vee_1$. Since no further mutations are permitted, 
the duality sequence ends at $\CT^\vee_2$ giving a triality, as shown in \figref{IRdual-Ex6b}. Note that we have 
a 2-parameter family of trialities in this case.\\

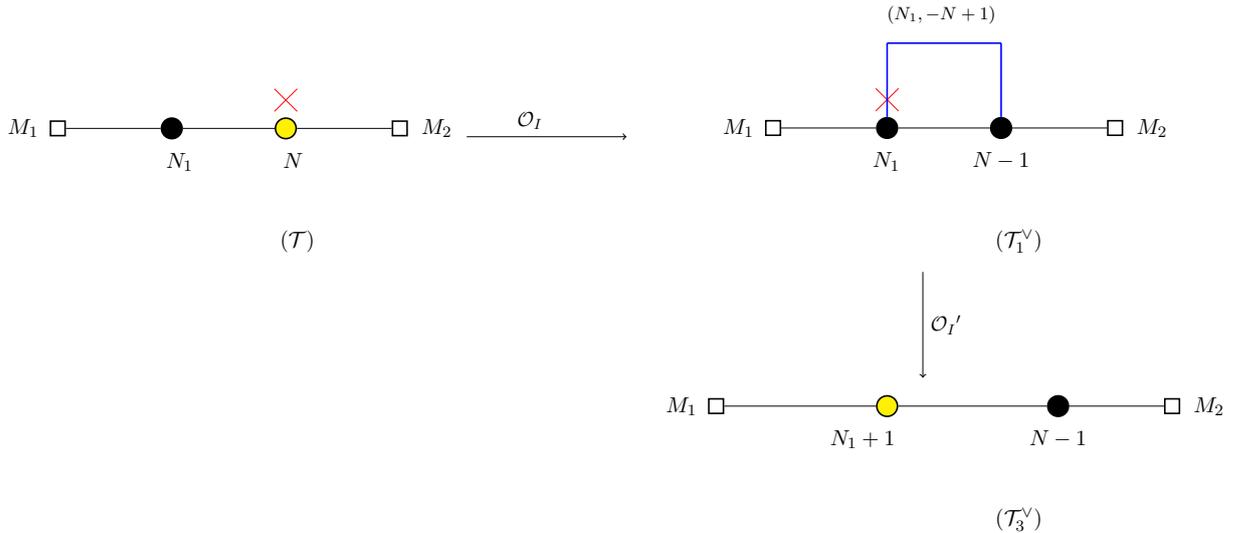
\begin{figure}[htbp]
\begin{center}
\begin{tabular}{ccc}
\scalebox{0.75}{\begin{tikzpicture}
\node[fnode] (1) at (0,0){};
\node[unode] (2) at (2,0){};
\node[sunode] (3) at (4,0){};
\node[fnode] (4) at (6,0){};
\node[cross, red] (5) at (4,0.5){};
\draw[-] (1) -- (2);
\draw[-] (2)-- (3);
\draw[-] (3) -- (4);
\node[text width=.1cm](20) [left=0.5 cm of 1]{$M_1$};
\node[text width=.2cm](21) [below=0.1cm of 2]{$N_1$};
\node[text width=.1cm](22) [below=0.1cm of 3]{$N$};
\node[text width=.1cm](23) [right=0.1cm of 4]{$M_2$};
\node[text width=.2cm](20) at (4,-2){$(\CT)$};
\end{tikzpicture}}
& \scalebox{.7}{\begin{tikzpicture}
\draw[->] (4,0) -- (7, 0);
\node[text width=0.1cm](29) at (5, 0.3) {$\CO_{I}$};
\node[](30) at (5, -2.2) {};
\end{tikzpicture}}
&\scalebox{0.75}{\begin{tikzpicture}
\node[fnode] (1) at (0,0){};
\node[unode] (2) at (2,0){};
\node[unode] (3) at (4,0){};
\node[fnode] (4) at (6,0){};
\node[cross, red] (5) at (2,0.5){};
\draw[-] (1) -- (2);
\draw[-] (2)-- (3);
\draw[-] (3) -- (4);
\draw[-, thick, blue] (2)--(2,1.5);
\draw[-, thick, blue] (3)--(4,1.5);
\draw[-, thick, blue] (2,1.5)--(4,1.5);
\node[text width=2 cm](10) at (3, 2){\footnotesize{$(N_1, -N+1)$}};
\node[text width=.1cm](20) [left=0.5 cm of 1]{$M_1$};
\node[text width=0.5 cm](21) [below=0.1cm of 2]{$N_1$};
\node[text width=1 cm](22) [below=0.1cm of 3]{$N-1$};
\node[text width=.1cm](23) [right=0.1cm of 4]{$M_2$};
\node[text width=.2cm](20) at (4,-2){$(\CT^\vee_1)$};
\end{tikzpicture}}\\
\qquad 
& \qquad
& \scalebox{.7}{\begin{tikzpicture}
\draw[->] (4,-1) -- (4, -3);
\node[text width=0.1cm](29) at (4.2, -2) {$\CO_{I}$$'$};
\end{tikzpicture}}\\
\qquad
& \qquad
&\scalebox{0.75}{\begin{tikzpicture}
\node[fnode] (1) at (-1,0){};
\node[sunode] (2) at (2,0){};
\node[unode] (3) at (5,0){};
\node[fnode] (4) at (7,0){};
\draw[-] (1) -- (2);
\draw[-] (2)-- (3);
\draw[-] (3) -- (4);
\node[text width=.1cm](20) [left=0.5 cm of 1]{$M_1$};
\node[text width=2 cm](21) [below=0.1cm of 2]{$N_1+1$};
\node[text width=1 cm](22) [below=0.1cm of 3]{$N-1$};
\node[text width=.1cm](23) [right=0.1cm of 4]{$M_2$};
\node[text width=.2cm](20) at (4,-2){$(\CT^\vee_3)$};
\end{tikzpicture}}
\end{tabular}
\end{center}
\caption{\footnotesize{Triality for the case $e=2$. The gauge node at which a mutation acts is marked by a red cross at each step.
The triality implies a duality between a pair of linear quivers with no Abelian multiplet.}}
\label{IRdual-Ex6c}
\end{figure}

Now consider the case of $e=2$, as shown in \figref{IRdual-Ex6c}. The first step again involves implementing 
mutation I at the $SU(N)$ gauge node in $\CT$ which gives the quiver $\CT^\vee_1$. Since $e=2$, the 
$U(N_1)$ gauge node of $\CT^\vee_1$ now has $2N_1+1$ fundamental/bifundamental hypers plus a 
single Abelian hyper. One can then implement the mutation I$'$ at this node which replaces the 
$U(N_1)$ node by an $SU(N_1+1)$ gauge node, leading to the dual theory $\CT^\vee_3$ shown 
in \figref{IRdual-Ex6c}.

The mutation can again be realized in terms of the partition function (the details can be found in \Appref{2node-pf-ob}),
and lead to the following partition function relation:
\be
\boxed{Z^{\CT}(\vec m; 0)=  Z^{\CT^\vee_1}(\vec m, m_{\rm{Ab}}=\tr \vec m^2; 0, 0) = Z^{\CT^\vee_3}(\vec m'; 0),}
\ee
where the duality map for the real masses can be read off from the above identity:
\begin{align}
\vec m'^1= \vec m^1 - \delta, \qquad \vec m'^2= \vec m^2, \qquad m'_{\rm bif}=\delta, \qquad \delta =\frac{1}{N_1+1}(\tr \vec m^1 + \tr \vec m^2).
\end{align}
Note that the number of independent real mass parameters are again $M_1 + M_2$ which live in the Cartan subalgebra of 
$\frg_{\rm H}= \frsu(M_1) \oplus \frsu(M_2) \oplus \fru(1) \oplus \fru(1)$. 
The $\fru(1) \oplus \fru(1)$ Coulomb branch symmetry is emergent in the theory $\CT$ as well as the theory $\CT^\vee_3$, 
while it is manifest in the theory $\CT^\vee_1$ as a $\fru(1) \oplus \fru(1)$ topological symmetry. 
This is manifest at the partition function level --  the FI parameters of $\CT^\vee_1$ have to be set to zero for the above identity to hold. 
Since no further mutation is allowed on the theory $\CT^\vee_3$, the duality sequence ends here, giving a 2-parameter family of trialities.

Note that the duality sequence in \figref{IRdual-Ex6c} implies a very interesting IR duality between the theories $\CT$ and 
$\CT^\vee_3$. Although the intermediate theory $\CT^\vee_1$ has a single Abelian multiplet, the theories $\CT$ and 
$\CT^\vee_3$ are both linear quivers with a unitary node and a special unitary node. The action of the duality is to simply 
flip a unitary node to a special unitary node of the same rank and vice-versa. Note that the overbalanced 
unitary node has $e=2$ for both $\CT$ and $\CT^\vee_3$.\\

\subsection{Balanced unitary node: Triality}\label{Bal-2node}

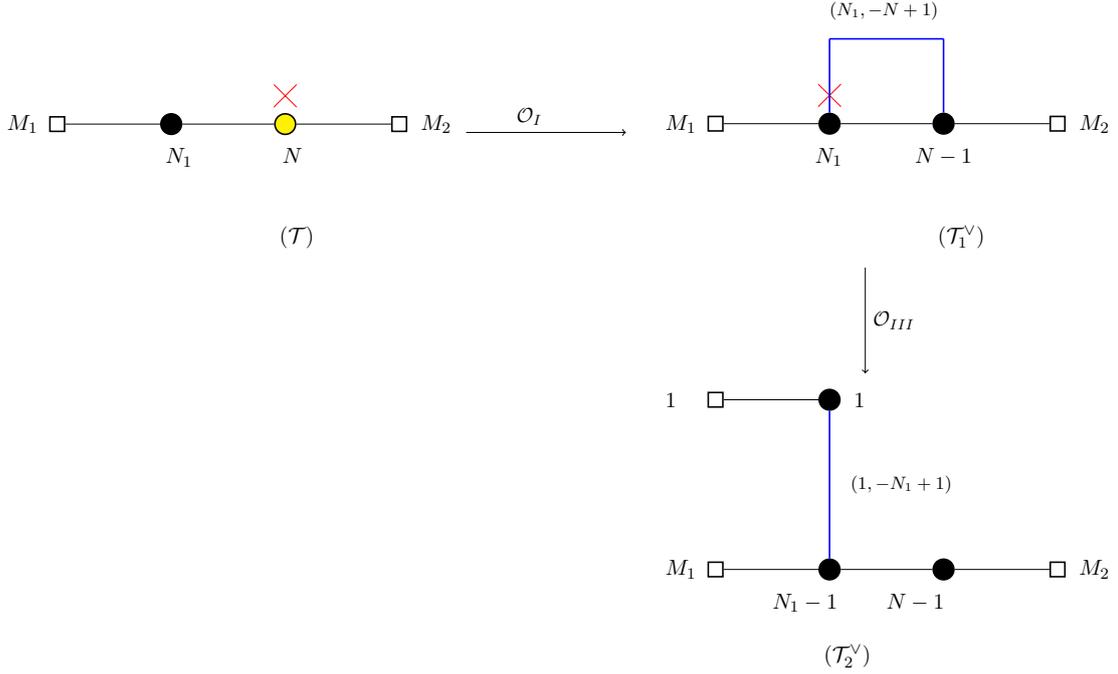
\begin{figure}[htbp]
\begin{center}
\begin{tabular}{ccc}
\scalebox{0.75}{\begin{tikzpicture}
\node[fnode] (1) at (0,0){};
\node[unode] (2) at (2,0){};
\node[sunode] (3) at (4,0){};
\node[fnode] (4) at (6,0){};
\node[cross, red] (5) at (4,0.5){};
\draw[-] (1) -- (2);
\draw[-] (2)-- (3);
\draw[-] (3) -- (4);
\node[text width=.1cm](20) [left=0.5 cm of 1]{$M_1$};
\node[text width=.2cm](21) [below=0.1cm of 2]{$N_1$};
\node[text width=.1cm](22) [below=0.1cm of 3]{$N$};
\node[text width=.1cm](23) [right=0.1cm of 4]{$M_2$};
\node[text width=.2cm](20) at (4,-2){$(\CT)$};
\end{tikzpicture}}
& \scalebox{.7}{\begin{tikzpicture}
\draw[->] (4,0) -- (7, 0);
\node[text width=0.1cm](29) at (5, 0.3) {$\CO_{I}$};
\node[](30) at (5, -2.2) {};
\end{tikzpicture}}
&\scalebox{0.75}{\begin{tikzpicture}
\node[fnode] (1) at (0,0){};
\node[unode] (2) at (2,0){};
\node[unode] (3) at (4,0){};
\node[fnode] (4) at (6,0){};
\node[cross, red] (5) at (2,0.5){};
\draw[-] (1) -- (2);
\draw[-] (2)-- (3);
\draw[-] (3) -- (4);
\draw[-, thick, blue] (2)--(2,1.5);
\draw[-, thick, blue] (3)--(4,1.5);
\draw[-, thick, blue] (2,1.5)--(4,1.5);
\node[text width=2 cm](10) at (3, 2){\footnotesize{$(N_1, -N+1)$}};
\node[text width=.1cm](20) [left=0.5 cm of 1]{$M_1$};
\node[text width=0.5 cm](21) [below=0.1cm of 2]{$N_1$};
\node[text width=1 cm](22) [below=0.1cm of 3]{$N-1$};
\node[text width=.1cm](23) [right=0.1cm of 4]{$M_2$};
\node[text width=.2cm](20) at (4,-2){$(\CT^\vee_1)$};
\end{tikzpicture}}\\
\qquad 
& \qquad
& \scalebox{.7}{\begin{tikzpicture}
\draw[->] (4,-1) -- (4, -3);
\node[text width=0.1cm](29) at (4.2, -2) {$\CO_{III}$};
\end{tikzpicture}}\\
\qquad
& \qquad
&\scalebox{0.75}{\begin{tikzpicture}
\node[fnode] (1) at (0,0){};
\node[unode] (2) at (2,0){};
\node[unode] (3) at (4,0){};
\node[fnode] (4) at (6,0){};
\node[unode] (6) at (2,3){};
\node[fnode] (7) at (0,3){};
\node[](8) at (2, 1.5){};
\draw[-] (1) -- (2);
\draw[-] (2)-- (3);
\draw[-] (3) -- (4);
\draw[-] (6) --(7);
\draw[-, thick, blue] (2)--(6);
\node[text width=.1cm](20) [left=0.5 cm of 1]{$M_1$};
\node[text width=2 cm](21) [below=0.1cm of 2]{$N_1-1$};
\node[text width=2 cm](22) [below=0.1cm of 3]{$N-1$};
\node[text width=.1cm](23) [right=0.1cm of 4]{$M_2$};
\node[text width=.1cm](24) [right=0.1cm of 6]{1};
\node[text width=.1cm](25) [left=0.5 cm of 7]{1};
\node[text width=2 cm](26) [right=0.1 cm of 8]{\footnotesize{$(1, -N_1+1)$}};
\node[text width=.2cm](40) [below=1cm of 2]{$(\CT^\vee_2)$};
\end{tikzpicture}}
\end{tabular}
\end{center}
\caption{\footnotesize{Triality for the case $e=0$. The gauge node at which a mutation acts is marked by a red cross at every step.}}
\label{IRdual-Ex6d}
\end{figure}

Let us now consider the case where the $U(N_1)$ node in $\CT$ is balanced i.e. $N + M_1 = 2N_1 +e$ with $e=0$. 
The first step is precisely the same as the duality sequences studied in \Secref{Unbal-2node} -- the balanced $SU(N)$ node 
is replaced by a $U(N-1)$ node plus an Abelian hypermultiplet by mutation I, which leads to the theory $\CT^\vee_1$ in \figref{IRdual-Ex6d}.  
The $U(N_1)$ gauge node in $\CT^\vee_1$ has $M_1 + (N-1)= 2N_1-1$ fundamental hypers as well as a single Abelian hyper. Therefore, one 
can implement the mutation III to this node, which leads to the theory $\CT^\vee_2$ in \figref{IRdual-Ex6d}. This gives a 2-parameter 
family of trialities involving the quiver gauge theories $\CT$, $\CT^\vee_1$ and $\CT^\vee_2$.

The duality sequence can be realized in terms of the sphere partition function as follows. The first step involving mutation I is exactly the same as 
in \Secref{Unbal-2node}. Starting from the partition function of $\CT^\vee_1$, one can implement mutation III using the identity \eref{Id-3} 
(see \Appref{2node-pf-b} for details) which leads to the following partition function relation:
\be
\boxed{Z^{\CT}(\vec m; 0)= Z^{(\CT^\vee_1)}(\vec m, m_{\rm{Ab}}=\tr \vec m^2; \eta, 0) = Z^{(\CT^\vee_2)}(\vec m', m'_{\rm{Ab}}; -\eta, \eta, \eta).}
\ee
The duality map for the masses are given as:
\begin{align}
& \vec m'^1 = \vec m^1, \qquad \vec m'^2 = \vec m^2, \qquad m'^3 = \tr \vec m^1, \qquad m'_{\rm{Ab}}= \tr \vec m^2,
\end{align}
where $m'^3$ denotes the mass for the fundamental hyper of the $U(1)$ gauge node in \figref{IRdual-Ex6d}.
As before, the number of independent real mass parameters are $M_1 + M_2$ living in the Cartan subalgebra of  
 $\frg_{\rm H}= \frsu(M_1) \oplus \frsu(M_2) \oplus \fru(1) \oplus \fru(1)$. \\

Let us discuss the matching of the Coulomb branch global symmetry in this case. 
In \Secref{MO-2node}, we discussed that the Coulomb branch global symmetry algebra for the theory $\CT$ with the balanced unitary node 
gets enhanced to $\frg_{\rm C}= \frsu(2) \oplus \frsu(2) \oplus \fru(1)$ in the IR, while only a $\fru(1)$ symmetry is visible in the IR. 
The first mutation produces the theory $\CT^\vee_2$, for which the symmetry visible in the UV
is $\fru(1) \oplus \fru(1)$. In the next step along the duality sequence, the second mutation produces the theory 
$\CT^\vee_2$. The symmetry manifest in the UV for $\CT^\vee_2$ is $\fru(1) \oplus \fru(1) \oplus \fru(1)$ 
which has the same rank as $\frg_{\rm C}$.
In addition, the enhanced global symmetry algebra can be read off from the quiver gauge theory $\CT^\vee_2$ in this particular case. 
Note that the quiver $\CT^\vee_2$ has two balanced gauge nodes -- $U(1)$ and $U(N-1)$, and a single 
overbalanced $U(N_1-1)$ gauge node. The balanced nodes are expected to contribute an $\frsu(2)$ factor each 
while the overbalanced node contributes a $\fru(1)$ factor, thereby reproducing the correct global symmetry algebra. 
One can check this more directly by computing the Coulomb branch Hilbert Series of the theory $\CT^\vee_2$.

\section{N-ality in a three-node quiver}\label{N-al-3node}

In this section, we will construct another explicit example of the duality sequence starting from a quiver gauge theory $\CT$ 
with three gauge nodes of the following form:\\
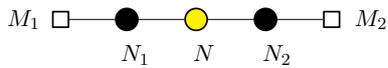
\begin{center}
\scalebox{0.8}{\begin{tikzpicture}
\node[fnode] (1) {};
\node[unode] (2) [right=.75cm  of 1]{};
\node[sunode] (3) [right=.75cm of 2]{};
\node[unode] (4) [right=0.75 cm of 3]{};
\node[fnode] (5) [right=0.75 cm of 4]{};
\draw[-] (1) -- (2);
\draw[-] (2)-- (3);
\draw[-] (3) -- (4);
\draw[-] (4) -- (5);
\node[text width=.1cm](10) [left=0.5 cm of 1]{$M_1$};
\node[text width=.2cm](11) [below=0.1cm of 2]{$N_1$};
\node[text width=.1cm](12) [below=0.1cm of 3]{$N$};
\node[text width=.1cm](13) [below=0.1cm of 4]{$N_2$};
\node[text width=.1cm](14) [right=0.1cm of 5]{$M_2$};
\end{tikzpicture}}
\end{center}

The $SU(N)$ gauge node is balanced i.e. $N_1 + N_2=2N-1$, while the unitary gauge nodes can be either balanced or overbalanced, 
where $M_1 + N =2N_1 +e_1$ and $M_2 + N =2N_2 +e_2$ such that $e_1, e_2 \geq 0$. The above quiver therefore represents a 
4-parameter family of quiver gauge theories. We will first discuss the Coulomb branch global symmetry of the quivers 
for different ranges of $e_1$ and $e_2$ in \Secref{MO-3node}, pointing out the emergent symmetry that appears in the IR for the respective ranges.
We will then discuss the duality sequence, which depends on the specific values of $e_1,e_2$, for the cases $\{e_1,e_2 \neq 0\}$, $\{e_1=0, e_2 \neq 0\}$ and $\{e_1,e_2=0\}$ in \Secref{Unbal-3node}, \Secref{Bal1-3node} and \Secref{Bal2-3node} respectively. For each case, we will discuss how the emergent IR symmetry of $\CT$ becomes manifest in one or more theories of the N-al set.

\subsection{Monopole operators and global symmetry enhancement}\label{MO-3node}

One can show that the quiver $\CT$ with the constraints mentioned above is a 
good theory in the Gaiotto-Witten sense, and therefore the Coulomb branch global symmetry algebra 
can again be read off by enumerating the monopole operators with conformal dimension 1. 
This can be worked out in a fashion similar to the 2-node quiver in \Secref{MO-2node}. Below, we summarize the results:

\begin{enumerate}

\item If both unitary nodes are overbalanced, the Coulomb branch global symmetry algebra is $\frg_{\rm C}= \fru(1)\oplus \fru(1)\oplus \fru(1)$, 
where two of the $\fru(1)$ factors correspond to the topological symmetries of the unitary gauge nodes, while the third one emerges in the IR and is associated to a monopole operator charged only under the $SU(N)$ gauge node. This case is similar to the case of a two-node quiver with the unitary node overbalanced. 

\item If one of the unitary nodes is balanced, the global symmetry algebra is enhanced to 
$\frg_{\rm C}=\frsu(2) \oplus \frsu(2) \oplus \fru(1)\oplus \fru(1)$. 
One of the $\frsu(2)$ factors arise from the balanced unitary node in a standard fashion i.e. the generator of the topological symmetry and 
the conserved currents associated with a pair of monopole operators charged only under the balanced node generate an $\frsu(2)$ algebra. The second is generated by currents associated with three monopole operators which are charged under the balanced unitary node as well as the $SU(N)$ node but not under the overbalanced unitary node. The two $\fru(1)$ factors correspond to the generator of the topological symmetry of the overbalanced node and the monopole operator charged only under the $SU(N)$ node respectively.

\item If both the unitary factors are balanced, the Coulomb branch global symmetry is enhanced further to $\frg_{\rm C}= \frsu(2) \oplus \frsu(2) \oplus \frsu(4) \oplus \fru(1)$. The $\frsu(2)$ factors arise from the two balanced unitary nodes in the standard fashion. The currents generating the
$\frsu(4)$ factor correspond to the monopole operators that are charged under one of the two balanced unitary nodes in addition to the $SU(N)$ node, and the monopole operators that are charged under all three gauge nodes. The $\fru(1)$ factor corresponds to the monopole operator charged only under the $SU(N)$ gauge node. 

\end{enumerate}

The Higgs branch global symmetry algebra can be directly read off from the quiver, i.e. 
$\frg_{\rm H} = \frsu(M_1) \oplus \frsu(M_2) \oplus \fru(1) \oplus \fru(1)$, which arise from 
the fundamental hypermultiplets of the respective gauge nodes.\\

\subsection{Overbalanced unitary nodes : a duality}\label{Unbal-3node}

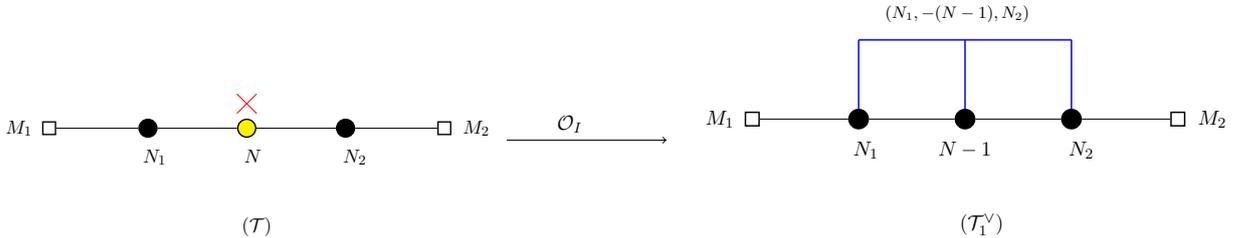
\begin{figure}[htbp]
\begin{center}
\begin{tabular}{ccc}
\scalebox{0.65}{\begin{tikzpicture}
\node[fnode] (1) at (0,0){};
\node[unode] (2) at (2,0){};
\node[sunode] (3) at (4,0){};
\node[unode] (4) at (6,0){};
\node[fnode] (5) at (8,0){};
\node[cross, red] (6) at (4,0.5){};
\draw[-] (1) -- (2);
\draw[-] (2)-- (3);
\draw[-] (3) -- (4);
\draw[-] (4) --(5);
\node[text width=.1cm](10) [left=0.5 cm of 1]{$M_1$};
\node[text width=.2cm](11) [below=0.1cm of 2]{$N_1$};
\node[text width=.1cm](12) [below=0.1cm of 3]{$N$};
\node[text width=.1cm](13) [below=0.1cm of 4]{$N_2$};
\node[text width=.1cm](14) [right=0.1cm of 5]{$M_2$};
\node[text width=.2cm](20) at (4,-2){$(\CT)$};
\end{tikzpicture}}
& \scalebox{.7}{\begin{tikzpicture}
\draw[->] (0,0) -- (3,0);
\node[text width=0.1cm](29) at (1, 0.3) {$\CO_{I}$};
\node[](30) at (1, -1.8) {};
\end{tikzpicture}}
&\scalebox{0.7}{\begin{tikzpicture}
\node[fnode] (1) at (0,0){};
\node[unode] (2) at (2,0){};
\node[unode] (3) at (4,0){};
\node[unode] (4) at (6,0){};
\node[fnode] (5) at (8,0){};
\draw[-] (1) -- (2);
\draw[-] (2)-- (3);
\draw[-] (3) -- (4);
\draw[-] (4) --(5);
\draw[-, thick, blue] (2)--(2,1.5);
\draw[-, thick, blue] (3)--(4,1.5);
\draw[-, thick, blue] (4)--(6,1.5);
\draw[-, thick, blue] (2,1.5)--(4,1.5);
\draw[-, thick, blue] (4,1.5)--(6,1.5);
\node[text width=3 cm](10) at (4, 2){\footnotesize{$(N_1, -(N-1), N_2)$}};
\node[text width=.1cm](10) [left=0.5 cm of 1]{$M_1$};
\node[text width=.2cm](11) [below=0.1cm of 2]{$N_1$};
\node[text width= 1cm](12) [below=0.1cm of 3]{$N-1$};
\node[text width=.1cm](13) [below=0.1cm of 4]{$N_2$};
\node[text width=.1cm](14) [right=0.1cm of 5]{$M_2$};
\node[text width=.2cm](20) at (4,-2){$(\CT^\vee_1)$};
\end{tikzpicture}}
\end{tabular}
\end{center}
\caption{\footnotesize{Duality for the case $e_1, e_2 >0$. The duality sequence terminates at this point for $e_1>2$ and $e_2 >2$.}}
\label{IRdual-Ex4a}
\end{figure}

Let us first consider the case where both unitary nodes in $\CT$ are overbalanced, i.e. $e_1 >0, e_2 >0$. The first step is to 
apply mutation I to the $SU(N)$ gauge node which leads to the quiver gauge theory $\CT^\vee_1$, as shown in \figref{IRdual-Ex4a}.
 
Implementing the mutation in terms of the sphere partition function, one arrives at the following identity (see \Appref{3node-pf-ob} for details):
\be \label{pf-duality-1}
\boxed{Z^{\CT}(\vec m^1, \vec m^2;\eta_1, \eta_3)=  Z^{\CT^\vee}(\vec m^1, \vec m^2, m_{\rm{Ab}}=0;\eta_1,0, \eta_3).}
\ee
As discussed earlier, the Coulomb branch global symmetry of the IR SCFT in this case is $\fru(1) \oplus \fru(1) \oplus \fru(1)$. 
For the theory $\CT$, only a $\fru(1) \oplus \fru(1)$ global symmetry is visible in the UV, and the other $\fru(1)$ factor appears as an emergent symmetry in the IR. For the theory $\CT^\vee_1$, the full $\fru(1) \oplus \fru(1) \oplus \fru(1)$ global symmetry is visible in the UV.

For generic values of the integers $e_1>2, e_2 >2$, no other mutation can be implemented on the quiver $\CT^\vee_1$. 
The N-ality in this case is simply a duality. However, for special values of $e_i=1,2$, the quiver admits additional mutations, 
as we found in the case of the 2-node quiver. These cases can be worked in a fashion similar to the 2-node quiver.

\subsection{A single balanced unitary node : a triality}\label{Bal1-3node}

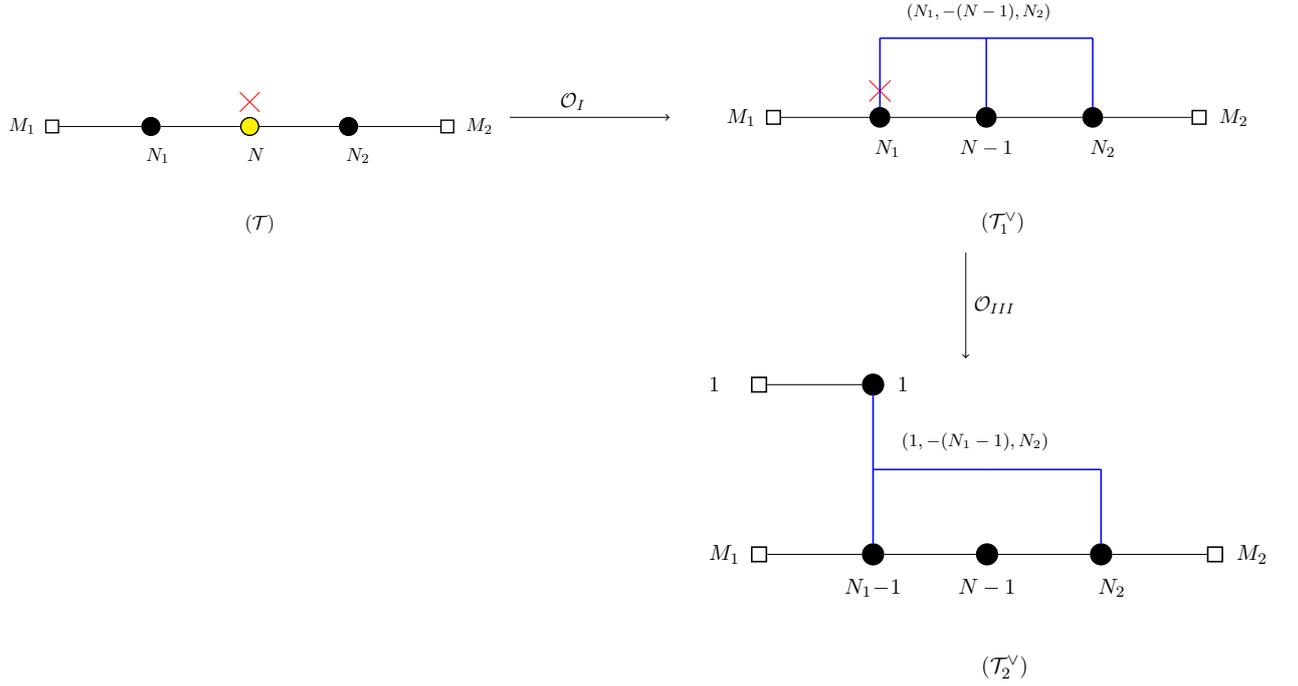
\begin{figure}[htbp]
\begin{center}
\begin{tabular}{ccc}
\scalebox{0.65}{\begin{tikzpicture}
\node[fnode] (1) at (0,0){};
\node[unode] (2) at (2,0){};
\node[sunode] (3) at (4,0){};
\node[unode] (4) at (6,0){};
\node[fnode] (5) at (8,0){};
\node[cross, red] (6) at (4,0.5){};
\draw[-] (1) -- (2);
\draw[-] (2)-- (3);
\draw[-] (3) -- (4);
\draw[-] (4) --(5);
\node[text width=.1cm](10) [left=0.5 cm of 1]{$M_1$};
\node[text width=.2cm](11) [below=0.1cm of 2]{$N_1$};
\node[text width=.1cm](12) [below=0.1cm of 3]{$N$};
\node[text width=.1cm](13) [below=0.1cm of 4]{$N_2$};
\node[text width=.1cm](14) [right=0.1cm of 5]{$M_2$};
\node[text width=.2cm](20) at (4,-2){$(\CT)$};
\end{tikzpicture}}
& \scalebox{.7}{\begin{tikzpicture}
\draw[->] (4,0) -- (7, 0);
\node[text width=0.1cm](29) at (5, 0.3) {$\CO_{I}$};
\node[](30) at (5, -2.2) {};
\end{tikzpicture}}
&\scalebox{0.7}{\begin{tikzpicture}
\node[fnode] (1) at (0,0){};
\node[unode] (2) at (2,0){};
\node[unode] (3) at (4,0){};
\node[unode] (4) at (6,0){};
\node[fnode] (5) at (8,0){};
\node[cross, red] (6) at (2,0.5){};
\draw[-] (1) -- (2);
\draw[-] (2)-- (3);
\draw[-] (3) -- (4);
\draw[-] (4) --(5);
\draw[-, thick, blue] (2)--(2,1.5);
\draw[-, thick, blue] (3)--(4,1.5);
\draw[-, thick, blue] (4)--(6,1.5);
\draw[-, thick, blue] (2,1.5)--(4,1.5);
\draw[-, thick, blue] (4,1.5)--(6,1.5);
\node[text width=3 cm](10) at (4, 2){\footnotesize{$(N_1, -(N-1), N_2)$}};
\node[text width=.1cm](10) [left=0.5 cm of 1]{$M_1$};
\node[text width=.2cm](11) [below=0.1cm of 2]{$N_1$};
\node[text width= 1cm](12) [below=0.1cm of 3]{$N-1$};
\node[text width=.1cm](13) [below=0.1cm of 4]{$N_2$};
\node[text width=.1cm](14) [right=0.1cm of 5]{$M_2$};
\node[text width=.2cm](20) at (4,-2){$(\CT^\vee_1)$};
\end{tikzpicture}}\\
\qquad 
& \qquad
& \scalebox{.7}{\begin{tikzpicture}
\draw[->] (4,-1) -- (4, -3);
\node[text width=0.1cm](29) at (4.2, -2) {$\CO_{III}$};
\end{tikzpicture}}\\
\qquad
& \qquad
&\scalebox{0.75}{\begin{tikzpicture}
\node[fnode] (1) at (0,0){};
\node[unode] (2) at (2,0){};
\node[unode] (3) at (4,0){};
\node[unode] (4) at (6,0){};
\node[fnode] (5) at (8,0){};
\node[unode] (6) at (2,3){};
\node[fnode] (7) at (0,3){};
\draw[-] (1) -- (2);
\draw[-] (2)-- (3);
\draw[-] (3) -- (4);
\draw[-] (4) --(5);
\draw[-] (6) --(7);
\draw[-, thick, blue] (2)--(6);
\draw[-, thick, blue] (4)--(6,1.5);
\draw[-, thick, blue] (2,1.5)--(4,1.5);
\draw[-, thick, blue] (4,1.5)--(6,1.5);
\node[text width=3 cm](40) at (4, 2){\footnotesize{$(1, -(N_1-1), N_2)$}};
\node[text width=.1cm](20) [left=0.5 cm of 1]{$M_1$};
\node[text width=1 cm](21) [below=0.1cm of 2]{$N_1-1$};
\node[text width=1 cm](22) [below=0.1cm of 3]{$N-1$};
\node[text width=0.1 cm](23) [below=0.1cm of 4]{$N_2$};
\node[text width=.1cm](24) [right=0.1cm of 5]{$M_2$};
\node[text width=.1cm](25) [right=0.1cm of 6]{1};
\node[text width=.1cm](26) [left=0.5 cm of 7]{1};
\node[text width=.2cm](20) at (4,-2){$(\CT^\vee_2)$};
\end{tikzpicture}}
\end{tabular}
\end{center}
\caption{\footnotesize{A triality that arises when $e_1=0$ and $e_2>2$.} }
\label{IRdual-Ex4b}
\end{figure}

Let us consider the case where the gauge node $U(N_1)$ is balanced, while the node $U(N_2)$ is overbalanced -- this will 
lead to a triality of the quiver gauge theories shown in \figref{IRdual-Ex4b}.
Starting from the theory $\CT$, the first step is the same as in the previous case, i.e.
one uses mutation I to replace the $SU(N)$ gauge node with a $U(N-1)$ gauge node 
and an Abelian hypermultiplet, giving the theory $\CT^\vee_1$.
The $U(N_1)$ gauge node in this theory has $N_f = M_1 + N -1 =2N_1-1$ fundamental flavors 
plus an Abelian hypermultiplet, and one can therefore implement a mutation III at this node, which 
gives the theory $\CT^\vee_2$.\\

The duality sequence can be realized at the partition function level by first implementing mutation I as we did 
in \Secref{Unbal-3node}. Starting from the partition function of $\CT^\vee_1$, one can implement mutation III using the identity \eref{Id-3} 
(see \Appref{3node-pf-b-1} for details) which leads to the following partition function relation:
\be \label{pf-triality-1}
\boxed{Z^{\CT}(\vec m;\eta_1, \eta_3)=  Z^{\CT^\vee_1}(\vec m, m_{\rm{Ab}}=0;\eta_1,0, \eta_3) =Z^{\CT^\vee_2}(\vec m, m_{\rm{Ab}}=0; \eta_1,-\eta_1, \eta_1, \eta_3) .}
\ee

The Coulomb branch global symmetry algebra of $\CT$ is $\frg_{\rm{C}}= \frsu(2) \oplus \frsu(2) \oplus \fru(1)\oplus \fru(1)$, 
of which only a $\fru(1)\oplus \fru(1)$ subalgebra is visible in the UV. For the theory $\CT^\vee_1$, a $\fru(1)\oplus \fru(1)\oplus \fru(1)$ subalgebra 
of the full global symmetry algebra is visible. In contrast, the correct rank of $\frg_{\rm{C}}$ can be read off 
from the quiver gauge theory $\CT^\vee_2$. In fact, one can also read off $\frg_{\rm{C}}$ by noting that the $U(1)$ and 
the $U(N-1)$ gauge nodes are balanced (thereby contributing an $\frsu(2)$ factor each), while the $U(N_1-1)$ and the $U(N_2)$ nodes 
are overbalanced (thereby contributing a $\fru(1)$ factor each). 

We can construct the triality in terms of the sphere partition function (see \Appref{3node-pf-b-1} for details) leading to the following 
identity:
\be \label{pf-triality-1}
\boxed{Z^{\CT}(\vec m;\eta_1, \eta_3)=  Z^{\CT^\vee_1}(\vec m, m_{\rm{Ab}}=0;\eta_1,0, \eta_3) =Z^{\CT^\vee_2}(\vec m, m_{\rm{Ab}}=0; \eta_1,-\eta_1, \eta_1, \eta_3) .}
\ee

For a generic integer $e_2>2$, the duality sequence ends with $\CT^\vee_2$. For $e_2=1,2$, the quiver $\CT^\vee_2$ will allow further mutations 
and these can be worked in a fashion analogous to the treatment for the case of a two-node quiver.

\subsection{Two balanced unitary nodes : a hexality}\label{Bal2-3node}

\begin{figure}[htbp]
\begin{center}
\begin{tabular}{ccc}
\scalebox{0.5}{\begin{tikzpicture}
\node[fnode] (1) at (0,0){};
\node[unode] (2) at (2,0){};
\node[sunode] (3) at (4,0){};
\node[unode] (4) at (6,0){};
\node[fnode] (5) at (8,0){};
\node[cross, red] (6) at (4,0.5){};
\draw[-] (1) -- (2);
\draw[-] (2)-- (3);
\draw[-] (3) -- (4);
\draw[-] (4) --(5);
\node[text width=.1cm](10) [left=0.5 cm of 1]{$M_1$};
\node[text width=.2cm](11) [below=0.1cm of 2]{$N_1$};
\node[text width=.1cm](12) [below=0.1cm of 3]{$N$};
\node[text width=.1cm](13) [below=0.1cm of 4]{$N_2$};
\node[text width=.1cm](14) [right=0.1cm of 5]{$M_2$};
\node[text width=.2cm](20) at (4,-2){$(\CT)$};
\end{tikzpicture}}
& \qquad 
& \qquad \\
\scalebox{.6}{\begin{tikzpicture}
\draw[->] (4,-1) -- (4, -2);
\node[text width=0.1cm](29) at (4.2, -1.5) {$\CO_{I}$};
\end{tikzpicture}}
& \qquad 
& \qquad \\
\scalebox{0.5}{\begin{tikzpicture}
\node[fnode] (1) at (0,0){};
\node[unode] (2) at (2,0){};
\node[unode] (3) at (4,0){};
\node[unode] (4) at (6,0){};
\node[fnode] (5) at (8,0){};
\node[cross, red] (6) at (2,0.5){};
\node[cross, red] (7) at (6,0.5){};
\draw[-] (1) -- (2);
\draw[-] (2)-- (3);
\draw[-] (3) -- (4);
\draw[-] (4) --(5);
\draw[-, thick, blue] (2)--(2,1.5);
\draw[-, thick, blue] (3)--(4,1.5);
\draw[-, thick, blue] (4)--(6,1.5);
\draw[-, thick, blue] (2,1.5)--(4,1.5);
\draw[-, thick, blue] (4,1.5)--(6,1.5);
\node[text width=3 cm](10) at (4, 2){\footnotesize{$(N_1, -(N-1), N_2)$}};
\node[text width=.1cm](10) [left=0.5 cm of 1]{$M_1$};
\node[text width=.2cm](11) [below=0.1cm of 2]{$N_1$};
\node[text width= 1cm](12) [below=0.1cm of 3]{$N-1$};
\node[text width=.1cm](13) [below=0.1cm of 4]{$N_2$};
\node[text width=.1cm](14) [right=0.1cm of 5]{$M_2$};
\node[text width=.2cm](20) at (4,-2){$(\CT^\vee_1)$};
\end{tikzpicture}}
&\scalebox{.6}{\begin{tikzpicture}
\draw[->] (4,0) -- (7, 0);
\node[text width=0.2cm](29) at (4.5, 0.3) {$\CO_{III}$};
\node[](30) at (5, -2.1) {};
\end{tikzpicture}}
&\scalebox{0.5}{\begin{tikzpicture}
\node[fnode] (1) at (0,0){};
\node[unode] (2) at (2,0){};
\node[unode] (3) at (4,0){};
\node[unode] (4) at (6,0){};
\node[fnode] (5) at (8,0){};
\node[unode] (6) at (2,3){};
\node[fnode] (7) at (0,3){};
\node[cross, red] (8) at (6,0.5){};
\draw[-] (1) -- (2);
\draw[-] (2)-- (3);
\draw[-] (3) -- (4);
\draw[-] (4) --(5);
\draw[-] (6) --(7);
\draw[-, thick, blue] (2)--(6);
\draw[-, thick, blue] (4)--(6,1.5);
\draw[-, thick, blue] (2,1.5)--(4,1.5);
\draw[-, thick, blue] (4,1.5)--(6,1.5);
\node[text width=3 cm](40) at (4, 2){\footnotesize{$(1, -(N_1-1), N_2)$}};
\node[text width=.1cm](20) [left=0.5 cm of 1]{$M_1$};
\node[text width=1 cm](21) [below=0.1cm of 2]{$N_1-1$};
\node[text width=1 cm](22) [below=0.1cm of 3]{$N-1$};
\node[text width=0.1 cm](23) [below=0.1cm of 4]{$N_2$};
\node[text width=.1cm](24) [right=0.1cm of 5]{$M_2$};
\node[text width=.1cm](25) [right=0.1cm of 6]{1};
\node[text width=.1cm](26) [left=0.5 cm of 7]{1};
\node[text width=.2cm](20) at (4,-2){$(\CT^\vee_2)$};
\end{tikzpicture}}\\
\qquad & \qquad & \qquad \\
\scalebox{.6}{\begin{tikzpicture}
\draw[->] (4,-1) -- (4, -2);
\node[text width=0.1cm](29) at (4.2, -1.5) {$\CO_{III}$};
\end{tikzpicture}}
&\qquad
&\scalebox{.6}{\begin{tikzpicture}
\draw[->] (4,-1) -- (4, -2);
\node[text width=0.1cm](29) at (4.2, -1.5) {$\CO_{III}$};
\end{tikzpicture}}\\
\qquad & \qquad & \qquad \\
\scalebox{0.6}{\begin{tikzpicture}
\node[fnode] (1) at (0,0){};
\node[unode] (2) at (2,0){};
\node[unode] (3) at (4,0){};
\node[unode] (4) at (6,0){};
\node[fnode] (5) at (8,0){};
\node[unode] (6) at (6,3){};
\node[fnode] (7) at (8,3){};
\node[cross, red] (8) at (2,0.5){};
\draw[-] (1) -- (2);
\draw[-] (2)-- (3);
\draw[-] (3) -- (4);
\draw[-] (4) --(5);
\draw[-] (6) --(7);
\draw[-, thick, blue] (4)--(6);
\draw[-, thick, blue] (2)--(2,1.5);
\draw[-, thick, blue] (2,1.5)--(4,1.5);
\draw[-, thick, blue] (4,1.5)--(6,1.5);
\node[text width=3 cm](40) at (4, 2){\footnotesize{$(1, N_1, -(N_2-1))$}};
\node[text width=.1cm](20) [left=0.5 cm of 1]{$M_1$};
\node[text width=0.1 cm](21) [below=0.1cm of 2]{$N_1$};
\node[text width=1 cm](22) [below=0.1cm of 3]{$N-1$};
\node[text width=1 cm](23) [below=0.1cm of 4]{$N_2-1$};
\node[text width=.1cm](24) [right=0.1cm of 5]{$M_2$};
\node[text width=.1cm](25) [left=0.1cm of 6]{1};
\node[text width=.1cm](26) [right=0.5 cm of 7]{1};
\node[text width=.2cm](20) at (4,-2){$(\CT^\vee_3)$};
\end{tikzpicture}}
&\scalebox{.6}{\begin{tikzpicture}
\draw[->] (4,0) -- (7, 0);
\node[text width=0.1cm](29) at (4.5, 0.3) {$\CO_{III}$};
\node[](30) at (5, -2.1) {};
\end{tikzpicture}}
& \scalebox{0.6}{\begin{tikzpicture}
\node[fnode] (1) at (0,0){};
\node[unode] (2) at (2,0){};
\node[unode] (3) at (4,0){};
\node[unode] (4) at (6,0){};
\node[fnode] (5) at (8,0){};
\node[unode] (6) at (2,3){};
\node[fnode] (7) at (0,3){};
\node[unode] (8) at (6,3){};
\node[fnode] (9) at (8,3){};
\node[cross, red] (10) at (4,0.5){};
\draw[-] (1) -- (2);
\draw[-] (2)-- (3);
\draw[-] (3) -- (4);
\draw[-] (4) --(5);
\draw[-] (6) --(7);
\draw[-] (8) --(9);
\draw[-, thick, blue] (6)--(4,1.5);
\draw[-, thick, blue] (3)--(4,1.5);
\draw[-, thick, blue] (8)--(4,1.5);
\draw[-, thick, blue] (2)--(2,1.5);
\draw[-, thick, blue] (4)--(6,1.5);
\draw[-, thick, blue] (2,1.5)--(4,1.5);
\draw[-, thick, blue] (4,1.5)--(6,1.5);
\node[text width=4.5cm](19) at (7.1, 2){\footnotesize{$(1,1, -N_1+1,N-1,-N_2+1)$}};
\node[text width=.2cm](12) at (4,-2){$(\CT^\vee_4)$};
\node[text width=.1cm](20) [left=0.5 cm of 1]{$M_1$};
\node[text width=1 cm](21) [below=0.1cm of 2]{$N_1-1$};
\node[text width=1 cm](22) [below=0.1cm of 3]{$N-1$};
\node[text width=1 cm](23) [below=0.1cm of 4]{$N_2-1$};
\node[text width=.1cm](24) [right=0.1cm of 5]{$M_2$};
\node[text width=.1cm](25) [right=0.1cm of 6]{1};
\node[text width=.1cm](26) [left=0.5 cm of 7]{1};
\node[text width=.1cm](27) [left=0.5 cm of 8]{1};
\node[text width=.1cm](28) [right=0.1 cm of 9]{1};
\end{tikzpicture}}\\
\qquad
& \qquad 
& \scalebox{.6}{\begin{tikzpicture}
\draw[->] (4,-1) -- (4, -2);
\node[text width=0.1cm](29) at (4.2, -1.5) {$\CO_{III}$};
\end{tikzpicture}}\\
\scalebox{0.6}{\begin{tikzpicture}
\node[fnode] (1) at (0,0){};
\node[unode] (2) at (2,0){};
\node[unode] (3) at (4,0){};
\node[unode] (4) at (6,0){};
\node[fnode] (5) at (8,0){};
\node[unode] (6) at (4,2){};
\node[unode] (7) at (2,2){};
\node[unode] (8) at (0,2){};
\node[fnode] (9) at (-2,2){};
\draw[-] (1) -- (2);
\draw[-] (2)-- (3);
\draw[-] (3) -- (4);
\draw[-] (4) --(5);
\draw[-] (6) --(7);
\draw[-] (8) --(7);
\draw[-] (8) --(9);
\draw[thick, blue] (6)-- (3);
\node[text width=.1cm](40) [left=0.5 cm of 1]{$M_1$};
\node[text width=1 cm](41) [below=0.1cm of 2]{$N_1-1$};
\node[text width=1 cm](42) [below=0.1cm of 3]{$N-2$};
\node[text width=1 cm](43) [below=0.1cm of 4]{$N_2-1$};
\node[text width=.1cm](44) [right=0.1cm of 5]{$M_2$};
\node[text width=.1cm](45) [above=0.1cm of 6]{1};
\node[text width=.1cm](46) [above=0.1 cm of 7]{1};
\node[text width=.1cm](47) [above=0.1 cm of 8]{1};
\node[text width=.1cm](48) [left=0.5 cm of 9]{1};
\node[text width=2cm](21) at (5.1, 1){$(1,-N+2)$};
\node[text width=.2cm](20) at (4,-2){$(\CT^\vee_5)$};
\end{tikzpicture}}
&\scalebox{.6}{\begin{tikzpicture}
\draw[->] (7,0) -- (4, 0);
\node[text width=1 cm](29) at (5, 0.3) {Redef.};
\node[](30) at (5, -2.1) {};
\end{tikzpicture}}
& \scalebox{0.6}{\begin{tikzpicture}
\node[fnode] (1) at (0,0){};
\node[unode] (2) at (2,0){};
\node[unode] (3) at (4,0){};
\node[unode] (4) at (6,0){};
\node[fnode] (5) at (8,0){};
\node[unode] (6) at (2,3){};
\node[fnode] (7) at (0,3){};
\node[unode] (8) at (6,3){};
\node[fnode] (9) at (8,3){};
\node[unode] (22) at (4,3){};
\node[fnode] (23) at (4,5){};
\draw[-] (1) -- (2);
\draw[-] (2)-- (3);
\draw[-] (3) -- (4);
\draw[-] (4) --(5);
\draw[-] (6) --(7);
\draw[-] (8) --(9);
\draw[-] (22) --(23);
\draw[-, thick, blue] (6)--(4,1.5);
\draw[-, thick, blue] (3)--(4,1.5);
\draw[-, thick, blue] (8)--(4,1.5);
\draw[-, thick, blue] (22)--(4,1.5);
\node[text width=4.5cm](19) at (6.5, 1.5){\footnotesize{$(1,1, 1, N-2)$}};
\node[text width=.1cm](40) [left=0.5 cm of 1]{$M_1$};
\node[text width=1 cm](41) [below=0.1cm of 2]{$N_1-1$};
\node[text width=1 cm](42) [below=0.1cm of 3]{$N-2$};
\node[text width=1 cm](43) [below=0.1cm of 4]{$N_2-1$};
\node[text width=.1cm](44) [right=0.1cm of 5]{$M_2$};
\node[text width=.1cm](45) [above=0.1cm of 6]{1};
\node[text width=.1cm](46) [left=0.5 cm of 7]{1};
\node[text width=.1cm](47) [above=0.1 cm of 8]{1};
\node[text width=.1cm](48) [right=0.1 cm of 9]{1};
\node[text width=.1cm](49) [above=0.1 cm of 22]{1};
\node[text width=.1cm](50) [right=0.1 cm of 23]{1};
\node[text width=.2cm](20) at (4,-2){$(\CT^\vee_5)$};
\end{tikzpicture}}
\end{tabular}
\end{center}
\caption{\footnotesize{A hexality of quiver gauge theories which arises when $e_1=0$ and $e_2=0$. The final step involves a field redefinition.}}
\label{IRdual-Ex4c}
\end{figure}
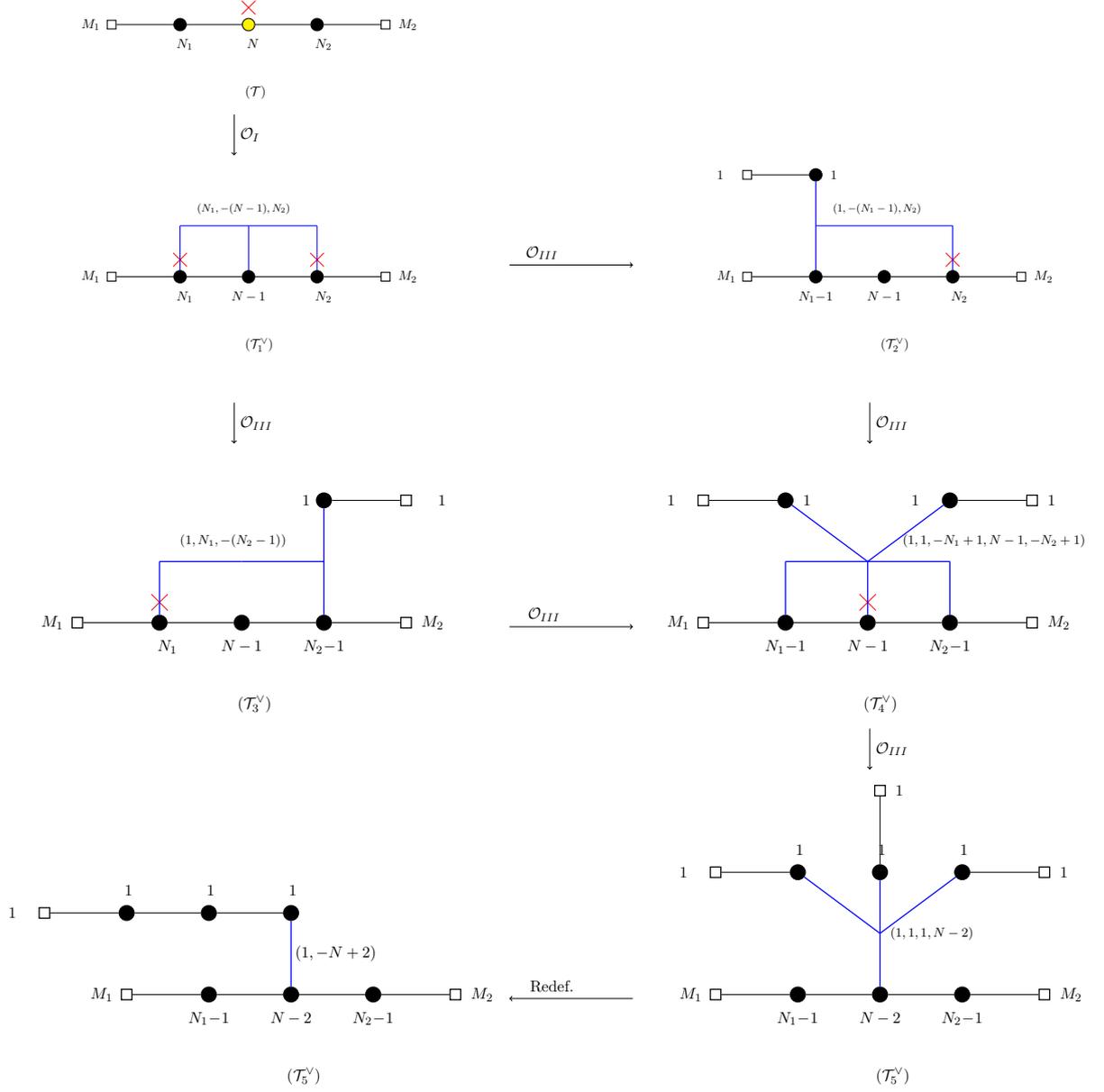

Let us now consider the case where both the gauge nodes $U(N_1)$ and $U(N_2)$ in $\CT$ are balanced. 
Starting from the theory $\CT$, the first step is to use mutation I to replace the $SU(N)$ gauge node 
with a $U(N-1)$ gauge node and an Abelian hypermultiplet, leading to the theory $\CT^\vee_1$. In the 
theory $\CT^\vee_1$, the $U(N_1)$ node has $2N_1 -1$ fundamental/bifundamental hypers plus an 
Abelian hypermultiplet. Similarly, the $U(N_2)$ node has $2N_2 -1$ fundamental/bifundamental hypers 
plus an Abelian hypermultiplet. One can either implement mutation III at the $U(N_1)$ node in the next step 
to obtain the quiver $\CT^\vee_2$, or mutation III at the $U(N_2)$ node instead to obtain 
the quiver $\CT^\vee_3$. In the next step, performing mutation III at the $U(N_2)$ node of $\CT^\vee_2$, 
or the $U(N_1)$ node of $\CT^\vee_3$ leads to the quiver $\CT^\vee_4$. 

Since the original $SU(N)$ node was exactly balanced, the central 
$U(N-1)$ in $\CT^\vee_4$ has $N_f =2(N-1)-1$ fundamental/bifundamental hypers in addition to the Abelian hyper, and 
therefore admits another mutation III, giving the quiver $\CT^\vee_5$. One therefore arrives at the hexality of 
quiver gauge theories shown in \figref{IRdual-Ex4c}. It is convenient to perform a field redefinition on $\CT^\vee_5$ 
to recast the quiver in the final form given in the bottom left corner of  \figref{IRdual-Ex4c}. 
The construction of this hexality in terms of the sphere partition function is worked out in detail in \Appref{3node-pf-b-2}. \\

The Coulomb branch global symmetry algebra of $\CT$ is $\frg_{\rm{C}}= \frsu(2) \oplus \frsu(2) \oplus \frsu(4) \oplus \fru(1)$, of which only a 
$\fru(1) \oplus \fru(1)$ subalgebra is visible in the UV. 
One observes that along the duality sequence starting from $\CT$, the rank of the global symmetry algebra manifest in the UV increases 
by 1 at every step, until the final mutation leading to the quiver $\CT^\vee_5$, where the full rank of $\frg_{\rm C}$ is manifest. 
In addition, one can again read off $\frg_{\rm C}$ by using the standard balanced node argument. 
The two $\frsu(2)$ factors arise from the balanced $U(N_1-1)$ 
and the balanced $U(N_2-1)$ nodes in $\CT^\vee_5$, while the $\fru(1)$ factor arises from the $U(N-2)$ gauge node. 
Finally, the $\frsu(4)$ factor arises from the three consecutive balanced $U(1)$ gauge nodes in $\CT^\vee_5$. \\

\section{Three dimensional mirror for an N-al set}\label{3dmirr}

In this section, we discuss the construction of the 3d mirror associated with a given set of N-al 
theories. After introducing the general procedure in \Secref{gen-mirr}, we construct the 3d mirror 
explicitly for the case of the two-node quiver in \Secref{2node-mirr} and the three-node quiver in 
\Secref{3node-mirr}.

\subsection{The general recipe : $S$-type operations}\label{gen-mirr}

The basic technology for this construction involves the $S$-type operation \cite{Dey:2020hfe}, which is a generalized  
version of Witten's $S \in SL(2,\BZ)$ generator acting on a 3d $\CN=4$ theory. Consider a 3d mirror pair $(X,Y)$ such 
that both theories admit weakly-coupled descriptions in the UV and the theory $X$ has a global symmetry 
subalgebra $\frg_{\rm sub} \subset \frg_{\rm H}$ of the form $\frg_{\rm sub}=\oplus_\gamma \, \fru(M_\gamma)$.
In addition, we demand that the theories are good in the Gaiotto-Witten sense. Given such a 3d mirror pair $(X,Y)$,
 an $S$-type operation generates a new dual pair $(X',Y')$ where $X'$ has a weakly-coupled description 
by construction but $Y'$ is not in general guaranteed to have a Lagrangian description. 
There is however an important subclass of $S$-type operations -- the Abelian $S$-type operations \cite{Dey:2020hfe, Dey:2021rxw}, 
for which $Y'$ can be shown to always have a Lagrangian and this Lagrangian can be explicitly constructed. 
We refer the reader to Section 2.2 and Section 2.3 of \cite{Dey:2021rxw} for a concise proof.\\

There are two types of Abelian $S$-type operations that will be relevant for the construction of 3d mirrors 
in this section -- the \textit{flavoring-gauging} and the \textit{identification-flavoring-gauging} 
operations, which we review next. The former acts on a generic quiver $X$ of the type mentioned above 
to give another quiver $X'$ in the following fashion:

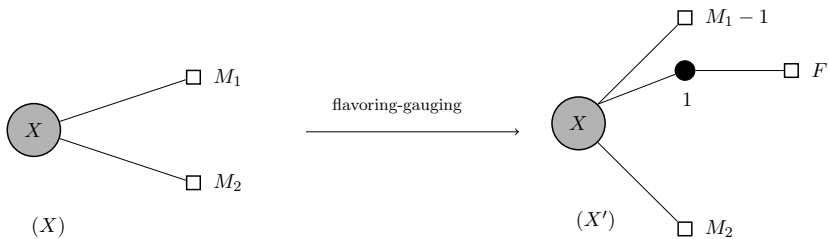
\begin{center}
\begin{tabular}{ccc}
\scalebox{0.7}{\begin{tikzpicture}[bnode/.style={circle,draw, thick, fill=black!30,minimum size=1cm}]
\node[bnode] (1) at (0,0){$X$} ;
\node[fnode] (2) at (3,-1) {} ;
\node[fnode] (3) at (3, 1) {} ;
\draw[-] (1)--(2);
\draw[-] (1)--(3);
\node[text width=1cm](11) [right=0.1cm of 2]{$M_2$};
\node[text width=1cm](12) [right=0.1cm of 3]{$M_1$};
\node[text width=0.1cm](13) [below=1cm of 1]{$(X)$};
\end{tikzpicture}}
& \scalebox{0.7}{\begin{tikzpicture}
\draw[->] (5,0) -- (9,0); 
\node[text width=3cm](10) at (7,0.5){\footnotesize{flavoring-gauging}};
\node[] at (6, -2){};
\end{tikzpicture}}
&\scalebox{0.7}{\begin{tikzpicture}[bnode/.style={circle,draw, thick, fill=black!30,minimum size=1cm}]
\node[bnode] (3) at (10,0){$X$} ;
\node[fnode] (4) at (12, 2){} ;
\node[unode] (5) at (12,1){} ;
\node[fnode] (6) at (12,-2){} ;
\node[fnode] (7) at (14,1){};
\draw[-] (3)--(4);
\draw[-] (3.north east) -- (5);
\draw[-] (3)--(6);
\draw[-] (5)--(7);
\node[text width=2cm](11) [right=0.1cm of 4]{$M_1-1$};
\node[text width=0.1cm](12) [below=0.1cm of 5]{$1$};
\node[text width=2cm](13) [right=0.1cm of 6]{$M_2$};
\node[text width=1cm](14) [right=0.1cm of 7]{$F$};
\node[text width=0.1cm](13) [below=1cm of 3]{$(X')$};
\end{tikzpicture}}
\end{tabular}
\end{center}

In the quiver diagram for $X$, we have only shown two flavor nodes labelled $M_1$ and $M_2$ manifestly, 
while the vector multiplets and the remaining hypermultiplets of the theory are collectively represented 
by the grey blob. The flavoring-gauging operation involves splitting the flavor node labelled $M_1$ into two flavor nodes 
labelled $M_1-1$ and 1, introducing $F$ hypermultiplets of charge 1 at the latter node and then promoting 
it to a gauge node. Similarly, the identification-flavoring-gauging operation acts on the quiver $X$ 
to give another quiver $X'$ as follows:

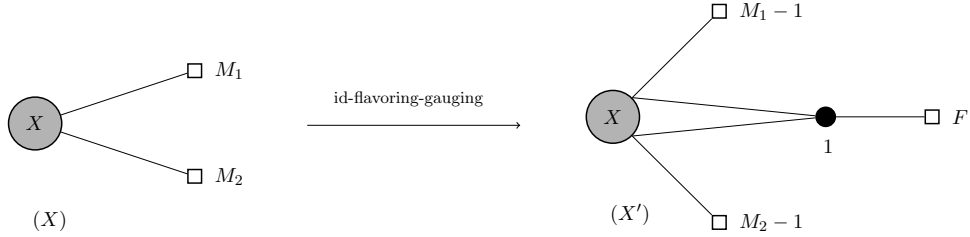
\begin{center}
\begin{tabular}{ccc}
\scalebox{0.7}{\begin{tikzpicture}[bnode/.style={circle,draw, thick, fill=black!30,minimum size=1cm}]
\node[bnode] (1) at (0,0){$X$} ;
\node[fnode] (2) at (3,-1) {} ;
\node[fnode] (3) at (3, 1) {} ;
\draw[-] (1)--(2);
\draw[-] (1)--(3);
\node[text width=1cm](11) [right=0.1cm of 2]{$M_2$};
\node[text width=1cm](12) [right=0.1cm of 3]{$M_1$};
\node[text width=0.1cm](13) [below=1cm of 1]{$(X)$};
\end{tikzpicture}}
& \scalebox{0.7}{\begin{tikzpicture}
\draw[->] (5,0) -- (9,0);
\node[text width=4 cm](6) at (7.5,0.5){\footnotesize{id-flavoring-gauging}};
\node[] at (6, -2){};
\end{tikzpicture}}
& \scalebox{0.7}{\begin{tikzpicture}[bnode/.style={circle,draw, thick, fill=black!30,minimum size=1cm}]
\node[bnode] (3) at (10,0){$X$} ;
\node[fnode] (4) at (12, 2){} ;
\node[unode] (5) at (14,0){} ;
\node[fnode] (6) at (12,-2){} ;
\node[fnode] (7) at (16,0){};
\draw[-] (3)--(4);
\draw[-] (3.north east) to  (5);
\draw[-] (3.south east) to (5);
\draw[-] (3)--(6);
\draw[-] (5)--(7);
\node[text width=2cm](11) [right=0.1cm of 4]{$M_1-1$};
\node[text width=0.1cm](12) [below=0.1cm of 5]{$1$};
\node[text width=2cm](13) [right=0.1cm of 6]{$M_2- 1$};
\node[text width=1cm](14) [right=0.1cm of 7]{$F$};
\node[text width=0.1cm](13) [below=1cm of 3]{$(X')$};
\end{tikzpicture}}
\end{tabular}
\end{center}

In this case, the operation involves splitting both the flavor nodes labelled $M_i$ ($i=1,2$) into flavor nodes $M_i -1$ and 1, 
and then identifying the two flavor nodes labelled 1. This is followed by attaching $F$ hypermultiplets of charge 1 to the identified 
flavor node, which is then promoted to a gauge node. \\

The corresponding operation on the quiver $Y$ in each case produces a quiver $Y'$ which is the 3d mirror of $X'$ by construction. 
As discussed in \cite{Dey:2021rxw}, the quiver $Y'$ can be explicitly determined in both cases and has the following general form:

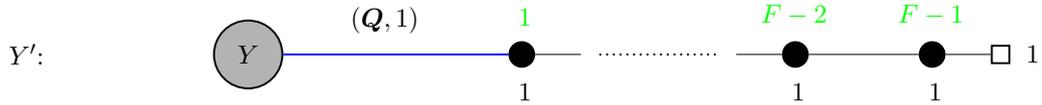
\begin{center}
\scalebox{0.9}{\begin{tikzpicture}[bnode/.style={circle,draw, thick, fill=black!30,minimum size=1cm}]
\node[text width=1 cm](34) at (-3,0) {$Y'$:};
\node[bnode] (1) at (0,0){$Y$} ;
\node[unode] (2) at (4,0){};
\node[] (3) at (5,0){};
\node[] (4) at (7,0){};
\node[unode] (5) at (8,0){};
\node[unode] (6) at (10,0){};
\node[fnode] (7) at (11,0){};
\draw[-] (2) -- (3);
\draw[thick, dotted] (3) -- (4);
\draw[-] (4) -- (5);
\draw[-] (5) -- (6);
\draw[-] (6) -- (7);
\draw[thick,blue] (1) -- (2);
\node[text width=0.1cm](31) [below=0.1 of 2] {1};
\node[text width=0.1 cm](32) [below=0.1 of 5] {1};
\node[text width=0.1cm](33) [below=0.1 of 6] {1};
\node[text width=0.1cm](34) [right=0.1 of 7] {1};
\node[text width=0.1cm, green](41) [above=0.1 of 2] {1};
\node[text width=1cm, green](42) [above=0.1 of 5] {$F-2$};
\node[text width=1cm, green](43) [above=0.1 of 6] {$F-1$};
\node[text width=1 cm](35) at (2,0.5) {$(\vec Q, 1)$};
\end{tikzpicture}} 
\end{center}

The quiver $Y'$ is constituted of two subquivers - the quiver $Y$ and a linear chain of $F-1$ $U(1)$ gauge nodes connected 
by bifundamental hypers and a single fundamental hyper at the far end. The subquivers are
connected by a single Abelian hypermultiplet which is charged under one of the $U(1)$ gauge nodes of the 
linear chain (as shown in the quiver diagram above) and various unitary gauge nodes of the quiver $Y$. The 
Abelian hypermultiplet has charge 1 under the $U(1)$ node and its charge under the unitary nodes of the 
quiver $Y$ is collectively denoted by $\vec Q$. The precise nodes and the charges depend on the 
quivers $(X,Y)$ and the details of the $S$-type operation itself. \\

We can now write down the general recipe for the construction of the 3d mirror for a given 
duality sequence generated from a theory in class $\CX$:

\begin{enumerate}

\item The first step is to identify in the N-al set a quiver which firstly does not have any 
balanced special unitary nodes and secondly gives a good quiver in the Gaiotto-Witten sense 
if one deletes the Abelian multiplets as well as the quiver tails introduced by the various mutations. 
Let us call the N-al theory $\CT'$ and the theory obtained after stripping off the Abelian multiplets and the quiver tails 
$\CT_{\rm good}$. 

\item The second step is to find the 3d mirror of $\CT_{\rm good}$. A systematic 
procedure for constructing the mirror theory is to start from a pair of mirror dual 
linear quivers with unitary gauge nodes $(X,Y)$ and implement a sequence of $S$-type 
operations \cite{Dey:2020hfe} to arrive at $\CT_{\rm good}$ and its 3d mirror.

\item If the 3d mirror of $\CT_{\rm good}$ has a weakly-coupled description $\wt{\CT}_{\rm good}$, 
then the final step is to perform a certain sequence of Abelian flavoring-gauging and/or identification-flavoring-gauging 
operations on $\wt{\CT}_{\rm good}$. On the mirror side, such an operation will generically amount to attaching quiver tails 
with Abelian hypermultiplets to $\CT_{\rm good}$, as we have discussed above. The appropriate sequence of Abelian $S$-type 
operations on $\wt{\CT}_{\rm good}$ is the one that produces the quiver $\CT'$ from $\CT_{\rm good}$ on the mirror side. The quiver 
gauge theory $\wt{\CT}'$ which is obtained from $\wt{\CT}_{\rm good}$ by the above sequence of Abelian 
operations is therefore the mirror dual of $\CT'$. Since $\CT'$ is IR dual to all the other theories 
in the N-al set, the theory $\wt{\CT}'$ is the 3d mirror associated with the duality sequence.

\end{enumerate}

A few comments are in order. Note that the first step specifies a non-trivial set of requirements. For example, given the duality
sequence of the 3-node quiver in the $e_1=e_2=0$ case in \figref{IRdual-Ex4c}, the only quiver that 
satisfies these requirements is the quiver $\CT^\vee_5$. In this case, the only choice for the quiver $\wt{\CT}' $ is 
$\wt{\CT}' = \CT^\vee_5$. The second step, which involves finding the 3d mirror of $\CT_{\rm good}$, is arguably 
the hardest step if $\CT_{\rm good}$ is a generic quiver gauge theory. In general, the 3d mirror may not admit a 
Lagrangian description. However, for the class of quivers 
discussed in this paper, $\CT_{\rm good}$ is always a linear quiver with unitary gauge nodes for which the 3d mirror 
$\wt{\CT}_{\rm good}$ is known. The final step involving the Abelian $S$-type operations can be performed in terms 
of the sphere partition function or the superconformal index and we refer the reader to the papers \cite{Dey:2020hfe,Dey:2021rxw} 
for the computational details.

\subsection{The two-node quiver sequence}\label{2node-mirr}

Let us now implement the recipe given in \Secref{gen-mirr} to find the 3d mirror for the duality sequence 
generated from the two-node quiver\footnote{The 3d mirrors of linear unitary-special unitary quivers have also been constructed using the technology of magnetic quivers. See \cite{Bourget:2021jwo} for details.}. We will focus on the case $e=0$ i.e. the quiver for which the 
unitary node is balanced. The $e\neq 0$ cases can be treated in an similar fashion.
The duality sequence for $e=0$ is given by \figref{IRdual-Ex6d}. To begin with, we observe that the only choice for the theory $\CT'$ is 
$\CT' = \CT^\vee_2$, since in this case the theory obtained after stripping off the Abelian hyper 
and the attached quiver tail is good in the Gaiotto-Witten sense. The theory $\CT_{\rm good}$ 
is then given by the linear quiver:

\begin{center}
\scalebox{0.65}{\begin{tikzpicture}
\node[fnode] (1) {};
\node[unode] (2) [right=2cm  of 1]{};
\node[unode] (3) [right=2 cm of 2]{};
\node[fnode] (4) [right=2 cm of 3]{};
\draw[-] (1) -- (2);
\draw[-] (2)-- (3);
\draw[-] (3) -- (4);
\node[text width=.1cm](10) [left=0.5 cm of 1]{$M_1$};
\node[text width= 1cm](11) [below=0.1cm of 2]{$N_1-1$};
\node[text width=1cm](12) [below=0.1cm of 3]{$N-1$};
\node[text width=.1cm](13) [right=0.1cm of 4]{$M_2$};
\end{tikzpicture}}
\end{center}

The labels in the quiver obey the constraints: $M_1 +N =2N_1$ and $N_1+M_2=2N-1$, 
which implies that the $U(N_1-1)$ gauge node is overbalanced while the $U(N-1)$ gauge node 
is balanced. The 3d mirror $\wt{\CT}_{\rm good}$ is also a linear quiver which can be found using the 
standard Hanany-Witten construction \cite{Hanany:1996ie} for generic labels $\{M_1, M_2,N_1,N\}$ subject to the 
constraints. One can then implement the third and final step of the construction by performing the 
Abelian $S$-type operations on the linear quiver $\wt{\CT}_{\rm good}$.\\

For the sake of simplifying the discussion of the final step, we will consider the quiver $\CT_{\rm good}$ 
with the following labels: $M_1=2$, $N_1=3$, $N=4$, and $M_2=4$. The theory $\wt{\CT}_{\rm good}$ 
then has the following form:

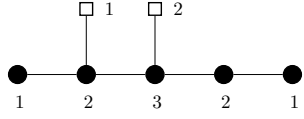
\begin{center}
\scalebox{0.65}{\begin{tikzpicture}
\node[unode] (1) {};
\node[unode] (2) [right=1cm  of 1]{};
\node[unode] (3) [right=1 cm of 2]{};
\node[unode] (4) [right=1 cm of 3]{};
\node[unode] (5) [right=1 cm of 4]{};
\node[fnode] (6) [above=1 cm of 2]{};
\node[fnode] (7) [above=1 cm of 3]{};
\draw[-] (1) -- (2);
\draw[-] (2)-- (3);
\draw[-] (3) -- (4);
\draw[-] (4) -- (5);
\draw[-] (2) -- (6);
\draw[-] (3) -- (7);
\node[text width=.1cm](10) [below=0.1 cm of 1]{$1$};
\node[text width= .1cm](11) [below=0.1cm of 2]{$2$};
\node[text width= .1cm](12) [below=0.1cm of 3]{$3$};
\node[text width=.1cm](13) [below=0.1cm of 4]{$2$};
\node[text width=.1cm](14) [below=0.1cm of 5]{$1$};
\node[text width=.1cm](15) [right=0.1cm of 6]{$1$};
\node[text width=.1cm](16) [right=0.1cm of 7]{$2$};
\end{tikzpicture}}
\end{center}

One can then implement a flavoring-gauging operation $\CO$ on the quiver $\wt{\CT}_{\rm good}$ 
and its 3d mirror as follows:

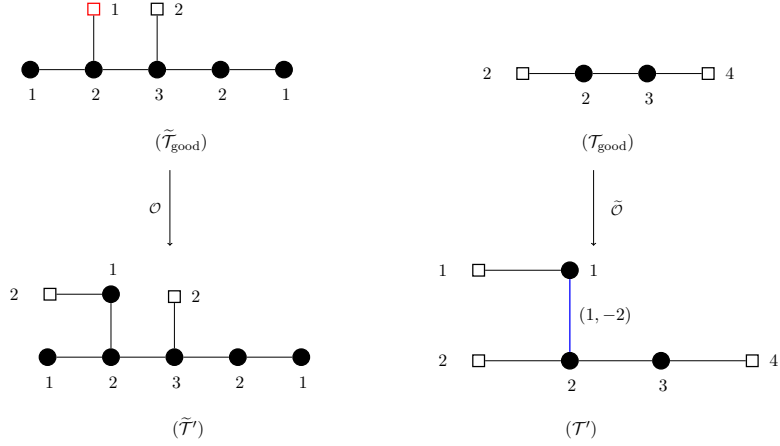
\begin{center}
\begin{tabular}{ccc}
 \scalebox{0.6}{\begin{tikzpicture}
\node[unode] (1) {};
\node[unode] (2) [right=1cm  of 1]{};
\node[unode] (3) [right=1 cm of 2]{};
\node[unode] (4) [right=1 cm of 3]{};
\node[unode] (5) [right=1 cm of 4]{};
\node[fnode, red] (6) [above=1 cm of 2]{};
\node[fnode] (7) [above=1 cm of 3]{};
\draw[-] (1) -- (2);
\draw[-] (2)-- (3);
\draw[-] (3) -- (4);
\draw[-] (4) -- (5);
\draw[-] (2) -- (6);
\draw[-] (3) -- (7);
\node[text width=.1cm](10) [below=0.1 cm of 1]{$1$};
\node[text width= .1cm](11) [below=0.1cm of 2]{$2$};
\node[text width= .1cm](12) [below=0.1cm of 3]{$3$};
\node[text width=.1cm](13) [below=0.1cm of 4]{$2$};
\node[text width=.1cm](14) [below=0.1cm of 5]{$1$};
\node[text width=.1cm](15) [right=0.1cm of 6]{$1$};
\node[text width=.1cm](16) [right=0.1cm of 7]{$2$};
\node[text width=0.1cm](30)[below=1 cm of 3]{$(\wt{\CT}_{\rm good})$};
\end{tikzpicture}}
& \qquad  
&\scalebox{.6}{\begin{tikzpicture}
\node[fnode] (1) {};
\node[unode] (2) [right=1cm  of 1]{};
\node[unode] (3) [right=1 cm of 2]{};
\node[fnode] (4) [right=1 cm of 3]{};
\draw[-] (1) -- (2);
\draw[-] (2)-- (3);
\draw[-] (3) -- (4);
\node[text width=.1cm](10) [left=0.5 cm of 1]{$2$};
\node[text width= .1cm](11) [below=0.1cm of 2]{$2$};
\node[text width=.1cm](12) [below=0.1cm of 3]{$3$};
\node[text width=.1cm](13) [right=0.1cm of 4]{$4$};
\node[text width=0.1cm](30)[below=1 cm of 2]{$({\CT}_{\rm good})$};
\end{tikzpicture}}\\
 \scalebox{.5}{\begin{tikzpicture}
\draw[->] (15,-3) -- (15,-5);
\node[text width=0.1cm](20) at (14.5, -4) {$\CO$};
\end{tikzpicture}}
&\qquad \qquad 
& \scalebox{.5}{\begin{tikzpicture}
\draw[->] (15,-3) -- (15,-5);
\node[text width=0.1cm](29) at (15.5, -4) {$\wt{\CO}$};
\end{tikzpicture}}\\
\scalebox{0.6}{\begin{tikzpicture}
\node[unode] (1) {};
\node[unode] (2) [right=1cm  of 1]{};
\node[unode] (3) [right=1 cm of 2]{};
\node[unode] (4) [right=1 cm of 3]{};
\node[unode] (5) [right=1 cm of 4]{};
\node[unode] (6) [above=1 cm of 2]{};
\node[fnode] (7) [above=1 cm of 3]{};
\node[fnode] (8) [left= 1 cm of 6]{};
\draw[-] (1) -- (2);
\draw[-] (2)-- (3);
\draw[-] (3) -- (4);
\draw[-] (4) -- (5);
\draw[-] (2) -- (6);
\draw[-] (3) -- (7);
\draw[-] (6) -- (8);
\node[text width=.1cm](10) [below=0.1 cm of 1]{$1$};
\node[text width= .1cm](11) [below=0.1cm of 2]{$2$};
\node[text width= .1cm](12) [below=0.1cm of 3]{$3$};
\node[text width=.1cm](13) [below=0.1cm of 4]{$2$};
\node[text width=.1cm](14) [below=0.1cm of 5]{$1$};
\node[text width=.1cm](15) [above=0.1cm of 6]{$1$};
\node[text width=.1cm](16) [right=0.1cm of 7]{$2$};
\node[text width=.1cm](17) [left=0.5cm of 8]{$2$};
\node[text width=0.1cm](20)[below=1 cm of 3]{$(\wt{\CT}')$};
\end{tikzpicture}}
&\qquad \qquad 
& \scalebox{.6}{\begin{tikzpicture}
\node[fnode] (1) at (0,0){};
\node[unode] (2) at (2,0){};
\node[unode] (3) at (4,0){};
\node[fnode] (4) at (6,0){};
\node[unode] (6) at (2,2){};
\node[fnode] (7) at (0,2){};
\draw[-] (1) -- (2);
\draw[-] (2)-- (3);
\draw[-] (3) -- (4);
\draw[-] (6) --(7);
\draw[-, thick, blue] (2)--(6);
\node[text width=2cm](10) at (3.2, 1){$(1,-2)$};
\node[text width=.1cm](20) [left=0.5 cm of 1]{$2$};
\node[text width=0.1 cm](21) [below=0.1cm of 2]{$2$};
\node[text width=0.1 cm](22) [below=0.1cm of 3]{$3$};
\node[text width=.1cm](23) [right=0.1cm of 4]{$4$};
\node[text width=.1cm](24) [right=0.1cm of 6]{1};
\node[text width=.1cm](25) [left=0.5 cm of 7]{1};
\node[text width=.2cm](40) [below=1 cm of 2]{$({\CT}')$};
\end{tikzpicture}}
\end{tabular}
\end{center}

The operation acts on the flavor node labelled 1 (marked in red) in the quiver $\wt{\CT}_{\rm good}$. 
On the mirror side, this operation involves attaching  to the $U(2)$ gauge node of ${\CT}_{\rm good}$ 
an Abelian hypermultiplet which in turn is connected to a quiver tail of a single $U(1)$ gauge node 
plus a fundamental hypermultiplet. Note that this is consistent with the form of the quiver $Y'$ in 
\Secref{gen-mirr} for $F=2$. Since the operation correctly reproduces the theory $\CT' = \CT^\vee_2$, 
the quiver $\wt{\CT}'$ can be identified as the 3d mirror associated with the duality sequence \figref{IRdual-Ex6d} for the chosen labels. 

One can check that the basic data of the mirror symmetry in the following fashion. Firstly, one can check that the moduli space 
quaternionic dimensions agree:
\be
{\rm dim}\,\CM^{(\wt{\CT}')}_{\rm C} = {\rm dim}\,\CM^{({\CT}')}_{\rm H} =10, \qquad 
{\rm dim}\,\CM^{(\wt{\CT}')}_{\rm H} = {\rm dim}\,\CM^{({\CT}')}_{\rm C} =6.
\ee
The theory $\wt{\CT}'$ has a $\frsu(2)\oplus \frsu(4) \oplus \fru(1) \oplus \fru(1)$ Coulomb branch global symmetry algebra 
which is realized as the Higgs branch global symmetry algebra for the theory $\CT'$. The Higgs branch symmetry of 
$\wt{\CT}'$ is $\frsu(2)\oplus \frsu(2) \oplus \fru(1)$ which is realized as the Coulomb branch symmetry of the theory 
${\CT}'$. Therefore the emergent Coulomb symmetry of the unitary-special unitary linear quiver $\CT$ in \figref{IRdual-Ex6d}
is realized as the Higgs branch symmetry of the 3d mirror $\wt{\CT}'$, where it is manifest in the UV Lagrangian. 

\subsection{The three-node quiver sequence}\label{3node-mirr}

Let us now construct the 3d mirror for the duality sequence 
generated from the three-node quiver. For the sake of concreteness, we will focus on the case $e_1=e_2=0$ i.e. 
the quiver for which both unitary nodes are balanced. The duality sequence which gives a hexality of quiver gauge 
theories is shown by \figref{IRdual-Ex4c}. 
There is a single choice for the theory $\CT'$ i.e.
$\CT' = \CT^\vee_5$, since only in this case the theory obtained after stripping off the Abelian hyper 
and the attached quiver tail is good in the Gaiotto-Witten sense. The theory $\CT_{\rm good}$ 
is then given by the linear quiver:

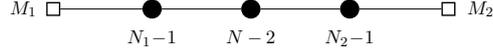
\begin{center}
\scalebox{0.65}{\begin{tikzpicture}
\node[fnode] (1) at (0,0){};
\node[unode] (2) at (2,0){};
\node[unode] (3) at (4,0){};
\node[unode] (4) at (6,0){};
\node[fnode] (5) at (8,0){};
\draw[-] (1) -- (2);
\draw[-] (2)-- (3);
\draw[-] (3) -- (4);
\draw[-] (4) --(5);
\node[text width=.1cm](10) [left=0.5 cm of 1]{$M_1$};
\node[text width=1cm](11) [below=0.1cm of 2]{$N_1-1$};
\node[text width=1cm](12) [below=0.1cm of 3]{$N-2$};
\node[text width=1cm](13) [below=0.1cm of 4]{$N_2-1$};
\node[text width=.1cm](14) [right=0.1cm of 5]{$M_2$};
\end{tikzpicture}}
\end{center}

The labels in the quiver obey the constraints: $M_1 +N =2N_1$, $N_1+N_2=2N-1$, and $N+M_2=2N_2$, 
which implies that the $U(N_1-1)$ and $U(N_2-1)$ gauge nodes are balanced while the $U(N-2)$ gauge node 
is overbalanced. The 3d mirror $\wt{\CT}_{\rm good}$ is also a linear quiver which can be found using the 
standard Hanany-Witten construction \cite{Hanany:1996ie} for generic labels $\{M_1, M_2,N_1,N_2,N\}$ subject to the 
constraints. One can then implement the third and final step of the construction by performing the 
Abelian $S$-type operations on the linear quiver $\wt{\CT}_{\rm good}$.\\

For the sake of simplifying the discussion of the final step, we will consider the quiver $\CT_{\rm good}$ 
with the following labels: $M_1=2$, $N_1=3$, $N=4$, $N_2=4$, and $M_2=4$. The theory $\wt{\CT}_{\rm good}$ 
then has the following form:

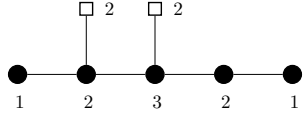
\begin{center}
\scalebox{0.65}{\begin{tikzpicture}
\node[unode] (1) {};
\node[unode] (2) [right=1cm  of 1]{};
\node[unode] (3) [right=1 cm of 2]{};
\node[unode] (4) [right=1 cm of 3]{};
\node[unode] (5) [right=1 cm of 4]{};
\node[fnode] (6) [above=1 cm of 2]{};
\node[fnode] (7) [above=1 cm of 3]{};
\draw[-] (1) -- (2);
\draw[-] (2)-- (3);
\draw[-] (3) -- (4);
\draw[-] (4) -- (5);
\draw[-] (2) -- (6);
\draw[-] (3) -- (7);
\node[text width=.1cm](10) [below=0.1 cm of 1]{$1$};
\node[text width= .1cm](11) [below=0.1cm of 2]{$2$};
\node[text width= .1cm](12) [below=0.1cm of 3]{$3$};
\node[text width=.1cm](13) [below=0.1cm of 4]{$2$};
\node[text width=.1cm](14) [below=0.1cm of 5]{$1$};
\node[text width=.1cm](15) [right=0.1cm of 6]{$2$};
\node[text width=.1cm](16) [right=0.1cm of 7]{$2$};
\end{tikzpicture}}
\end{center}

Finally, one has to implement an identification-flavoring-gauging operation $\CO$ on the quiver $\wt{\CT}_{\rm good}$ 
and its 3d mirror as follows:

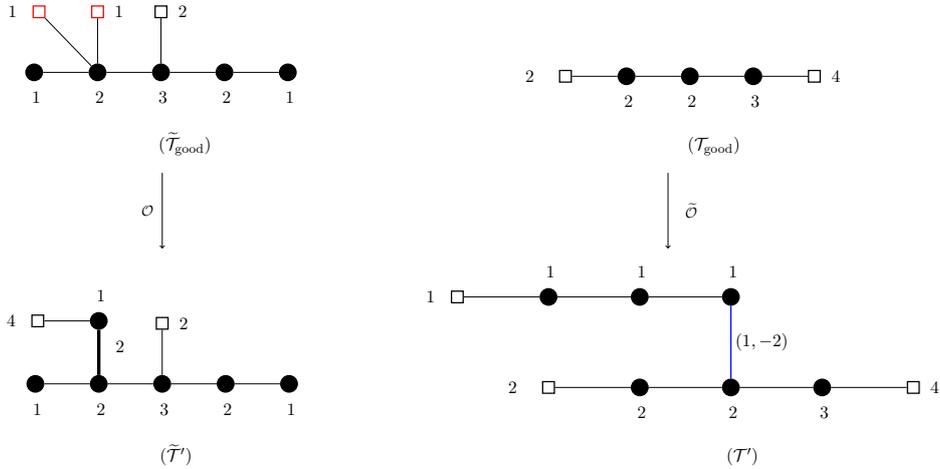
\begin{center}
\begin{tabular}{ccc}
 \scalebox{0.6}{\begin{tikzpicture}
\node[unode] (1) {};
\node[unode] (2) [right=1cm  of 1]{};
\node[unode] (3) [right=1 cm of 2]{};
\node[unode] (4) [right=1 cm of 3]{};
\node[unode] (5) [right=1 cm of 4]{};
\node[fnode, red] (6) [above=1 cm of 2]{};
\node[fnode, red] (8) [left=1 cm of 6]{};
\node[fnode] (7) [above=1 cm of 3]{};
\draw[-] (1) -- (2);
\draw[-] (2)-- (3);
\draw[-] (3) -- (4);
\draw[-] (4) -- (5);
\draw[-] (2) -- (6);
\draw[-] (2) -- (8);
\draw[-] (3) -- (7);
\node[text width=.1cm](10) [below=0.1 cm of 1]{$1$};
\node[text width= .1cm](11) [below=0.1cm of 2]{$2$};
\node[text width= .1cm](12) [below=0.1cm of 3]{$3$};
\node[text width=.1cm](13) [below=0.1cm of 4]{$2$};
\node[text width=.1cm](14) [below=0.1cm of 5]{$1$};
\node[text width=.1cm](15) [right=0.1cm of 6]{$1$};
\node[text width=.1cm](16) [right=0.1cm of 7]{$2$};
\node[text width=.1cm](17) [left=0.3 cm of 8]{$1$};
\node[text width=0.1cm](30)[below=1 cm of 3]{$(\wt{\CT}_{\rm good})$};
\end{tikzpicture}}
& \qquad  
&\scalebox{.6}{\begin{tikzpicture}
\node[fnode] (1) {};
\node[unode] (2) [right=1cm  of 1]{};
\node[unode] (3) [right=1 cm of 2]{};
\node[unode] (4) [right=1 cm of 3]{};
\node[fnode] (5) [right=1 cm of 4]{};
\draw[-] (1) -- (2);
\draw[-] (2)-- (3);
\draw[-] (3) -- (4);
\draw[-] (4) -- (5);
\node[text width=.1cm](10) [left=0.5 cm of 1]{$2$};
\node[text width= 0.1cm](11) [below=0.1cm of 2]{$2$};
\node[text width=0.1cm](12) [below=0.1cm of 3]{$2$};
\node[text width=.1cm](13) [below=0.1cm of 4]{$3$};
\node[text width=.1cm](14) [right=0.1cm of 5]{$4$};
\node[text width=0.1cm](30)[below=1 cm of 3]{$({\CT}_{\rm good})$};
\end{tikzpicture}}\\
 \scalebox{.5}{\begin{tikzpicture}
\draw[->] (15,-3) -- (15,-5);
\node[text width=0.1cm](20) at (14.5, -4) {$\CO$};
\end{tikzpicture}}
&\qquad \qquad 
& \scalebox{.5}{\begin{tikzpicture}
\draw[->] (15,-3) -- (15,-5);
\node[text width=0.1cm](29) at (15.5, -4) {$\wt{\CO}$};
\end{tikzpicture}}\\
\scalebox{0.6}{\begin{tikzpicture}
\node[unode] (1) {};
\node[unode] (2) [right=1cm  of 1]{};
\node[unode] (3) [right=1 cm of 2]{};
\node[unode] (4) [right=1 cm of 3]{};
\node[unode] (5) [right=1 cm of 4]{};
\node[unode] (6) [above=1 cm of 2]{};
\node[fnode] (7) [above=1 cm of 3]{};
\node[fnode] (8) [left= 1 cm of 6]{};
\node[](9) [above=0.5 cm of 2]{};
\draw[-] (1) -- (2);
\draw[-] (2)-- (3);
\draw[-] (3) -- (4);
\draw[-] (4) -- (5);
\draw[line width=0.75mm] (2) -- (6);
\draw[-] (3) -- (7);
\draw[-] (6) -- (8);
\node[text width=.1cm](10) [below=0.1 cm of 1]{$1$};
\node[text width= .1cm](11) [below=0.1cm of 2]{$2$};
\node[text width= .1cm](12) [below=0.1cm of 3]{$3$};
\node[text width=.1cm](13) [below=0.1cm of 4]{$2$};
\node[text width=.1cm](14) [below=0.1cm of 5]{$1$};
\node[text width=.1cm](15) [above=0.1cm of 6]{$1$};
\node[text width=.1cm](16) [right=0.1cm of 7]{$2$};
\node[text width=.1cm](17) [left=0.3cm of 8]{$4$};
\node[text width=.1cm](18) [right=0.1cm of 9]{$2$};
\node[text width=0.1cm](20)[below=1 cm of 3]{$(\wt{\CT}')$};
\end{tikzpicture}}
&\qquad \qquad 
& \scalebox{.6}{\begin{tikzpicture}
\node[fnode] (1) at (0,0){};
\node[unode] (2) at (2,0){};
\node[unode] (3) at (4,0){};
\node[unode] (4) at (6,0){};
\node[fnode] (5) at (8,0){};
\node[unode] (6) at (4,2){};
\node[unode] (7) at (2,2){};
\node[unode] (8) at (0,2){};
\node[fnode] (9) at (-2,2){};
\draw[-] (1) -- (2);
\draw[-] (2)-- (3);
\draw[-] (3) -- (4);
\draw[-] (4) --(5);
\draw[-] (6) --(7);
\draw[-] (8) --(7);
\draw[-] (8) --(9);
\draw[thick, blue] (6)-- (3);
\node[text width=.1cm](40) [left=0.5 cm of 1]{$2$};
\node[text width=.1 cm](41) [below=0.1cm of 2]{$2$};
\node[text width=.1 cm](42) [below=0.1cm of 3]{$2$};
\node[text width=.1 cm](43) [below=0.1cm of 4]{$3$};
\node[text width=.1cm](44) [right=0.1cm of 5]{$4$};
\node[text width=.1cm](45) [above=0.1cm of 6]{1};
\node[text width=.1cm](46) [above=0.1 cm of 7]{1};
\node[text width=.1cm](47) [above=0.1 cm of 8]{1};
\node[text width=.1cm](48) [left=0.3 cm of 9]{1};
\node[text width=2cm](21) at (5.1, 1){$(1,-2)$};
\node[text width=.2cm](40) [below=1 cm of 3]{$({\CT}')$};
\end{tikzpicture}}
\end{tabular}
\end{center}

The operation involves splitting the flavor node labelled 2 (attached to the $U(2)$ gauge node of  $\wt{\CT}_{\rm good}$) 
into two flavor nodes labelled 1 (marked in red), and identifying them into a single flavor node. This is followed by 
attaching 4 hypermultiplets of charge 1 to the identified flavor node, which is then promoted to a $U(1)$ gauge node. 
On the mirror side, the operation amounts to attaching to the central $U(2)$ gauge node of ${\CT}_{\rm good}$ 
an Abelian hypermultiplet which in turn is connected to a quiver tail of three $U(1)$ gauge nodes 
plus a fundamental hypermultiplet at the far end. Note that this is consistent with the form of the quiver $Y'$ in 
\Secref{gen-mirr} for $F=4$. Since the operation correctly reproduces the theory $\CT' = \CT^\vee_5$, 
the quiver $\wt{\CT}'$ can be identified as the 3d mirror associated with the duality sequence \figref{IRdual-Ex4c} 
for the chosen labels. 

One can check that the basic data of the mirror symmetry in the following fashion. Firstly, one can check that the moduli space 
quaternionic dimensions agree:
\be
{\rm dim}\,\CM^{(\wt{\CT}')}_{\rm C} = {\rm dim}\,\CM^{({\CT}')}_{\rm H} =10, \qquad 
{\rm dim}\,\CM^{(\wt{\CT}')}_{\rm H} = {\rm dim}\,\CM^{({\CT}')}_{\rm C} =10.
\ee
The theory $\wt{\CT}'$ has a $\frsu(2)\oplus \frsu(4) \oplus \fru(1) \oplus \fru(1)$ Coulomb branch global symmetry algebra 
which is realized as the Higgs branch global symmetry algebra for the theory $\CT'$. The Higgs branch symmetry of 
$\wt{\CT}'$ is $\frsu(4)\oplus\frsu(2)\oplus \frsu(2) \oplus \fru(1)$ which is realized as the Coulomb branch symmetry of the theory 
${\CT}'$. Therefore the emergent Coulomb symmetry of the unitary-special unitary linear quiver $\CT$ in \figref{IRdual-Ex6d}
is realized as the Higgs branch symmetry of the 3d mirror $\wt{\CT}'$, where it is manifest in the UV Lagrangian.

\section*{Acknowledgements}
The author would like to thank Ibrahima Bah, Stefano Cremonesi, Amihay Hanany, Zohar Komargodski, Andrew Neitzke and 
Jaewon Song for discussion at different stages of the project. The author was supported in part at the Johns Hopkins 
University by the NSF grant PHY-2112699, and also by the Simons Collaboration on Global Categorical Symmetries, while 
this work was in preparation.

\appendix

\section{Sphere partition function computation: 2 node quiver} \label{2node-pf}

In this section, we summarize the details of the sphere partition function computation for the duality sequences that 
arise from the 2-node quiver.

\subsection{Overbalanced unitary node: $e=1$ and $e=2$}\label{2node-pf-ob}

For the case $e=1$, the first step involves implementing mutation I at the $SU(N)$ node of $\CT$ to 
give the quiver $\CT^\vee_1$ -- the partition function realization of this was discussed in \Secref{Unbal-2node}. 
The next step involves implementing mutation II at the $U(N_1)$ gauge node of $\CT^\vee_1$, which we will 
now discuss. The starting point is the matrix integral for the partition function of the theory $\CT^\vee_1$ given in \eref{PF-CT-01a}:
\begin{align}
Z^{(\CT^\vee_1)}(\vec m^1, \vec m^2, m_{\rm ab}=\tr \vec m^2; \eta,0) = & \int\, \Big[d\vec s^{1}\Big]\, \Big[d  \vec \s^2 \Big] \,Z_{\rm FI}(\vec s^1, \eta) \, Z^{\rm vec}_{\rm 1-loop}(\vec s^1)\,Z^{\rm vec}_{\rm 1-loop}(\vec \s^2)\,Z^{\rm fund}_{\rm 1-loop}(\vec s^1, \vec m^1) \nn \\
 \times & \,Z^{\rm bif}_{\rm 1-loop}(\vec s^1, \vec \s^2, 0)\, Z^{\rm fund}_{\rm 1-loop}(\vec \s^2, \vec m^2)\, Z^{\rm hyper}_{\rm 1-loop}(\vec s^1, \vec \s^2, \vec m^2).
\end{align}

Let us shift the $U(N_1)$ integration variables $\vec s^1$ as $s^1_i \to s^1_i + \delta, \, (i=1,\ldots,N_1)$, where $\delta =\frac{1}{N_1}(\tr \vec m^1 + \tr \vec m^2)$. This transformation allows one to rewrite the partition function in a form where the identity \eref{Id-2b} can be used, i.e.
\begin{align}
Z^{(\CT^\vee_1)}(\vec m, \vec \eta) = e^{2\pi i \eta N_1 \delta}\, & \int\, \Big[d\vec s^{1}\Big]\, \Big[d  \vec \s^2 \Big]\, \Big[\ldots \Big] \,Z_{\rm FI}(\vec s^1, \eta) \, Z^{\rm vec}_{\rm 1-loop}(\vec s^1)\, Z^{\rm fund}_{\rm 1-loop}(\vec s^1, \vec m^1-\delta) \nn \\
& \times Z^{\rm bif}_{\rm 1-loop}(\vec s^1, \vec \s^2-\delta,0)\cdot \frac{1}{\ch{(\tr \vec s^1 - \tr \vec \s^2 - \tr \vec m^1 + 2N_1\delta)}} \nn \\
= e^{2\pi i \eta N_1 \delta}\, & \int\, \Big[d\vec \s^{1}\Big]\, \Big[d  \vec \s^2 \Big]\, \Big[\ldots \Big] \, Z_{\rm FI}(\vec \s^1, -\eta)\, 
Z^{\rm vec}_{\rm 1-loop}(\vec \s^1)\, Z^{\rm fund}_{\rm 1-loop}(\vec \s^1, \vec m^1-\frac{\tr \vec \s^2}{N_1} + \xi) \nn \\
& \times Z^{\rm bif}_{\rm 1-loop}(\vec \s^1, \vec \s^2- \frac{\tr \vec \s^2}{N_1} + \xi,0)\,\cdot \frac{1}{\ch{(\tr \vec \s^1 + \tr \vec \s^2 - \tr \vec m^1-2\tr \vec m^2)}},
\end{align}
where $\vec \s^1$ denotes the integration variables for a $U(N_1)$ gauge group and the parameter $\xi =\frac{1}{N_1}\tr \vec m^2$. 
Finally, we implement a change of variables $\s^1_i \to \s^1_i - \frac{\tr \vec \s^2}{N_1} + \delta, \, (i=1,\ldots,N_1)$, we arrive at the following form 
of the partition function:
\begin{align}
Z^{(\CT^\vee_1)}(\vec m, \vec \eta) = & \int\, \Big[d\vec \s^{1}\Big]\, \Big[d  \vec \s^2 \Big]\, \Big[\ldots \Big] \, Z_{\rm FI}(\vec \s^1, -\eta)\, 
Z_{\rm FI}(\vec \s^2, \eta)\,
Z^{\rm vec}_{\rm 1-loop}(\vec \s^1)\, Z^{\rm fund}_{\rm 1-loop}(\vec \s^1, \vec m^1- \frac{\tr \vec m^1}{N_1}) \nn \\
& \times Z^{\rm bif}_{\rm 1-loop}(\vec \s^1, \vec \s^2, \frac{\tr \vec m^1}{N_1})\,\cdot \frac{1}{\ch{(\tr \vec \s^1- \tr \vec m^2)}}.
\end{align}
The matrix model integral on the RHS can be readily identified with the partition function of the theory $\CT^\vee_2$ in \figref{IRdual-Ex6b}. 
Therefore, we have the partition function relation:
\be
\boxed{Z^{\CT}(\vec m; 0)= Z^{(\CT^\vee_1)}(\vec m, m_{\rm{Ab}}=\tr \vec m^2; \eta, 0) = Z^{(\CT^\vee_2)}(\vec m', m'_{\rm{Ab}}; -\eta, \eta),}
\ee
where the duality map for the masses are given as:
\begin{align}
& \vec m'^1 = \vec m^1 - \frac{\tr \vec m^1}{N_1}, \qquad \vec m'^2 = \vec m^2, \\
& m'_{\rm bif}= \frac{\tr \vec m^1}{N_1}, \qquad m'_{\rm{Ab}} = \tr \vec m^2.
\end{align}

For the case $e=2$, we again start from the partition function of $\CT^\vee_1$ given above. Let us first set the 
FI parameter of the $U(N)$ gauge node to zero, and then shift the integration variables $\vec s^1$ as 
$s^1_i \to s^1_i + \delta, \, (i=1,\ldots,N_1)$, where $\delta =\frac{1}{N_1+1}(\tr \vec m^1 + \tr \vec m^2)$. 
This transformation allows one to rewrite the partition function in a form where the identity \eref{Id-1} can now be used:
\begin{align}
Z^{(\CT^\vee_1)}(\vec m, \vec \eta) = & \int\, \Big[d\vec s^{1}\Big]\, \Big[d  \vec \s^2 \Big]\, \Big[\ldots \Big] \, \frac{Z^{\rm vec}_{\rm 1-loop}(\vec s^1)\, Z^{\rm fund}_{\rm 1-loop}(\vec s^1, \vec m^1-\delta)\,Z^{\rm bif}_{\rm 1-loop}(\vec s^1, \vec \s^2-\delta,0)}{\ch{(\tr \vec s^1 - \tr \vec \s^2 - \tr \vec m^1 + (2N_1+1)\delta)}} \nn \\
= & \int\, \Big[d\vec \s^{1}\Big]\, \Big[d  \vec \s^2 \Big]\, \Big[\ldots \Big] \, \delta( \tr \vec \s^1)\, 
Z^{\rm vec}_{\rm 1-loop}(\vec \s^1)\, Z^{\rm fund}_{\rm 1-loop}(\vec \s^1, \vec m^1- \delta)\,Z^{\rm bif}_{\rm 1-loop}(\vec \s^1, \vec \s^2, \delta),
\end{align}
where $\vec \s^1$ denotes the integration variables for a $U(N_1)$ gauge group and $\delta$ is defined above. 
The matrix model integral on the RHS can be identified as the partition function of the theory $\CT^\vee_2$ in \figref{IRdual-Ex6c}, 
and we have the partition function relation:
\be
\boxed{Z^{\CT}(\vec m; 0)=  Z^{\CT^\vee_1}(\vec m, m_{\rm{Ab}}=\tr \vec m^2; 0, 0) = Z^{\CT^\vee_2}(\vec m'; 0),}
\ee
where the duality map for the masses $\vec m'$ are given as:
\begin{align}
\vec m'^1= \vec m^1 - \delta, \qquad \vec m'^2= \vec m^2, \qquad m'_{\rm bif}=\delta.
\end{align}

\subsection{Balanced unitary node: $e=0$}\label{2node-pf-b}

The triality sequence in \figref{IRdual-Ex6d} can be realized in terms of the sphere partition function as follows. The first step involving mutation I is exactly the same as treated above in the $e \neq 0$ cases. Starting from the partition function of $\CT^\vee_1$, one can implement mutation III using the identity \eref{Id-3}:
\begin{align}
Z^{(\CT^\vee_1)}(\vec m, \vec \eta) =  & \int\, \Big[d\vec s^{1}\Big]\, \Big[d  \vec \s^2 \Big]\, \Big[\ldots \Big] \,\frac{Z_{\rm FI}(\vec s^1, \eta) \, Z^{\rm vec}_{\rm 1-loop}(\vec s^1)\, Z^{\rm fund}_{\rm 1-loop}(\vec s^1, \vec m^1)\,Z^{\rm bif}_{\rm 1-loop}(\vec s^1, \vec \s^2,0)}{\ch{(\tr \vec s^1 - \tr \vec \s^2 + \tr \vec m^2)}} \nn \\
= & \int\, \Big[d\vec \s^{1}\Big]\, \Big[d  \vec \s^2 \Big]\, du\,\Big[\ldots \Big] \, Z_{\rm FI}(u, \eta)\, Z_{\rm FI}(\vec \s^1, -\eta)\, 
Z^{\rm vec}_{\rm 1-loop}(\vec \s^1)\, Z^{\rm fund}_{\rm 1-loop}(\vec \s^1, \vec m^1) \nn \\
& \times Z^{\rm bif}_{\rm 1-loop}(\vec \s^1, \vec \s^2,0)\,\cdot \frac{1}{\ch{(u -\tr \vec m^1-\tr \vec \s^2)}\,\ch{(u- \tr \vec \s^1 - \tr \vec \s^2 +\tr \vec m^2)}},
\end{align}
where $u$ and $\vec \s^1$ denote the matrix integral variables for the $U(1)$ and the $U(N_1-1)$ gauge groups respectively. Finally, 
implementing the change of variables $u \to u + \tr \vec \s^2$, we obtain:
\begin{align}
Z^{(\CT^\vee_1)}(\vec m, \vec \eta) = & \int\, \Big[d\vec \s^{1}\Big]\, \Big[d  \vec \s^2 \Big]\, du\,\Big[\ldots \Big] \, Z_{\rm FI}(u, \eta)\, Z_{\rm FI}(\vec \s^1, -\eta)\, 
Z_{\rm FI}(\vec \s^2, \eta)\, Z^{\rm vec}_{\rm 1-loop}(\vec \s^1)\, Z^{\rm fund}_{\rm 1-loop}(\vec \s^1, \vec m^1) \nn \\
& \times Z^{\rm bif}_{\rm 1-loop}(\vec \s^1, \vec \s^2,0)\,\cdot \frac{1}{\ch{(u -\tr \vec m^1)}\,\ch{(u- \tr \vec \s^1 +\tr \vec m^2)}},
\end{align}
where the RHS can readily be identified as the sphere partition function of the theory $\CT^\vee_2$ in \figref{IRdual-Ex6d}. 
We therefore have the following partition function relation:
\be
\boxed{Z^{\CT}(\vec m; 0)= Z^{(\CT^\vee_1)}(\vec m, m_{\rm{Ab}}=\tr \vec m^2; \eta, 0) = Z^{(\CT^\vee_2)}(\vec m', m'_{\rm{Ab}}; -\eta, \eta, \eta).}
\ee
The duality map for the masses are given as:
\begin{align}
& \vec m'^1 = \vec m^1, \qquad \vec m'^2 = \vec m^2, \qquad m'^3 = \tr \vec m^1, \qquad m'_{\rm{Ab}}= \tr \vec m^2,
\end{align}
where $m'^3$ denotes the mass for the fundamental hyper of the $U(1)$ gauge node in \figref{IRdual-Ex6d}.

\section{Sphere partition function computation: 3 node quiver} \label{3node-pf}

In this section, we summarize the details of the sphere partition function computation for the duality sequences that 
arise from the 3-node quiver.

\subsection{Overbalanced unitary node: $e_1,e_2 > 2$}\label{3node-pf-ob}

The duality can be realized at the level of the sphere partition function as follows. The partition function of the quiver 
$\CT$ in \figref{IRdual-Ex4a} is given as:

\begin{align}\label{PF-CT-3A}
Z^{(\CT)}(\vec m, \vec \eta) = & \int\, \prod^{3}_{\gamma=1}\,\Big[d\vec s^{\gamma}\Big] \, \prod^{3}_{\gamma=1}\,Z_{\rm FI}(\vec\s^\gamma, \eta_\gamma)|_{\gamma \neq 2}\,\delta(\tr \vec s^2)\, \prod^3_{\gamma=1}\, Z^{\rm vec}_{\rm 1-loop}(\vec s^\gamma) \nn \\
& \times \, Z^{\rm fund}_{\rm 1-loop}(\vec s^1, \vec m^1)\, \prod^2_{\gamma=1}\,Z^{\rm bif}_{\rm 1-loop}(\vec s^\gamma, \vec s^{\gamma+1}, 0)\,
Z^{\rm fund}_{\rm 1-loop}(\vec s^3, \vec m^2), 
\end{align}
where $\{\vec s^\gamma\}_{\gamma \neq 2}$ denote the integration variables associated with the unitary gauge groups, while $\vec s^2$ denotes 
the integration variable associated with the $SU(N)$ gauge group. Isolating the $\vec s^2$-dependent part of the matrix integral, and using the 
relation \eref{Id-1}, we have the following identity:
\begin{align}
& \int \, \Big[d \vec s^2 \Big] \, \delta(\tr \vec s^2)\, Z^{\rm vec}_{\rm 1-loop}(\vec s^2)\, Z^{\rm bif}_{\rm 1-loop}(\vec s^2, \vec s^1, 0)\, Z^{\rm bif}_{\rm 1-loop}(\vec s^2, \vec s^3,0) \nn \\
&=   \int \, \Big[d  \vec \s^2 \Big] \, \frac{1}{\ch{(\tr \vec \s^2 - \tr \vec s^1 - \tr \vec s^3)}}\,Z^{\rm vec}_{\rm 1-loop}(\vec \s^2)\, Z^{\rm bif}_{\rm 1-loop}(\vec \s^2, \vec s^1, 0)\,Z^{\rm bif}_{\rm 1-loop}(\vec \s^2, \vec s^3, 0),
\end{align}
where $\vec \s^2$ are the integration variables in the Cartan subalgebra of a $U(N-1)$ group. 
Substituting the above identity in \eref{PF-CT-3A}, the partition function for $\CT$ can be written as
\begin{align}\label{PF-CT-4A}
Z^{(\CT)}(\vec m, \vec \eta) = & \int\, \prod_{\gamma=1,3}\,\Big[d\vec s^{\gamma}\Big]\, \Big[d  \vec \s^2 \Big] \,\prod_{\gamma=1,3}\,Z_{\rm FI}(\vec s^\gamma, \eta_\gamma)|_{\gamma \neq 2} \, \prod_{\gamma=1,3}\, Z^{\rm vec}_{\rm 1-loop}(\vec s^\gamma)\,Z^{\rm vec}_{\rm 1-loop}(\vec \s^2) \nn \\
 \times & \,Z^{\rm fund}_{\rm 1-loop}(\vec s^1, \vec m^1)\,Z^{\rm bif}_{\rm 1-loop}(\vec s^1, \vec \s^2, 0)\,Z^{\rm bif}_{\rm 1-loop}(\vec \s^2, \vec s^3,0) \,Z^{\rm fund}_{\rm 1-loop}(\vec s^3, m^2)\, Z^{\rm hyper}_{\rm 1-loop}(\vec s^1, \vec \s^2, \vec s^3), 
\end{align}
where $Z^{\rm hyper}_{\rm 1-loop}$ is the contribution of a single Abelian hypermultiplet with charge $(N_1, -(N-1), N_2)$ under the gauge group 
$U(N_1) \times U(N-1) \times U(N_2)$, i.e.
\be
Z^{\rm hyper}_{\rm 1-loop}(\vec s^1, \vec \s^2, \vec s^3) = \frac{1}{\ch{(-\tr \vec \s^2 +\tr \vec s^1 + \tr \vec s^3)}}. 
\ee
The matrix integral on the RHS of the above equation can be evidently identified with the sphere partition function of the quiver gauge 
theory $\CT^\vee_1$ in \figref{IRdual-Ex4a}. We therefore have the following relation between the two partition functions:
\be \label{pf-duality-1A}
\boxed{Z^{(\CT)}(\vec m;\eta_1, \eta_3)=  Z^{(\CT^\vee_1)}(\vec m, m_{\rm{Ab}}=0;\eta_1,0, \eta_3).}
\ee

\subsection{A single balanced unitary node: $e_1=0$ and $e_2 > 2$}\label{3node-pf-b-1}

Let us construct the triality in terms of the sphere partition function. The quiver gauge theory $\CT^\vee_1$ can be obtained from 
the theory $\CT^\vee$, by following the same steps as above. Starting from the partition function of the theory $\CT^\vee_1$:
\begin{align}\label{PF-CT-5A}
Z^{(\CT^\vee_1)}(\vec m, \vec \eta) = & \int\, \prod_{\gamma=1,3}\,\Big[d\vec s^{\gamma}\Big]\, \Big[d  \vec \s^2 \Big] \,\prod_{\gamma=1,3}\,Z_{\rm FI}(\vec s^\gamma, \eta_\gamma)|_{\gamma \neq 2} \, \prod_{\gamma=1,3}\, Z^{\rm vec}_{\rm 1-loop}(\vec s^\gamma)\,Z^{\rm vec}_{\rm 1-loop}(\vec \s^2) \nn \\
 \times & \,Z^{\rm fund}_{\rm 1-loop}(\vec s^1, \vec m^1)\,Z^{\rm bif}_{\rm 1-loop}(\vec s^1, \vec \s^2, 0)\,Z^{\rm bif}_{\rm 1-loop}(\vec \s^2, \vec s^3,0) \,Z^{\rm fund}_{\rm 1-loop}(\vec s^3, \vec m^2)\, Z^{\rm hyper}_{\rm 1-loop}(\vec s^1, \vec \s^2, \vec s^3), 
\end{align}
we isolate the $\vec s^1$-dependent part of the matrix integral and rewrite it using the identity \eref{Id-3} as follows:
\begin{align}
&\int\,\Big[d\vec s^{1}\Big]\, Z_{\rm FI}(\vec s^1, \eta_1) \, Z^{\rm vec}_{\rm 1-loop}(\vec s^1)\,Z^{\rm fund}_{\rm 1-loop}(\vec s^1, \vec m^1)\,Z^{\rm bif}_{\rm 1-loop}(\vec s^1, \vec \s^2, 0)\, Z^{\rm hyper}_{\rm 1-loop}(\vec s^1, \vec \s^2, \vec s^3) \nn \\
=& \int\,\Big[d\vec \s^{1}\Big]\,du_1 \,Z_{\rm FI}(\vec \s^1, -\eta_1)\,Z_{\rm FI}(u_1, \eta_1)\,Z^{\rm vec}_{\rm 1-loop}(\vec \s^1)\,Z^{\rm fund}_{\rm 1-loop}(\vec \s^1, \vec m^1)\,Z^{\rm bif}_{\rm 1-loop}(\vec \s^1, \vec \s^2, 0)\, \nn \\
& \qquad \qquad \times \,Z^{\rm fund}_{\rm 1-loop}(u_1, \tr \vec m^1 + \tr \vec \s^2) \cdot \frac{1}{\ch{(u_1 -\tr \vec \s^1 - \tr \vec\s^2 + \tr \vec s^3)}} \nn \\
= & \int\,\Big[d\vec \s^{1}\Big]\,du_1 \,Z_{\rm FI}(\vec \s^1, -\eta_1)\,Z_{\rm FI}(u_1, \eta_1)\,Z^{\rm vec}_{\rm 1-loop}(\vec \s^1)\,Z^{\rm fund}_{\rm 1-loop}(\vec \s^1, \vec m^1)\,Z^{\rm bif}_{\rm 1-loop}(\vec \s^1, \vec \s^2, 0)\, \nn \\
& \qquad \qquad \times \,Z^{\rm fund}_{\rm 1-loop}(u_1, \tr \vec m^1 + \tr \vec \s^2) \, Z'^{\rm hyper}_{\rm 1-loop}(u_1,\vec \s^1, \vec\s^2, \vec \s^3),
\end{align}
where $\vec \s^1$ and $u_1$ are the matrix integration variables in the Cartan subalgebra of $U(N_1-1)$ and $U(1)$ respectively, and 
$Z'^{\rm hyper}_{\rm 1-loop}$ is the contribution of the new Abelian hypermultiplet. After a change of variable $u_1 \to u_1 + \tr \vec \s^2$, 
and substituting the above identity in \eref{PF-CT-5A}, the partition function assumes the following form:
\begin{align}\label{PF-CT-6A}
Z^{(\CT^\vee_1)}(\vec m, \vec \eta) = & \int\, \prod_{\gamma=1,2}\,\Big[d\vec \s^{\gamma}\Big]\, \Big[d  \vec s^3 \Big]\,du_1 \,Z_{\rm FI}(\vec \s^1, -\eta_1)\, 
Z_{\rm FI}( u_1, \eta_1)\,Z_{\rm FI}(\vec \s^2, \eta_1)\, Z_{\rm FI}(\vec s^3, \eta_3)\, \nn \\
 \times & \prod_{\gamma=1,2}\, Z^{\rm vec}_{\rm 1-loop}(\vec \s^\gamma)\,Z^{\rm vec}_{\rm 1-loop}(\vec s^3)\,Z^{\rm fund}_{\rm 1-loop}(\vec \s^1, \vec m^1)\,  Z^{\rm fund}_{\rm 1-loop}(u_1, \tr \vec m^1)\,Z^{\rm bif}_{\rm 1-loop}(\vec \s^1, \vec \s^2, 0)\nn \\
 \times & Z^{\rm bif}_{\rm 1-loop}(\vec \s^2, \vec s^3,0) \,Z^{\rm fund}_{\rm 1-loop}(\vec s^3, \vec m^2)\, Z'^{\rm hyper}_{\rm 1-loop}(u_1, \vec \s^1, \vec s^3),
\end{align}
where $Z'^{\rm hyper}_{\rm 1-loop}$ is the contribution of the Abelian hypermultiplet, i.e.
\be
Z'^{\rm hyper}_{\rm 1-loop}(u_1, \vec \s^1, \vec s^3)=  \frac{1}{\ch{(u_1 -\tr \vec \s^1  + \tr \vec s^3)}}.
\ee
The matrix integral on the RHS of \eref{PF-CT-6A} can be manifestly identified as the partition function of the quiver $\CT^\vee_2$, 
as shown in \figref{IRdual-Ex4b}, and we therefore have the following relation between the three partition functions:
\be \label{pf-triality-1A}
\boxed{Z^{\CT}(\vec m;\eta_1, \eta_3)=  Z^{\CT^\vee_1}(\vec m, m_{\rm{Ab}}=0;\eta_1,0, \eta_3) =Z^{\CT^\vee_2}(\vec m, m_{\rm{Ab}}=0; \eta_1,-\eta_1, \eta_1, \eta_3) .}
\ee

\subsection{Two balanced unitary nodes: $e_1=0$ and $e_2 =0$}\label{3node-pf-b-2}

Let us construct the hexality in terms of the sphere partition function. The quiver gauge theory $\CT^\vee_1$ can be obtained from 
the theory $\CT^\vee$ by mutation I, and the theories $\CT^\vee_2$, $\CT^\vee_3$ can be obtained 
from the theory $\CT^\vee_1$ by mutation III at two different unitary nodes, in precisely the same way as we did for $e_1=0, e_2>0$ case.  
We can therefore start from the partition function of the theory $\CT^\vee_2$:
\begin{align}\label{PF-CT-7A}
Z^{(\CT^\vee_2)}(\vec m, \vec \eta) = & \int\, \prod_{\gamma=1,2}\,\Big[d\vec \s^{\gamma}\Big]\, \Big[d  \vec s^3 \Big]\,du_1 \,Z_{\rm FI}(\vec \s^1, -\eta_1)\, 
Z_{\rm FI}( u_1, \eta_1)\,Z_{\rm FI}(\vec \s^2, \eta_1)\, Z_{\rm FI}(\vec s^3, \eta_3)\, \nn \\
 \times & \prod_{\gamma=1,2}\, Z^{\rm vec}_{\rm 1-loop}(\vec \s^\gamma)\,Z^{\rm vec}_{\rm 1-loop}(\vec s^3)\,Z^{\rm fund}_{\rm 1-loop}(\vec \s^1, \vec m^1)\,  Z^{\rm fund}_{\rm 1-loop}(u_1, \tr \vec m^1)\,Z^{\rm bif}_{\rm 1-loop}(\vec \s^1, \vec \s^2, 0)\nn \\
 \times & Z^{\rm bif}_{\rm 1-loop}(\vec \s^2, \vec s^3,0) \,Z^{\rm fund}_{\rm 1-loop}(\vec s^3, \vec m^2)\, Z'^{\rm hyper}_{\rm 1-loop}(u_1, \vec \s^1, \vec s^3),
\end{align}
where $Z'^{\rm hyper}_{\rm 1-loop}$ is the contribution of the Abelian hypermultiplet, i.e.
\be
Z'^{\rm hyper}_{\rm 1-loop}(u_1, \vec \s^1, \vec s^3)=  \frac{1}{\ch{(u_1 -\tr \vec \s^1  + \tr \vec s^3)}}.
\ee
We next isolate the $\vec s^3$-dependent part of the matrix integral and implement mutation III by rewriting it 
using the identity \eref{Id-3} as follows:
\begin{align}
&\int\,\Big[d\vec s^{3}\Big]\, Z_{\rm FI}(\vec s^3, \eta_3) \, Z^{\rm vec}_{\rm 1-loop}(\vec s^3)\,Z^{\rm fund}_{\rm 1-loop}(\vec s^3, \vec m^2)\,Z^{\rm bif}_{\rm 1-loop}(\vec s^3, \vec \s^2, 0)\, Z'^{\rm hyper}_{\rm 1-loop}(u^1, \vec \s^1, \vec s^3) \nn \\
=& \int\,\Big[d\vec \s^{3}\Big]\,du_2 \,Z_{\rm FI}(\vec \s^3, -\eta_3)\,Z_{\rm FI}(u_2, \eta_3)\,Z^{\rm vec}_{\rm 1-loop}(\vec \s^3)\,Z^{\rm fund}_{\rm 1-loop}(\vec \s^3, \vec m^2)\,Z^{\rm bif}_{\rm 1-loop}(\vec \s^3, \vec \s^2, 0)\, \nn \\
& \qquad \qquad \times \,Z^{\rm fund}_{\rm 1-loop}(u_2, \tr \vec m^2 + \tr \vec \s^2) \cdot \Big(\frac{1}{\ch{(u_2 +u_1 -\tr \vec \s^1 - \tr \vec \s^3)}}\Big), \nn \\
\end{align}
where $\vec \s^3$ and $u_2$ are the matrix integration variables in the Cartan subalgebra of $U(N_2-1)$ and $U(1)$ respectively. 
After a change of variable $u_1 \to u_1 + \tr \vec \s^2$, and substituting the above identity in \eref{PF-CT-7A}, the partition function 
reduces to that of the theory $(\CT^\vee_4)$ in \figref{IRdual-Ex4c}, i.e. :
\begin{align}\label{PF-CT-8A}
Z^{(\CT^\vee_2)} = & \int\, \prod^3_{\gamma=1}\,\Big[d\vec \s^{\gamma}\Big]\, \prod^2_{i=1}\,du_i \,Z_{\rm FI}(\vec \s^1, -\eta_1)\, 
\,Z_{\rm FI}(\vec \s^2, \eta_1+\eta_3)\, Z_{\rm FI}(\vec \s^3, -\eta_3)\,Z_{\rm FI}( u_1, \eta_1)\,Z_{\rm FI}( u_2, \eta_3) \nn \\
 \times & \prod^3_{\gamma=1}\, Z^{\rm vec}_{\rm 1-loop}(\vec \s^\gamma)\, Z^{\rm fund}_{\rm 1-loop}(\vec \s^1, \vec m^1)\,  
 \prod^2_{i=1}\,Z^{\rm fund}_{\rm 1-loop}(u_i, \tr \vec m^i)\,Z^{\rm bif}_{\rm 1-loop}(\vec \s^1, \vec \s^2, 0) \nn \\
 \times & Z^{\rm bif}_{\rm 1-loop}(\vec \s^2, \vec \s^3,0) \,Z^{\rm fund}_{\rm 1-loop}(\vec \s^3, \vec m^2)\, Z^{\rm hyper}_{\rm 1-loop}(\{u_i\}, \{ \vec \s^\gamma \})
 =: Z^{(\CT^\vee_3)},
\end{align}
where $Z^{\rm hyper}_{\rm 1-loop}$ is the contribution of the Abelian hypermultiplet, i.e.
\be
Z^{\rm hyper}_{\rm 1-loop}(\{u_i\}, \{ \vec \s^\gamma \}) =  \frac{1}{\ch{(u_1+u_2 -\tr \vec \s^1  - \tr \vec \s^3 + \tr \vec \s^2)}}.
\ee

Finally, the $U(N-1)$ node in the quiver $(\CT^\vee_4)$ admits a mutation III. Therefore, using \eref{Id-3} at the 
$U(N-1)$ node, the partition function for $(\CT^\vee_4)$ can be written in the following form:
\begin{align}\label{PF-CT-9A}
Z^{(\CT^\vee_4)} = & \int\, \prod_{\gamma=1,3}\,\Big[d\vec \s^{\gamma}\Big]\, \Big[d\vec \s'^{2}\Big]\,\prod^3_{i=1}\,du_i \,Z_{\rm FI}(\vec \s^1, -\eta_1)\, 
\,Z_{\rm FI}(\vec \s'^2, -\eta_1-\eta_3)\, Z_{\rm FI}(\vec \s^3, -\eta_3)\nn \\
& \times Z_{\rm FI}( u_1, \eta_1)\,Z_{\rm FI}( u_2, \eta_3)\, Z_{\rm FI}( u_3, \eta_1 + \eta_3)
 \, \prod_{\gamma=1,3}\, Z^{\rm vec}_{\rm 1-loop}(\vec \s^\gamma)\,Z^{\rm vec}_{\rm 1-loop}(\vec \s'^2) \nn \\
 & \times  Z^{\rm fund}_{\rm 1-loop}(\vec \s^1, \vec m^1)\,  
 \prod^2_{i=1}\,Z^{\rm fund}_{\rm 1-loop}(u_i, \tr \vec m^i)\, Z^{\rm fund}_{\rm 1-loop}(u_3, \tr \vec \s^1 + \tr \s^3)  \,Z^{\rm bif}_{\rm 1-loop}(\vec \s^1, \vec \s'^2, 0) \nn \\
 \times & Z^{\rm bif}_{\rm 1-loop}(\vec \s'^2, \vec \s^3,0) \,Z^{\rm fund}_{\rm 1-loop}(\vec \s^3, \vec m^2)\, 
 Z^{\rm hyper}_{\rm 1-loop}(\{u_i\}, \{ \vec \s^\gamma \}),
\end{align}
where $\vec \s'^2$ and $u_3$ denotes the matrix integral variables for a $U(N-2)$ and a $U(1)$ gauge group respectively, 
and $Z^{\rm hyper}_{\rm 1-loop}$ is the contribution of the Abelian hypermultiplet, i.e.
\be
Z^{\rm hyper}_{\rm 1-loop}(\{u_i\}, \{ \vec \s^\gamma \}) =  \frac{1}{\ch{(u_1+u_2+u_3 -\tr \vec \s^1  - \tr \vec \s^3 - \tr \vec \s'^2)}}.
\ee
After a change of variables $u_3 \to u_3 + \tr \vec \s^1 + \tr \vec \s^3$, the above partition function manifestly reduces to the 
partition function of the quiver $(\CT^\vee_5)$ as shown in the bottom-left of \figref{IRdual-Ex4c}, i.e.
\begin{align}\label{PF-CT-10}
Z^{(\CT^\vee_4)} = & \int\, \prod_{\gamma=1,3}\,\Big[d\vec \s^{\gamma}\Big]\, \Big[d\vec \s'^{2}\Big]\,\prod^3_{i=1}\,du_i \,Z_{\rm FI}(\vec \s^1, \eta_3)\, 
\,Z_{\rm FI}(\vec \s'^2, -\eta_1-\eta_3)\, Z_{\rm FI}(\vec \s^3, \eta_1)\nn \\
& \times Z_{\rm FI}( u_1, \eta_1)\,Z_{\rm FI}( u_2, \eta_3)\, Z_{\rm FI}( u_3, \eta_1 + \eta_3)
 \, \prod_{\gamma=1,3}\, Z^{\rm vec}_{\rm 1-loop}(\vec \s^\gamma)\,Z^{\rm vec}_{\rm 1-loop}(\vec \s'^2) \nn \\
 & \times  Z^{\rm fund}_{\rm 1-loop}(\vec \s^1, \vec m^1)\,  
 \prod^2_{i=1}\,Z^{\rm fund}_{\rm 1-loop}(u_i, \tr \vec m^i)\, Z^{\rm fund}_{\rm 1-loop}(u_3, 0)  \,Z^{\rm bif}_{\rm 1-loop}(\vec \s^1, \vec \s'^2, 0) \nn \\
 & \times  Z^{\rm bif}_{\rm 1-loop}(\vec \s'^2, \vec \s^3,0) \,Z^{\rm fund}_{\rm 1-loop}(\vec \s^3, \vec m^2)\, 
 Z^{\rm hyper}_{\rm 1-loop}(\{u_i\}, \{ \vec \s^\gamma \}) := Z^{(\CT^\vee_5)},
\end{align}
where the contribution of the Abelian hypermultiplet is now
\be
Z^{\rm hyper}_{\rm 1-loop}(\{u_i\}, \{ \vec \s^\gamma \}) =  \frac{1}{\ch{(u_1+u_2+u_3  - \tr \vec \s'^2)}}.
\ee
The final form of the quiver $(\CT^\vee_5)$ is obtained by the change of variables $u_3 \to u_2 -u_1$ and $u_2 \to u_3 -u_2$, 
leading to the partition function
\begin{align}\label{PF-CT-10}
Z^{(\CT^\vee_5)} = & \int\, \prod_{\gamma=1,3}\,\Big[d\vec \s^{\gamma}\Big]\, \Big[d\vec \s'^{2}\Big]\,\prod^3_{i=1}\,du_i \,Z_{\rm FI}(\vec \s^1, \eta_3)\, 
\,Z_{\rm FI}(\vec \s'^2, -\eta_1-\eta_3)\, Z_{\rm FI}(\vec \s^3, \eta_1)\nn \\
& \times Z_{\rm FI}( u_1, -\eta_3)\,Z_{\rm FI}( u_2, \eta_1)\, Z_{\rm FI}( u_3,  \eta_3)
 \, \prod_{\gamma=1,3}\, Z^{\rm vec}_{\rm 1-loop}(\vec \s^\gamma)\,Z^{\rm vec}_{\rm 1-loop}(\vec \s'^2) \nn \\
 & \times  Z^{\rm fund}_{\rm 1-loop}(\vec \s^1, \vec m^1)\,  
 Z^{\rm fund}_{\rm 1-loop}(u_1, \tr \vec m^1)\, Z^{\rm bif}_{\rm 1-loop}(u_1,u_2,0)\,Z^{\rm bif}_{\rm 1-loop}(u_2,u_3, \tr \vec m^2) \nn \\
 & \times  Z^{\rm bif}_{\rm 1-loop}(\vec \s^1, \vec \s'^2, 0)\, Z^{\rm bif}_{\rm 1-loop}(\vec \s'^2, \vec \s^3,0) \,Z^{\rm fund}_{\rm 1-loop}(\vec \s^3, \vec m^2)\, 
 Z^{\rm hyper}_{\rm 1-loop}(u_3, \vec \s'^2),
\end{align}
where the contribution of the Abelian hypermultiplet is now
\be
Z^{\rm hyper}_{\rm 1-loop}(u_3, \vec \s'^2) =  \frac{1}{\ch{(u_3  - \tr \vec \s'^2)}}.
\ee
In this final form, the Abelian hypermultiplet connects one of the $U(1)$ gauge nodes and the $U(N-2)$ gauge node, as 
shown in the bottom-right corner of \figref{IRdual-Ex4c}.

\section{Abelian $S$-type operations and 3d mirror}\label{Mirr-pf-S}

In this section, we perform the Abelian $S$-type operations discussed in \Secref{2node-mirr} and \Secref{3node-mirr} in terms of the 
sphere partition function and find the explicit forms of the 3d mirrors. 

It will be useful for the computations performed below to define a linear quiver of the following form:
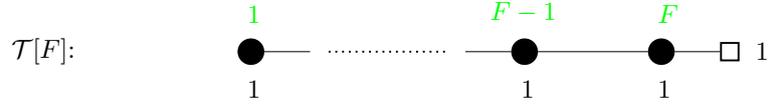
\begin{center}
\scalebox{0.9}{\begin{tikzpicture}[bnode/.style={circle,draw, thick, fill=black!30,minimum size=1cm}]
\node[text width=1 cm](34) at (1,0) {$\CT[F]$:};
\node[unode] (2) at (4,0){};
\node[] (3) at (5,0){};
\node[] (4) at (7,0){};
\node[unode] (5) at (8,0){};
\node[unode] (6) at (10,0){};
\node[fnode] (7) at (11,0){};
\draw[-] (2) -- (3);
\draw[thick, dotted] (3) -- (4);
\draw[-] (4) -- (5);
\draw[-] (5) -- (6);
\draw[-] (6) -- (7);
\node[text width=0.1cm](31) [below=0.1 of 2] {1};
\node[text width=0.1 cm](32) [below=0.1 of 5] {1};
\node[text width=0.1cm](33) [below=0.1 of 6] {1};
\node[text width=0.1cm](34) [right=0.1 of 7] {1};
\node[text width=0.1cm, green](41) [above=0.1 of 2] {1};
\node[text width=1cm, green](42) [above=0.1 of 5] {$F-1$};
\node[text width=0.1cm, green](43) [above=0.1 of 6] {$F$};
\end{tikzpicture}} 
\end{center}

This a linear quiver of $F$ $U(1)$ gauge nodes connected by bifundamental hypermultiplets and a single 
fundamental hypermultiplet at one end. The partition function of the quiver is given as:
\be
Z^{(\CT[F])}(m, \vec \xi) = \int \, \prod^{F}_{a=1} \, d\tau_a\, Z^{(\CT[F])}_{\rm int}(\vec \tau, m, \vec \xi)= \int \, \prod^{F}_{a=1} \, d\tau_a\, \frac{ \prod^{F}_{a=1}\,e^{2\pi i \tau_a\,(\xi_{a} - \xi_{a+1})}}{\prod^{F}_{a=1}\, \ch{(\tau_a - \tau_{a+1})}\, \ch{(\tau_{F}-m)}},
\ee
where $m$ is the real mass of the single fundamental hyper and $\{\vec \xi_a - \vec \xi_{a+1}\}_{a=1,\ldots,F}$ are the $F$ FI parameters.

\subsection{2-node quiver: Flavoring-gauging operation}

Consider the flavoring-gauging operation $\CO$ acting on the 3d mirror pair $(\wt{\CT}_{\rm good}, {\CT}_{\rm good})$ as follows:

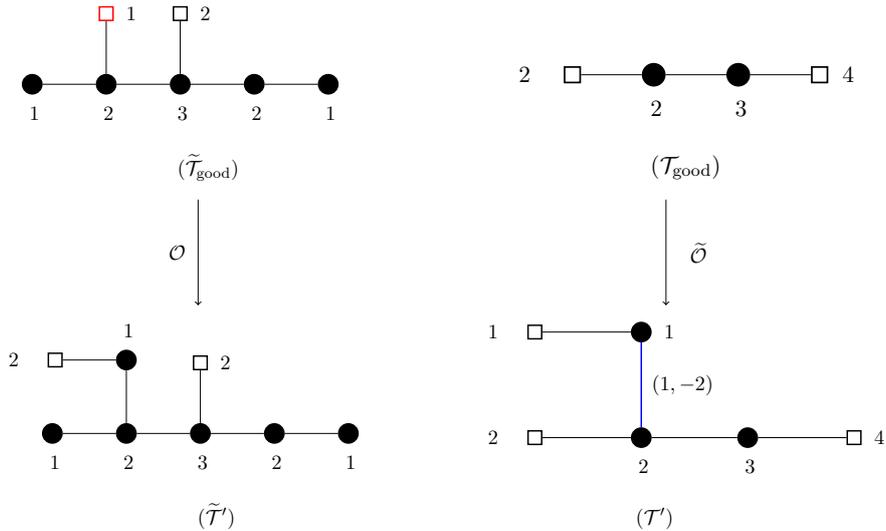
\begin{center}
\begin{tabular}{ccc}
 \scalebox{0.7}{\begin{tikzpicture}
\node[unode] (1) {};
\node[unode] (2) [right=1cm  of 1]{};
\node[unode] (3) [right=1 cm of 2]{};
\node[unode] (4) [right=1 cm of 3]{};
\node[unode] (5) [right=1 cm of 4]{};
\node[fnode, red] (6) [above=1 cm of 2]{};
\node[fnode] (7) [above=1 cm of 3]{};
\draw[-] (1) -- (2);
\draw[-] (2)-- (3);
\draw[-] (3) -- (4);
\draw[-] (4) -- (5);
\draw[-] (2) -- (6);
\draw[-] (3) -- (7);
\node[text width=.1cm](10) [below=0.1 cm of 1]{$1$};
\node[text width= .1cm](11) [below=0.1cm of 2]{$2$};
\node[text width= .1cm](12) [below=0.1cm of 3]{$3$};
\node[text width=.1cm](13) [below=0.1cm of 4]{$2$};
\node[text width=.1cm](14) [below=0.1cm of 5]{$1$};
\node[text width=.1cm](15) [right=0.1cm of 6]{$1$};
\node[text width=.1cm](16) [right=0.1cm of 7]{$2$};
\node[text width=0.1cm](30)[below=1 cm of 3]{$(\wt{\CT}_{\rm good})$};
\end{tikzpicture}}
& \qquad  
&\scalebox{.8}{\begin{tikzpicture}
\node[fnode] (1) {};
\node[unode] (2) [right=1cm  of 1]{};
\node[unode] (3) [right=1 cm of 2]{};
\node[fnode] (4) [right=1 cm of 3]{};
\draw[-] (1) -- (2);
\draw[-] (2)-- (3);
\draw[-] (3) -- (4);
\node[text width=.1cm](10) [left=0.5 cm of 1]{$2$};
\node[text width= .1cm](11) [below=0.1cm of 2]{$2$};
\node[text width=.1cm](12) [below=0.1cm of 3]{$3$};
\node[text width=.1cm](13) [right=0.1cm of 4]{$4$};
\node[text width=0.1cm](30)[below=1 cm of 2]{$({\CT}_{\rm good})$};
\end{tikzpicture}}\\
 \scalebox{.7}{\begin{tikzpicture}
\draw[->] (15,-3) -- (15,-5);
\node[text width=0.1cm](20) at (14.5, -4) {$\CO$};
\end{tikzpicture}}
&\qquad \qquad 
& \scalebox{.7}{\begin{tikzpicture}
\draw[->] (15,-3) -- (15,-5);
\node[text width=0.1cm](29) at (15.5, -4) {$\wt{\CO}$};
\end{tikzpicture}}\\
\scalebox{0.7}{\begin{tikzpicture}
\node[unode] (1) {};
\node[unode] (2) [right=1cm  of 1]{};
\node[unode] (3) [right=1 cm of 2]{};
\node[unode] (4) [right=1 cm of 3]{};
\node[unode] (5) [right=1 cm of 4]{};
\node[unode] (6) [above=1 cm of 2]{};
\node[fnode] (7) [above=1 cm of 3]{};
\node[fnode] (8) [left= 1 cm of 6]{};
\draw[-] (1) -- (2);
\draw[-] (2)-- (3);
\draw[-] (3) -- (4);
\draw[-] (4) -- (5);
\draw[-] (2) -- (6);
\draw[-] (3) -- (7);
\draw[-] (6) -- (8);
\node[text width=.1cm](10) [below=0.1 cm of 1]{$1$};
\node[text width= .1cm](11) [below=0.1cm of 2]{$2$};
\node[text width= .1cm](12) [below=0.1cm of 3]{$3$};
\node[text width=.1cm](13) [below=0.1cm of 4]{$2$};
\node[text width=.1cm](14) [below=0.1cm of 5]{$1$};
\node[text width=.1cm](15) [above=0.1cm of 6]{$1$};
\node[text width=.1cm](16) [right=0.1cm of 7]{$2$};
\node[text width=.1cm](17) [left=0.5cm of 8]{$2$};
\node[text width=0.1cm](20)[below=1 cm of 3]{$(\wt{\CT}')$};
\end{tikzpicture}}
&\qquad \qquad 
& \scalebox{.7}{\begin{tikzpicture}
\node[fnode] (1) at (0,0){};
\node[unode] (2) at (2,0){};
\node[unode] (3) at (4,0){};
\node[fnode] (4) at (6,0){};
\node[unode] (6) at (2,2){};
\node[fnode] (7) at (0,2){};
\draw[-] (1) -- (2);
\draw[-] (2)-- (3);
\draw[-] (3) -- (4);
\draw[-] (6) --(7);
\draw[-, thick, blue] (2)--(6);
\node[text width=2cm](10) at (3.2, 1){$(1,-2)$};
\node[text width=.1cm](20) [left=0.5 cm of 1]{$2$};
\node[text width=0.1 cm](21) [below=0.1cm of 2]{$2$};
\node[text width=0.1 cm](22) [below=0.1cm of 3]{$3$};
\node[text width=.1cm](23) [right=0.1cm of 4]{$4$};
\node[text width=.1cm](24) [right=0.1cm of 6]{1};
\node[text width=.1cm](25) [left=0.5 cm of 7]{1};
\node[text width=.2cm](40) [below=1 cm of 2]{$({\CT}')$};
\end{tikzpicture}}
\end{tabular}
\end{center}

The partition functions of the theories $(\wt{\CT}_{\rm good}, {\CT}_{\rm good})$  are given as
\begin{align}
& Z^{(\wt{\CT}_{\rm good})} (\vec m; \vec t)=\int \, \prod^5_{\gamma=1}\, \Big[d\vec s^{\gamma}\Big]\, Z^{(\wt{\CT}_{\rm good})}_{\rm int} (\vec s, \vec m, \vec t) = \int \, \prod^5_{\gamma=1}\, \Big[d\vec s^{\gamma}\Big]\,Z^{(\wt{\CT}_{\rm good})}_{\rm FI}(\vec s, \vec t) \,Z^{(\wt{\CT}_{\rm good})}_{\rm 1-loop}(\vec s, \vec m) , \label{pfX-1}\\
& Z^{({\CT}_{\rm good})} (\vec t; \vec m)= \int \, \prod^2_{\gamma'=1}\,\Big[d\vec \s^{\gamma'}\Big]\,Z^{({\CT}_{\rm good})}_{\rm int} (\vec \s, \vec t, \vec m) = \int \, \prod^2_{\gamma'=1}\, Z^{({\CT}_{\rm good})}_{\rm FI}(\vec \s, \vec m) \,Z^{({\CT}_{\rm good})}_{\rm 1-loop}(\vec \s, \vec t), \label{pfY-1}
\end{align}
where the FI terms are:
\be
Z^{(\wt{\CT}_{\rm good})}_{\rm FI}(\vec s, \vec t)=\prod^5_{\gamma=1}\,e^{2\pi i (t_\gamma - t_{\gamma+1})\,\tr \vec s^\gamma} , \qquad Z^{({\CT}_{\rm good})}_{\rm FI}(\vec \s, \vec m)= \prod^2_{\gamma'=1}\, e^{2\pi i (m_{\gamma'} - m_{\gamma'+1})\,\tr \vec \s^{\gamma'}}. \label{FI-1}
\ee
The gauge node labels $\gamma$ and $\gamma'$ increase from left to right.
Mirror symmetry implies that the partition function of these theories are related as:

\be \label{MS-id1}
Z^{(\wt{\CT}_{\rm good})} (\vec m; \vec t)= C(\vec m, \vec t)\,Z^{({\CT}_{\rm good})} (\vec t; -\vec m),
\ee
where $C(\vec m, \vec t)$ is a contact term. The Abelian operation $\CO$ can be implemented at the partition function level in the following 
fashion:
\begin{align} \label{SOp-1}
Z^{\CO(\wt{\CT}_{\rm good})} (\vec m_f, m_2, m_3; \vec t, \eta) =& \int\, du\, \frac{e^{2\pi i \eta\,u}}{\prod^2_{a=1}\,\ch{(u-m^a_f)}}\,Z^{(\wt{\CT}_{\rm good})} (u, m_2, m_3; \vec t) \nn \\
= & \int\,\,du\,\prod^5_{\gamma=1}\,\Big[d\vec s^{\gamma}\Big]\,\frac{e^{2\pi i \eta\,u}}{\prod^2_{a=1}\,\ch{(u-m^a_f)}}\, Z^{(\wt{\CT}_{\rm good})}_{\rm int} (\vec s, u, m_2,m_3, \vec t),
\end{align}
where $\vec m_f$ are the real masses for the two fundamental hypermultiplets added in the flavoring operation (associated with the flavor 
node labelled 2), and $\eta$ is the FI parameter for the new $U(1)$ gauge node. Unpacking the function $Z^{(\wt{\CT}_{\rm good})}_{\rm int}$, 
one can readily check that the above matrix integral is the partition function of the quiver $\wt{\CT}'$ given above. 

The partition function for the 3d mirror of $\CO(\wt{\CT}_{\rm good})= \wt{\CT}'$ can be obtained by using the identity \eref{MS-id1} 
to substitute the function $Z^{(\wt{\CT}_{\rm good})}$ in the first line of \eref{SOp-1} and changing the order of integration to perform the integral over $u$ first, i.e.
\begin{align} \label{pf-dual-1a}
Z^{\wt{\CO}({\CT}_{\rm good})} = \int \, \prod^2_{\gamma'=1}\,\Big[d\vec \s^{\gamma'}\Big]\,\int du\, \frac{e^{2\pi i \eta\,u}}{\prod^2_{a=1}\,\ch{(u-m^a_f)}}\, Z^{({\CT}_{\rm good})}_{\rm int} (\vec \s, \vec t, -u, -m_2, -m_3),
\end{align}
where $\wt{\CO}$ is the operation dual to $\CO$, and the function $Z^{({\CT}_{\rm good})}_{\rm int}$ can be read off from \eref{pfX-1}-\eref{FI-1}. 
It is useful at this stage to use the following identity:
\be \label{flavor-id}
\frac{1}{\prod^F_{a=1}\,\ch{(u-m^a_f)}} = \int \prod^{F-1}_{a=0} \, d\tau_a\, \frac{e^{2\pi i \tau_0(u - m^1_f)}\, \prod^{F-1}_{a=1}\,e^{2\pi i \tau_a\,(m^{a}_f - m^{a+1}_f)}}{\prod^{F-2}_{a=0}\, \ch{(\tau_a - \tau_{a+1})}\, \ch{\tau_{F-1}}}.
\ee
Plugging in the $F=2$ version of this identity in \eref{pf-dual-1a} and performing the integrals over $u$ and $\tau_0$, the partition function $Z^{\wt{\CO}({\CT}_{\rm good})}$ can be rewritten as:
\begin{align} \label{pf-dual-1b}
Z^{\wt{\CO}({\CT}_{\rm good})} =  \int \, \prod^2_{\gamma'=1}\,\Big[d\vec \s^{\gamma'}\Big]\, d\tau_1\, Z^{\CT[1]}_{\rm int}(\tau_1, 0, -\vec m_f)\, Z^{\rm hyper}_{\rm 1-loop}(\tau_1, \tr \vec \s^1, -\eta)\, Z^{({\CT}_{\rm good})}_{\rm int} (\vec \s, \vec t, -m^1_f, -m_2,-m_3),
\end{align}
where $Z^{\CT[1]}$ is the partition function of the linear chain $\CT[1]$ as defined above, and 
$Z^{\rm hyper}_{\rm 1-loop}=\frac{1}{\ch{(\tau_1 -\tr \vec \s^1 + \eta)}}$ is the 1-loop 
contribution of an Abelian hypermultiplet with charges $(1,-2)$ under the $U(1)$ gauge node of $\CT[1]$ 
and the $U(2)$ gauge node of ${\CT}_{\rm good}$ respectively. Therefore, one can read off the 
dual quiver from the matrix model of \eref{pf-dual-1b} and check that it reproduces the quiver $\CT'$.

\subsection{3-node quiver: Identification-flavoring-gauging operation}

Consider the identification-flavoring-gauging operation $\CO$ acting on the 3d mirror pair 
$(\wt{\CT}_{\rm good}, {\CT}_{\rm good})$ as follows:

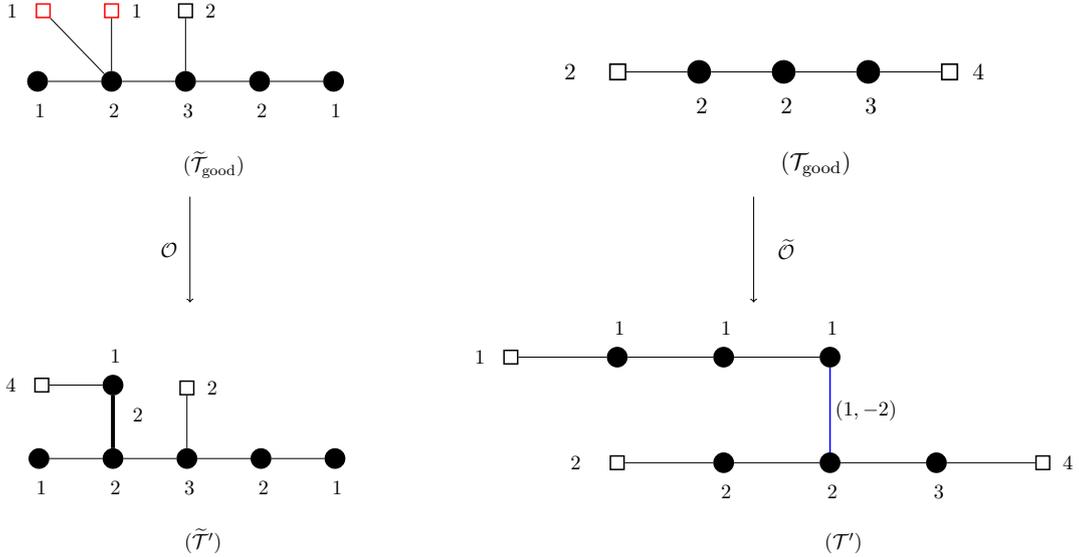
\begin{center}
\begin{tabular}{ccc}
 \scalebox{0.7}{\begin{tikzpicture}
\node[unode] (1) {};
\node[unode] (2) [right=1cm  of 1]{};
\node[unode] (3) [right=1 cm of 2]{};
\node[unode] (4) [right=1 cm of 3]{};
\node[unode] (5) [right=1 cm of 4]{};
\node[fnode, red] (6) [above=1 cm of 2]{};
\node[fnode, red] (8) [left=1 cm of 6]{};
\node[fnode] (7) [above=1 cm of 3]{};
\draw[-] (1) -- (2);
\draw[-] (2)-- (3);
\draw[-] (3) -- (4);
\draw[-] (4) -- (5);
\draw[-] (2) -- (6);
\draw[-] (2) -- (8);
\draw[-] (3) -- (7);
\node[text width=.1cm](10) [below=0.1 cm of 1]{$1$};
\node[text width= .1cm](11) [below=0.1cm of 2]{$2$};
\node[text width= .1cm](12) [below=0.1cm of 3]{$3$};
\node[text width=.1cm](13) [below=0.1cm of 4]{$2$};
\node[text width=.1cm](14) [below=0.1cm of 5]{$1$};
\node[text width=.1cm](15) [right=0.1cm of 6]{$1$};
\node[text width=.1cm](16) [right=0.1cm of 7]{$2$};
\node[text width=.1cm](17) [left=0.3 cm of 8]{$1$};
\node[text width=0.1cm](30)[below=1 cm of 3]{$(\wt{\CT}_{\rm good})$};
\end{tikzpicture}}
& \qquad  
&\scalebox{.8}{\begin{tikzpicture}
\node[fnode] (1) {};
\node[unode] (2) [right=1cm  of 1]{};
\node[unode] (3) [right=1 cm of 2]{};
\node[unode] (4) [right=1 cm of 3]{};
\node[fnode] (5) [right=1 cm of 4]{};
\draw[-] (1) -- (2);
\draw[-] (2)-- (3);
\draw[-] (3) -- (4);
\draw[-] (4) -- (5);
\node[text width=.1cm](10) [left=0.5 cm of 1]{$2$};
\node[text width= 0.1cm](11) [below=0.1cm of 2]{$2$};
\node[text width=0.1cm](12) [below=0.1cm of 3]{$2$};
\node[text width=.1cm](13) [below=0.1cm of 4]{$3$};
\node[text width=.1cm](14) [right=0.1cm of 5]{$4$};
\node[text width=0.1cm](30)[below=1 cm of 3]{$({\CT}_{\rm good})$};
\end{tikzpicture}}\\
 \scalebox{.7}{\begin{tikzpicture}
\draw[->] (15,-3) -- (15,-5);
\node[text width=0.1cm](20) at (14.5, -4) {$\CO$};
\end{tikzpicture}}
&\qquad \qquad 
& \scalebox{.7}{\begin{tikzpicture}
\draw[->] (15,-3) -- (15,-5);
\node[text width=0.1cm](29) at (15.5, -4) {$\wt{\CO}$};
\end{tikzpicture}}\\
\scalebox{0.7}{\begin{tikzpicture}
\node[unode] (1) {};
\node[unode] (2) [right=1cm  of 1]{};
\node[unode] (3) [right=1 cm of 2]{};
\node[unode] (4) [right=1 cm of 3]{};
\node[unode] (5) [right=1 cm of 4]{};
\node[unode] (6) [above=1 cm of 2]{};
\node[fnode] (7) [above=1 cm of 3]{};
\node[fnode] (8) [left= 1 cm of 6]{};
\node[](9) [above=0.5 cm of 2]{};
\draw[-] (1) -- (2);
\draw[-] (2)-- (3);
\draw[-] (3) -- (4);
\draw[-] (4) -- (5);
\draw[line width=0.75mm] (2) -- (6);
\draw[-] (3) -- (7);
\draw[-] (6) -- (8);
\node[text width=.1cm](10) [below=0.1 cm of 1]{$1$};
\node[text width= .1cm](11) [below=0.1cm of 2]{$2$};
\node[text width= .1cm](12) [below=0.1cm of 3]{$3$};
\node[text width=.1cm](13) [below=0.1cm of 4]{$2$};
\node[text width=.1cm](14) [below=0.1cm of 5]{$1$};
\node[text width=.1cm](15) [above=0.1cm of 6]{$1$};
\node[text width=.1cm](16) [right=0.1cm of 7]{$2$};
\node[text width=.1cm](17) [left=0.3cm of 8]{$4$};
\node[text width=.1cm](18) [right=0.1cm of 9]{$2$};
\node[text width=0.1cm](20)[below=1 cm of 3]{$(\wt{\CT}')$};
\end{tikzpicture}}
&\qquad \qquad 
& \scalebox{.7}{\begin{tikzpicture}
\node[fnode] (1) at (0,0){};
\node[unode] (2) at (2,0){};
\node[unode] (3) at (4,0){};
\node[unode] (4) at (6,0){};
\node[fnode] (5) at (8,0){};
\node[unode] (6) at (4,2){};
\node[unode] (7) at (2,2){};
\node[unode] (8) at (0,2){};
\node[fnode] (9) at (-2,2){};
\draw[-] (1) -- (2);
\draw[-] (2)-- (3);
\draw[-] (3) -- (4);
\draw[-] (4) --(5);
\draw[-] (6) --(7);
\draw[-] (8) --(7);
\draw[-] (8) --(9);
\draw[thick, blue] (6)-- (3);
\node[text width=.1cm](40) [left=0.5 cm of 1]{$2$};
\node[text width=.1 cm](41) [below=0.1cm of 2]{$2$};
\node[text width=.1 cm](42) [below=0.1cm of 3]{$2$};
\node[text width=.1 cm](43) [below=0.1cm of 4]{$3$};
\node[text width=.1cm](44) [right=0.1cm of 5]{$4$};
\node[text width=.1cm](45) [above=0.1cm of 6]{1};
\node[text width=.1cm](46) [above=0.1 cm of 7]{1};
\node[text width=.1cm](47) [above=0.1 cm of 8]{1};
\node[text width=.1cm](48) [left=0.3 cm of 9]{1};
\node[text width=2cm](21) at (5.1, 1){$(1,-2)$};
\node[text width=.2cm](40) [below=1 cm of 3]{$({\CT}')$};
\end{tikzpicture}}
\end{tabular}
\end{center}

The partition functions of the theories $(\wt{\CT}_{\rm good}, {\CT}_{\rm good})$  are given as
\begin{align}
& Z^{(\wt{\CT}_{\rm good})} (\vec m; \vec t)=\int \, \prod^5_{\gamma=1}\, \Big[d\vec s^{\gamma}\Big]\, Z^{(\wt{\CT}_{\rm good})}_{\rm int} (\vec s, \vec m, \vec t) = \int \, \prod^5_{\gamma=1}\, \Big[d\vec s^{\gamma}\Big]\,Z^{(\wt{\CT}_{\rm good})}_{\rm FI}(\vec s, \vec t) \,Z^{(\wt{\CT}_{\rm good})}_{\rm 1-loop}(\vec s, \vec m) ,  \label{pfX-2} \\
& Z^{({\CT}_{\rm good})} (\vec t; \vec m)= \int \, \prod^3_{\gamma'=1}\,\Big[d\vec \s^{\gamma'}\Big]\,Z^{({\CT}_{\rm good})}_{\rm int} (\vec \s, \vec t, \vec m) = \int \, \prod^3_{\gamma'=1}\, Z^{({\CT}_{\rm good})}_{\rm FI}(\vec \s, \vec m) \,Z^{({\CT}_{\rm good})}_{\rm 1-loop}(\vec \s, \vec t), \label{pfY-2}
\end{align}
where the FI terms are:
\be
Z^{(\wt{\CT}_{\rm good})}_{\rm FI}(\vec s, \vec t)=\prod^5_{\gamma=1}\,e^{2\pi i (t_\gamma - t_{\gamma+1})\,\tr \vec s^\gamma} , \qquad Z^{({\CT}_{\rm good})}_{\rm FI}(\vec \s, \vec m)= \prod^3_{\gamma'=1}\, e^{2\pi i (m_{\gamma'} - m_{\gamma'+1})\,\tr \vec \s^{\gamma'}}. \label{FI-2}
\ee

The gauge node labels $\gamma$ and $\gamma'$ increase from left to right. Mirror symmetry implies that the partition function of these theories are related as:

\be \label{MS-id2}
Z^{(\wt{\CT}_{\rm good})} (\vec m; \vec t)= C(\vec m, \vec t)\,Z^{({\CT}_{\rm good})} (\vec t; -\vec m),
\ee
where $C(\vec m, \vec t)$ is a contact term. The Abelian operation $\CO$ can be implemented at the partition function level in the following 
fashion:
\begin{align}\label{SOp-2}
Z^{\CO(\wt{\CT}_{\rm good})} (\vec m_f, m_3, m_4; \vec t, \eta) =& \int\, du\,\prod^2_{i=1}\,dm_i\, \frac{e^{2\pi i \eta\,u}\,\prod^2_{i=1}\,\delta(u-m_i + \mu_i)}{\prod^4_{a=1}\,\ch{(u-m^a_f)}}\,Z^{(\wt{\CT}_{\rm good})} (m_1, m_2, m_3,m_4; \vec t) \nn \\
= & \int\,\prod^5_{\gamma=1}\,\Big[d\vec s^{\gamma}\Big]\,\,du\,\frac{e^{2\pi i \eta\,u}}{\prod^4_{a=1}\,\ch{(u-m^a_f)}}\, Z^{(\wt{\CT}_{\rm good})}_{\rm int} (u+\mu_1, u+\mu_2, m_3,m_4, \vec t),
\end{align}
where $\vec m_f$ are the real masses for the four fundamental hypermultiplets added in the flavoring operation (associated with the flavor 
node labelled 4), $\vec \mu$ are real masses for the two bifundamental hypermultiplets for the $U(2) \times U(1)$ gauge group,
and $\eta$ is the FI parameter for the new $U(1)$ gauge node. Unpacking the function $Z^{(\wt{\CT}_{\rm good})}_{\rm int}$, 
one can readily check that the above matrix integral is the partition function of the quiver $\wt{\CT}'$ given above. 

The partition function for the 3d mirror of $\CO(\wt{\CT}_{\rm good})= \wt{\CT}'$ can be obtained by using the identity \eref{MS-id2} 
to substitute the function $Z^{(\wt{\CT}_{\rm good})}$ in the first line of \eref{SOp-2} and changing the order of integration to perform the integral over $u$ first, i.e.
\begin{align} \label{pf-dual-2a}
Z^{\wt{\CO}({\CT}_{\rm good})} = \int \, \prod^3_{\gamma'=1}\,\Big[d\vec \s^{\gamma'}\Big]\,\int du\, \frac{e^{2\pi i \eta\,u}}{\prod^4_{a=1}\,\ch{(u-m^a_f)}}\, Z^{({\CT}_{\rm good})}_{\rm int} (\vec \s, \vec t, -u-\mu_1, -u-\mu_2, -m_3, -m_4),
\end{align}
where $\wt{\CO}$ is the operation dual to $\CO$, and the function $Z^{({\CT}_{\rm good})}_{\rm int}$ can be read off from \eref{pfX-2}-\eref{FI-2}. 
Plugging in the $F=4$ version of the identity \eref{flavor-id} in \eref{pf-dual-1a} and performing the integrals over $u$ and $\tau_0$, the partition function $Z^{\wt{\CO}({\CT}_{\rm good})}$ can be rewritten as:
\begin{align} \label{pf-dual-2b}
Z^{\wt{\CO}({\CT}_{\rm good})} =  \int \, \prod^3_{\gamma'=1}\,\Big[d\vec \s^{\gamma'}\Big]\,  \prod^3_{a=1} d\tau_a\,\,& 
Z^{({\CT}_{\rm good})}_{\rm int} (\vec \s, \vec t, -m^1_f-\mu_1, -m^1_f-\mu_2, -m_3, -m_4) \nn \\
& \times Z^{\CT[3]}(\vec \tau, 0, -\vec m_f)\cdot Z^{\rm hyper}_{\rm 1-loop}(\tau_1, \vec \s^2, -\eta),
\end{align}
where $Z^{\CT[3]}$ is the partition function of the linear chain $\CT[3]$ as defined above, and 
$Z^{\rm hyper}_{\rm 1-loop}=\frac{1}{\ch{(\tau_1 -\tr \vec \s^2 + \eta)}}$ is the 1-loop 
contribution of an Abelian hypermultiplet with charges $(1,-2)$ under one of the $U(1)$ gauge nodes of $\CT[3]$ 
and the central $U(2)$ gauge node of ${\CT}_{\rm good}$ respectively. Therefore, one can read off the 
dual quiver from the matrix model of \eref{pf-dual-1b} and check that it reproduces the quiver $\CT'$.

\bibliography{cpn1-1}
\bibliographystyle{JHEP}

\end{document}